\documentstyle[12pt]{article}
\def\ie{{\em i.e.}}
\def\eg{{\em e.g.}}

\def\S{S\hskip-6pt/\hskip2pt}
\def\H{H\hskip-7.5pt/\hskip2pt}

\def\beq{\begin{equation}}
\def\eeq{\end{equation}}

\catcode`\@=11 % This allows us to modify PLAIN macros.
\def\coeff#1#2{{\textstyle{#1\over #2}}}
\def\bra#1{\left\langle #1\right|}
\def\ket#1{\left| #1\right\rangle}
\def\VEV#1{\left\langle #1\right\rangle}
\def\vev#1{\left\langle #1\right\rangle}
\def\lsim{\mathrel{\mathpalette\@versim<}}
\def\gsim{\mathrel{\mathpalette\@versim>}}
\def\@versim#1#2{\vcenter{\offinterlineskip
    \ialign{$\m@th#1\hfil##\hfil$\crcr#2\crcr\sim\crcr } }}
\def\etal{{\em et. al.}}
\def\JL{J. L. Lopez}
\def\DVN{D. V. Nanopoulos}

\def\t1{{\tilde 1}}

\def\GeV{\,{\rm GeV}}

\def\to{\rightarrow}

\def\Hz{\,{\rm Hz}}

\def\NPB#1#2#3{Nucl. Phys. B {\bf#1} (19#2) #3}
\def\PLB#1#2#3{Phys. Lett. B {\bf#1} (19#2) #3}

\def\PRL#1#2#3{Phys. Rev. Lett. {\bf#1} (19#2) #3}

\begin{document}
% TH format
\begin{flushright}
\baselineskip=13pt
ACT-08/95\\
CERN-TH/95-128\\
CTP-TAMU-22/95\\
{\tt hep-ph/9505374}
\end{flushright}
\begin{center}
{\Huge Theory of Brain Function, Quantum Mechanics and Superstrings$^\star$\\}
\vglue 0.75cm
{\large D.V. NANOPOULOS$^{\star\star}$\\}
\vglue 0.5cm
{ Center for Theoretical Physics, Department of Physics\\
Texas A\&M University, College Station, TX 77843--4242, USA;\\
and\\
Astroparticle Physics Group, Houston Advanced Research Center
(HARC),\\ The Mitchell Campus, The Woodlands, TX 77381, USA;\\
and\\
CERN Theory Division, 1211 Geneva 23, Switzerland\\}
\end{center}

\vglue 0.1cm
\begin{center}
\bf Abstract
\end{center}
\baselineskip=14pt
Recent developments/efforts to understand aspects of the brain function at the
{\em sub-neural} level are discussed. MicroTubules (MTs), protein polymers
constructing the cytoskeleton, participate in a wide variety of dynamical
processes in the cell. Of special interest to us is the MTs participation in
bioinformation processes such as {\em learning} and {\em memory}, by possessing
a well-known binary error-correcting code [$K_1(13,2^6,5)$] with 64 words.
In fact, MTs and DNA/RNA are {\em unique} cell structures  that possess a code
system. It seems that the MTs' code system is strongly related to a kind of
``{\em Mental Code}" in the following sense. The MTs' periodic paracrystalline
structure make them able to support a {\em superposition} of coherent quantum
states, as it has been recently conjectured by Hameroff and Penrose,
representing an {\em external} or {\em mental order}, for sufficient time
needed for {\em efficient} quantum computing. Then the quantum superposition
collapses spontaneously/dynamically through a new, string-derived mechanism
for collapse proposed recently by Ellis, Mavromatos, and myself.
%\medskip
\begin{flushleft}
\baselineskip=12pt
ACT-08/95\\
CERN-TH/95-128\\
CTP-TAMU-22/95\\
May 1995
\end{flushleft}

\hrule
\medskip
{\small\baselineskip=10pt\noindent
$^\star$ Extended version based on an invited talk presented at the ``{\em XV
Brazilian National Meeting on Particles and Fields}", Angra dos Reis, Brazil,
October 4-8, 1994, and on an invited talk at the ``{\em Physics without
frontiers Four Seas Conference}", Trieste, Italy, June 25-July 1, 1995.\\
$^{\star\star}$ E-mail: {\tt dimitri@phys.tamu.edu} or
{\tt nanopoud@cernvm.cern.ch}}

\newpage

\noindent  At the moment of collapse, organized {\em quantum exocytosis}
occurs, \ie, the simultaneous emission of neurotransmitter molecules by the
synaptic vesicles, embedded in the ``firing zone" of the presynaptic vesicular
grids. Since in the superposition of the quantum states only those participate
that are {\em related} to the ``initial signal", when collapse occurs, it only
enhances the probability for  ``firing" of the {\em relevant} neurotransmitter
molecules. That is how a ``{\em mental order}" may be translated into a ``{\em
physiological action}". Our equation for quantum collapse, tailored to the MT
system, predicts that it takes 10,000 neurons ${\cal O}(1\,{\rm sec})$ to
dynamically collapse, in other words to process and imprint information.
Different observations/experiments and various schools of thought are in
agreement with the above numbers concerning ``{\em conscious events}". If
indeed MTs, with their fine structure, vulnerable to our quantum collapse
mechanism may be considered as the {\em microsites of
consciousness}, then several, unexplained (at least to my knowledge) by
traditional neuroscience, properties of consciousness/awareness, get easily
explained, including ``{\em backward masking}", ``{\em referal backwards in
time}", etc. Furthermore, it is amusing to notice that the famous puzzle of why
the left (right) part of the brain coordinates the right (left) part of the
body, \ie, the signals travel maximal distance, is easily explained in our
picture. In order to have timely quantum collapse we need to excite as much
relevant material as possible, thus signals have to travel the maximal possible
distance. The {\em non-locality} in the cerebral cortex of neurons related to
particular missions, and the related {\em unitary sense of self} as well as
{\em non-deterministic free will} are consequences of the basic principles of
quantum mechanics, in sharp contrast to the ``sticks and balls" classical
approach of conventional neural networks. The proposed approach clearly belongs
to the {\em reductionist} school since quantum physics is an integrated part of
our physical world. It is highly amazing that string black-hole dynamics that
have led us to contemplate some modifications of standard quantum mechanics,
such that the quantum collapse becomes a detailed dynamical mechanism instead
of being an ``external" ad-hoc process, may find some application to some
quantum aspects of brain function. It looks like a big universality principle
is at work here, because both in the black hole and the brain we are struggling
with the way information is processed, imprinted, and retrieved.

\begin{quotation}
``...{\em the Astonishing Hypothesis -- that each of us is the behavior of a
vast, interacting set of neurons}."\\
\begin{flushright}
Francis Crick in\\
{\em The Astonishing Hypothesis}
\end{flushright}

\vspace{1cm}
``...{\em what will they think? -- What I tell them to think}."\\
\begin{flushright}
Orson Welles in\\
{\em Citizen Kane}
\end{flushright}
\end{quotation}
\newpage

\setcounter{page}{1}
\markright{}
\pagestyle{myheadings}
\pagenumbering{roman}
\section*{Prooimion}
Theory of brain function, quantum mechanics, and superstrings are three
fascinating topics, which at first look bear little, if any at all, relation
to each other. Trying to put them together in a cohesive way, as described
in this task, becomes a most demanding challenge and unique experience. The
main thrust of the present work is to put forward a, maybe, foolhardy attempt
at developing a new, general, but hopefully scientifically sound framework
of Brain Dynamics, based upon some recent developments, both in (sub)neural
science and in (non)critical string theory. I do understand that Microtubules
\cite{1,2} are not considered by all neuroscientists, to put it politely, as
the microsites of consciousnes, as has been recently conjectured by Hameroff
and Penrose \cite{3,4}. Also, I do know that, the {\em one interpretation}
of non-critical string theory, put forward by Ellis, Mavromatos, and myself
\cite{5,6}, which has led to {\em not} just an incremental change, but a
total rethinking of the Quantum Mechanics {\em doctrine(s)} from the ground
up, is not universally, to put it mildly, accepted. Leaving that aside, and
time will tell, the
emerging big picture ``{\em when microtubules meet density matrix mechanics}",
as the reader hopefully will be able to judge for her(him)self, is rather
astonishing. It looks like the modified quantum dynamics \cite{5,6} of
microtubules \cite{1,2} may indeed lead \cite{3,4} to a rather concise,
experimentally verifiable (presently and in the immediate future) theory of
brain function \cite{89}. Since this is a rather vast, multidisciplinary,
and multidimensional subject, I kept in mind that {\em potential} readers
may include (high-energy) physicists, biologists, biochemists, neuroscientists,
medical doctors, including psychiatrists, psychologists, and psychotherapists,
etc. Thus, I have tried my best to obey the ``{\em technical minimality}"
principle, and at the same time, to make it as self-contained and informative,
as possible, by not assuming that psychoanalysts know about ``{\em quantum
coherence}", or formal string theorists know about the Freudian
``{\em unconscious proper}", even if, in the latter case, they believe that
{\em they know everything, and so why bother}?!

A concrete, technically elaborated proposal materializing some of the general
ideas that I have tried to put forward here, has been worked out by Mavromatos
and myself \cite{89}, work that I strongly encourage the interested reader to
consult. I am fully aware of the rather speculative nature of the ideas
presented here {\em and} of the sometimes circumstancial looking experimental
evidence used to support them.  Nevertheless, the way that different
structures/mechanisms, from completely disconnected fields of knowledge,
{\em fit and bind} together to produce such a {\em coherent}, dynamical scheme
of the Brain function, makes it very hard to ignore the whole thing, by just
believing that it is all coincidental, and nothing more than a grand illusion!
It goes without saying that the responsibility for all views expressed here is
completely mine.

\newpage

{\baselineskip=17pt
\tableofcontents}
\newpage

\pagenumbering{arabic}

\section{Introduction}
\label{s1}

The brain is our most valuable asset. The workings of the brain enable us to
{\em think}, a fundamental function that, among other things, make us aware
of our own existence or self-aware: {\em cogito, ergo sum}. Our perceptions
of the universe, concerning its physical structure, form and function according
to the universal physical laws, emerge from processed-in-the-brain
representations of, hopefully, objective physical reality. Understanding the
way that the brain functions is the primordial prerequisite for a complete
physical understanding of the dynamic universe that we are part of.
Undoubtedly, the brain is a very complicated system and thus to understand its
function we need a coordinated effort involving several, if not all, branches
of natural sciences: biology, neuroscience, biochemistry, physics, information
theory/computer
science, medicine, pharmacology, etc. We may eventually need some
well-organized excursions to the realms of the science of mental life or
psychology, for some extra help. Alas, the compartmentalization of science
in our epoch, the highly technical jargon used in every field today, and the
endemic narrow mindness, expressed best by the dictum {\em scientific
conformity means intellectual stultification}, make the study of brain function
a titanic struggle. Nevertheless, we ought to try to figure out, as explicitly
as possible, as detailed as possible, and as predictive as possible, what are
the most fundamental brain constituents and how they interact, so that they
eventually
produce this miracle that is called brain function, or put it differently,
what makes the brain {\em tick}! This kind of {\em reductionist} approach
has turned out to be very successful in the past, both in biology and in
physics. The discovery of the double-helical structure of DNA, its
identification with the gene, and the subsequent breaking of the genetic code,
three bases for one aminoacid, in biology, as well as the discovery of
electroweak unification and its subsequent spontaneous breaking that led to the
Standard Model of the strong and electroweak interactions, in particle physics,
are glowing examples of applied reductionism. In the case of the brain
function, things are a bit more complicated and delicate extra care is needed,
because the {\em mind} pops into the picture and thus the workings of the
associated {\em Mental World} have to be addressed one way or another!

There are two extremes in handling the mental world problem:
\begin{itemize}
\item {\em Strong Artificial Intelligence} (SAI), purporting that the brain is
just a computer and the only thing we have to figure out is the algorithm.
\item {\em Cartesian or dualistic view}, assuming that brain and mind are
{\em two distinct entities}, in interaction with each other.
\end{itemize}
\begin{equation}
\begin{tabular}{ccc}
Brain&$\begin{array}{c}\longleftarrow\\ \longrightarrow\end{array}$&Mind\\
$\cap$&&$\cap$\\
(Attainable physical world)&&(Mental world)\\
$|||$&&$|||$\\
$W_1$&&$W_2$
\end{tabular}
\label{1}
\end{equation}
where the mental world contains perceptions, ideas, memories, feelings, acts
of volition, etc. I believe that both the above extremes are needlessly
exaggerated. Instead I would like to propose here a new {\em unified approach}
in which there is an ``effective" mental world {\em emerging} from the physical
world, but with distinct qualities
\begin{equation}
\begin{tabular}{ccccccc}
$W$&$\supset$&$W_1$&$\otimes$&$W_2$&$\longrightarrow$&$W_1$\\
$|||$&&&&$\downarrow$&$\uparrow$&\\
physical world&&&&causes &``collapse"&\\
\end{tabular}
\label{2}
\end{equation}
Hard-core {\em materialists} are very welcome to be {\em $W_2$-world ``blind"}
and just concentrate on the transition $W\to W_1$, in a kind of ``just the
facts ma'am" attitude! The present approach combines two new ideas/mechanisms
developed recently, one in biology/neuroscience and one in superstring theory:
\begin{description}
\item i. It has been suggested by Hameroff (for some time now) \cite{1} that
{\em MicroTubules} (MT), cytosceletal protein polymer paracrystalline
structures within the neurons \cite{1,2}, may be the fundamental units or
microsites where most of the brain function originates. Furthermore, Hameroff
and Penrose argued very recently \cite{3,4} that quantum effects may play a
central role in the MT functioning and they were {\em desperately} looking for
an explicit ``{\em collapse of the wavefunction}" mechanism, that would
validate their claims.
\item ii. It has been suggested by Ellis, Mavromatos and myself that, in
{\em one interpretation} of non-critical string theory, one gets naturally
modifications of Quantum Mechanics, leading among other things to a new
explicit ``{\em collapse of the wavefunction}" {\em mechanism} and {\em a
microscopic arrow of time} \cite{5,6}.
\end{description}
The present proposal is to combine (i) and (ii).

It is remarkable that the string-derived collapse of the wavefunction mechanism
fits ``hand-in-glove" to the MT hypothesis. Thus, by complementing (i) with
(ii), a rather detailed and spelled out scenario of brain function {\em
emerges}. Namely, because the stringy collapse of the wavefunction is due to
the truncation of the unattainable global degrees of freedom, the scheme
depicted
as Eq.~(\ref{2}) naturally emerges. $W_2$ should be identified with the
physical ``global state space" {\em isomorphic} to the ``effective" mental
world in the following sense: the {\em collapse of the wavefunction} is what
causes the system to ``decide" its course of action, thus being completely
identifiable with the Jamesian\footnote{William James (1842--1910), the father
of American (physiological) psychology, observed that consciousness is not a
thing or a substance, but rather a {\em process} \cite{7}.} view of {\em
consciousness}, as a {\em selecting agency}. The $W_2$ global states are the
agents of collapse!

	In this approach, the ``collapse of the wavefunction" will result in
well-coordinated, organized {\em exocytosis}, \ie, the {\em simultaneous}
emission of neurotransmitter molecules by the synaptic vesicles, embedded
in the ``firing zone" of the presynaptic vesicular grids \cite{8}. From then
on, standard neurophysiology applies, \eg, setting the motor in action, etc.
Clearly, the strong correlation between the ``effective" mental world and the
``collapse of the wavefunction" (through the ``global state space") makes it
clear how a mental intention (\eg, I wish to bend my index finger) is {\em
physically} and {\em causally} related to the motor action (\eg, bending my
index finger). Eventually, we may even be able to develop a ``{\em mental
code}", \ie, a dictionary that would translate feelings, intentions, etc
directly into specific neurochemical states charting out detailed
neurotransmitter molecule topologies. Actually, even if this statement sounds
extremely far-fetched speculation and off-the-wall, the {\em universality}
of the ``effective" mental world for all humans, with of course all its
diversity, cries out for an objective mapping between mental and specific
neurochemical processes. A good analogy here is the ``{\em genetic code}",
a well-tabulated dictionary between ``base" sequencing in DNA and aminoacid,
thus protein, production on the ribosomes \cite{9}. Proteins, of course, are
our basic building blocks that are responsible for the way we look, move, etc.

In section~\ref{s2}, I will discuss {\em Brain Mechanics}, \ie, some very
general arguments of what the brain is supposed to do and how it does it, while
in section~\ref{s3}, I will present some elements of {\em Quantum Mechanics},
useful in our subsequent discussions. Section~\ref{s4} provides a view of some
{\em Brain morphology and modeling}, based upon classical notions and some
criticism and problems they are facing. Section~\ref{s5} provides some elements
of {\em string}-derived {\em density matrix mechanics}, an extension of
orthodox Quantum Mechanics, while sections~\ref{s6} and \ref{s7} discuss the
{\em biochemical} and
{\em physical profile} of {\em Microtubules} (MT) respectively, and their
potentially important role in brain function. Section~\ref{s8} shows how {\em
microtubule} dynamics, in a stringy-derived {\em density matrix mechanics}
framework, may yield a {\em unified model} of Brain and Mind, a {\em quantum
theory of brain function}, while the final Section~\ref{s9} covers the emerging
{\em quantum psychophysics}.

\section{Brain Mechanics}
\label{s2}
The brain is a rather complicated physical system in constant interaction with
the external world or environment. Very generically and in {\em grosso modo}
the brain functions as follows:
\begin{description}
\item (i) Imagine that the brain is in some state $\ket{A}$, when some {\em
external stimulus} is applied, for some given period of time, then
\item (ii) after the removal of the external stimulus, the brain is in some
state $\ket{B}$, which in principle should have in some way {\em coded} (or
recorded) the ``message" that was carried by the external stimulus, in such a
way that
\item (iii) ``later" it is possible to {\em retrieve} (or recall) the message
directly from the state $\ket{B}$, keeping in mind that
\item (iv) it is possible that the brain has not necessarily gone directly
from $\ket{A}$ to $\ket{B}$, but many intermediate steps may have occurred:
$\ket{A}\to\ket{A_1}\to\ket{A_2}\to\cdots\to\ket{B}$, \ie, the {\em
information} (or message) has been {\em processed} in the brain before it
was recorded.
\end{description}

There are some fundamental properties that characterize successful brain
function, namely: {\em long-term stability} and {\em non-locality}, as strongly
suggested by the plethora of experimental data. While the need for long-term
stability is rather obvious, non-locality, \ie, coherent neuronic activity
at spatially remote cortical locations, makes the classical
treatment of the brain function rather questionable. At the same time,
non-locality is strongly suggestive of {\em quantum treatment} \cite{10,11,12}.
Since we are
concerned here clearly with macroscopic states, and at the same time we need
to invoke quantum treatment, we have to look at the so-called {\em Macroscopic
Quantum States} (MQS), which are abundant in the quantum world.
Superconductivity, superfluidity, magnetization, etc are typical examples of
MQS with very specific characteristics:
\begin{description}
\item (i) For special ``structures" and ``conditions",
\item (ii) a critical degree of coherence may be achieved that leads to an
\item (iii) {\em ordered state}, that is highly stable.
\end{description}
Consider for example {\em Magnetization}: the special ``structures" are the
{\em Weiss regions}, small regions in a ferromagnet within which all electron
spins are polarized in a specific direction. Though, because there are many
small regions and polarizations, on the average there is no magnetization
visible in the ferromagnet. If we now apply a sufficiently strong magnetic
field $\vec B$ or we decrease sufficiently the temperature (below the P. Curie
point), \ie, the special ``conditions", the ferromagnet exhibits magnetization
because now all electron spins in the whole {\em macroscopic} crystal, are
polarized in the same direction, strongly correlated with each other, thus
leading to a highly stable {\em macroscopic coherent} (or quantum) {\em state},
the {\em ordered state}.

In a more physical language, the transition from an unordered state (\eg, many
Weiss-regions) to an ordered state (\eg, magnetization) is called a phase
transition. The value(s) of the crucial parameter(s) (\eg, the magnetic field
$\vec B$ or temperature $T$) at the transition point characterize the phase
transition and define the {\em critical point} (\eg, Curie temperature).
It should be apparent that an ordered state contains some {\em information}
(\eg, all electron spins polarized in the same direction) than the unordered
state (electron spins randomly polarized). On the other hand, the unordered
state is more {\em symmetric} (randomly distributed electron spins are
rotationally
invariant, \ie, there is no {\em preferred direction}), while the ordered
state exhibits less symmetry (polarized electron spins have chosen
spontaneously a specific direction, thus breaking the rotational symmetry).
Thus, ordered states are the net result of {\em spontaneous symmetry breaking}
that triggers the phase transition. There are certain characteristics of phase
transitions very useful for our subsequent discussions
\begin{description}
\item (i) {\em Universality}: many, qualitatively and quantitatively different,
systems can be described by the same phase transition.
\item (ii) {\em Attractor}: by varying suitably the system parameters, they can
be brought close to
their {\em critical values}, so as to cause a phase transition. It is not
necessary to be infinitesimally close to the critical point. The critical point
acts as an {\em attractor} for anything in its environment. In other words, we
don't really need a {\em fine-tuning} of our system parameters to reach an
ordered state.
\item (iii) {\em Evolution equations}: All the basic properties of phase
transitions
(including (i) and (ii) above) can be encoded in a set of evolution equations
called renormalization group equations (RGEs). They describe deviations (and
approach) from (to) criticality, as well as other characteristics of phase
transitions \cite{13}.
\end{description}

Macroscopic coherent (or quantum) states, or ordered states have some highly
exclusive characteristics:
\begin{description}
\item (i) {\em Long-range/term stability}: highly stable, long-range
correlations
between the fundamental elements are maintained by wave-like, self-propagating
excitation loops (\eg: phonons, spin-waves, magnons, etc.) that {\em Regulate}
the behavior of the ``other" fundamental elements and {\em Feedback} to the
original fundamental element that caused the ``disturbance". I will call this
the {\em R+F} property of MQS.
\item (ii) {\em Non-locality}: clearly MQS, as its very nature indicates may go
beyond microscopic locality.
\item (iii) {\em Emergence}: MQS have {\em new properties} that are {\em not}
present
at the fundamental elements level. The new properties characterize states at a
{\em hierarchical level above} the level where the fundamental interactions
among the fundamental constituents apply. For example, superconductivity is
a new property/phenomenon, {\ie, emerging} from a {\em collective treatment}
of electrons under special circumstances, while of course each electron follows
at the fundamental level the laws of quantum electrodynamics.
\end{description}

Let us use now the physical language of MQS and phase transitions to describe
{\em by analogy}, for the time being, the basic functions of the Brain:
\begin{description}
\item (I) {\em Uncoded Brain}: random signals, unattended perception are
some of the characteristics of this case. It corresponds to the random
polarizations in the many, small Weiss regions of the ferromagnet.
\item (II) {\em Learning}: An {\em external stimulus} is applied, say for a
few seconds, that ``straightens out" or ``puts an order" to the random neuronic
signals so that they are able to represent some {\em coherent piece} of
information. It corresponds, in the case of the ferromagnet, to applying for
some time an external magnetic field $\vec B$ or lowering the temperature below
the Curie point. They cause the {\em breaking} of the multi-domain small
structures
with their random polarizations,and thus they lead to the ordered state, where
all electron spins, throughout the whole ferromagnet, are strongly correlated
to all point in the {\em same direction}. We are talking about a phase
transition or, in the spirit of the previous discussion, a spontaneous breaking
of some symmetry. Clearly, it depends on the nature of the external
stimulus with which specific fundamental elements will interact and set them
``straight", so that a corresponding MQS, or ordered state, is created.
Realistically, in order to be able to {\em encode} all qualitatively different
signals and create a coherent {\em unitary
sense of self}, a tremendous number of qualitatively different ordered states
is needed, \ie, practically an infinite number of qualitatively different
spontaneously broken symmetries. Furthermore, these symmetries should be
accompanied by a set of {\em selection rules}, thus providing a physical {\em
filter} against undesirable, irrelevant ``stray" signals. A very tall order
indeed, if one recalls the fact that, until now, the only ``known" (observable)
spontaneously broken symmetry, at the fundamental level, is the one describing
the electroweak interactions. Just {\em one}, which is kind of short with
respect to the desirable {\em infinity} of spontaneously broken symmetries!
We will see later how string theory takes care of this problem.
\item (III) {\em Coded Brain} or {\em Memory}: the resulting, highly stable,
coherent ``{\em firing}" of a bunch of involved neurons, not necessarily
localized, corresponding, in the case of the ferromagnet, to the stability
and macroscopic nature (including non-locality) of the {\em emerging}
magnetization ({\em ordered}) state. Such a kind of naturally organized,
coherent neuron firing, not necessarily localized, may provide the solution to
the so-called ``{\bf binding problem}". More later.
\item (IV) {\em Recall Process}: In this picture, a {\em replication} {\em
weak} signal, {\em sufficiently} resembling the {\em learning} signal, may
excite {\em momentarily} the ordered state, but, thanks to its R+F property, it
will relax back to its previous form. It is this,
ordered-state$\to$excitation$\to$ordered-state process that make us aware of
recalling something, \ie, we ``feel" it! It corresponds in the case of the
ferromagnet, to apply a {\em weak} magnetic
field $\vec B'$, not necessarily exactly parallel to the original $\vec B$,
which will force the electron spins to oscillate, {\em momentarily}, before
they relax back to their equilibrium, \ie, we recover the ordered state,
thanks of course to the R+F property of MQS. It should be stressed that it is
not necessary for the replication signal to be exactly identical to the
learning signal in order to recall full information, thanks to the {\em
attractor} property of the phase transitions, discussed above. In
phase-transition language, the recall memory process corresponds to the act
of an {\em irrelevant operator}. It should not escape our attention  that the
endemic, in the framework of phase transitions, R+F
and {\em attractor} properties fascilitate tremendously the retrieving of
information, without the need of {\em complete} identity of the replication
and learning signal. Otherwise, it would take extraneous {\em fine-tuning},
which here translates to very long time periods, in order to retrieve
information. Imagine what would happen if we need to see all the details
of a fast approaching, hungry lion, including say the length and shape of its
claws, before we run up a tree! Not very practical indeed.
\end{description}

The above presented generic picture for the brain function may sound plausible
and promising. But, is there any ``{\em experimental}" evidence for its
support? The answer is {\em yes}. The main observational tool is the
ElectroEncephaloGram (EEG). It is usually assumed that the EEG waveforms
emerge from the summation of local neuron firings, but things are a bit more
complicated. One would expect that {\em asynchronous} firing of randomly
distributed
neurons would produce a zero net effect on the scalp electrodes. By studying
electric potentials evoked during sensory stimulation and during learning
trials, E. R. John has been able to show that these evoked potentials arise
from the firing of large and disperse neural groups and that they are radically
different from those obtained by the spontaneous random cortical activity
\cite{14}. {\em Temporal rearrangement} within the neural groups characterizes
the externally evoked potentials. Furthermore, Sayers \etal \cite{15},
presented independent
evidence strengthening the temporal rearrangement case, by studying EEG {\em
phase coherence}. Frequency components of the EEG spectrum obtained during
spontaneous cortical activity show a random configuration of phase relations,
which shifts to a distinct pattern of {\em phase coherence} immediately
following sensory stimulation. Amazingly enough, imposing the {\em phase
characteristics} of the evoked potential on the spontaneous waveform, we can
reproduce the characteristic shape of the observed evoked waveform \cite{15}.
These findings support E. R. John's \cite{14} case for temporal rearrangement,
while at the same time it falsifies the kind of classical expectation that the
EEG arises from the  summation of neural firings, which would imply that just
the {\em amplitude characteristics} is the only difference between spontaneous
and evoked waveforms. Clearly, it seems that the external stimulus does not
just add energy to the brain, but it organizes it in a {\em coherent way}, in a
similar fashion that an external field $\vec B$ acts on a ferromagnet! It seems
that the analogy between brain function and critical phenomena dynamics may be
quite useful and fruitful.

In the {\em unified approach} suggested here (see Eq.~(\ref{2})) the
``effective" mental world ($W_2$) is {\em actively} interacting with the
emerging  MQS, and thus through
the R+F property of the MQS {\em and} the subsequent triggering by $W_2$ of the
collapse of the MQS, it provides the solution to the age-old problem of how
intentional/emotional acts are strongly correlated to body acts, as explained
in the Introduction. It should be stressed that {\em emergence} here has a
multi-valued meaning: it encompasses the natural (Darwinian \cite{16})
evolution and selection, the development of brain in specific subjects and
eventually the ``conscious" moment under consideration.

\section{Quantum Mechanics}
\label{s3}

The physical principles that govern the microworld, as provided by Quantum
Mechanics (QM), are {\em profoundly} different from the ones that the
macrocosmos obeys. The ``microworld" here denotes anything at and below
the molecular level: molecules, atoms, electrons, nuclei, protons, neutrons,
quarks. As Linus Pauling taught us, chemistry is nothing else but applied
quantum mechanics at the atomic and molecular level. Interestingly enough,
Molecular Biology holds a very intriguing position between
the macro and micro worlds in the following sense: {\em ab initio}, Molecular
Biology is concerned with the structure and function of the cell \cite{9},
which is
mainly composed of {\em macromolecular structures} (DNA, RNA, proteins, ...)
and as such, {\em most of the time} and for many purposes, are sufficiently
and accurately described by classical physics. Nevertheless, we should not be
carried away and discard QM from the picture by interpreting {\em most of the
times} as implying {\em at all times}! After all, as Watson and Crick \cite{17}
taught us, the double helical structure of DNA, which is the source of DNA's
fundamental genetic properties is due to the quantum mechanical H-bonds between
purines (A,G) and pyrimidines (T,C): always a double H-bond for A=T and a
triple H-bond for $\rm G\equiv C$. It is in the {\em stability} and {\em
universality} of these H-bonds, as verified experimentally by
Chargaff \cite{18}, that the secret of the genetic code lies! Since my central
thesis here, as emphasized earlier, is that quantum mechanics plays {\em also}
a very fundamental role in the emergence
of the mental world from the physical world, \ie, in the brain-mind relation,
I will discuss very briefly some elements of QM, that I will need later.

The central dogma of Quantum Mechanics is the {\em particle-wave duality}: it
depends on the particular circumstances if a {\em quantum state} is going to
express itself as a particle or as a wave \cite{19}. Consider for example a
particle travelling in spacetime. Its quantum state is described by a {\em
wavefunction} $\Psi(\vec x,t)$ obeying a Schr\"odinger-type equation of the
form
\begin{equation}
i\hbar{\partial\Psi\over\partial t}=H\Psi
\label{3}
\end{equation}
where $\hbar$ ($\equiv1$ in natural units) is the Planck constant, and
$H$ is a system-dependent operator, called the {\em Hamiltonian}
of the system. It provides the {\em unitary}, {\em time-evolution} of the
system, and with eigenvalues identifiable with the different energy levels
of the system. A fundamental, and {\em immensely crucial} for us here, property
of the quantum equation (\ref{3}) is its {\em linearity}. Imagine that
$\Psi_1,\Psi_2,\ldots,\Psi_n$ are different solutions of (\ref{3}), then
clearly the {\em linear superposition}
\begin{equation}
\Psi=\sum_{i=1}^n c_i(t)\Psi_i\ ,
\label{4}
\end{equation}
with $c_i$ arbitrary complex numbers, is also a solution of (\ref{3}). This is
the mathematical statement of {\em Quantum Superposition}. Let us discuss next
its physical meaning. Suppose that we would like to describe quantum
mechanically the following ``history" of a particle, say an electron: it
starts at some initial point around $(\vec x_0,t_0)$, it {\em goes through}
a wall that contains $n$ slits, say $1,2,\ldots,n$, {\em without us
knowing which specific one}, and it ends up at some final point around
$(\vec x_f,t_f)$. Let $\Psi_1,\Psi_2,\ldots,\Psi_n$ denote the wavefunctions
of the electron, referring to the case that the electron went through the
slit $1,2,\ldots,n$, respectively. Since we don't know the specific slit
that the electron went through, we are {\em obliged} to take as the
wavefunction of the electron, a linear superposition of
$\Psi_1,\Psi_2,\ldots,\Psi_n$, \ie, (\ref{4}). The physical meaning then of the
$c_i$'s becomes clear: $|c_i|^2$ is the {\em probability} that the electron
went through the slit $i$, and thus $c_i$ is referred to as the {\em the
probability amplitude}. Notice that conservation of probability entails that
at any time $t$
\begin{equation}
\sum_{i=1}^n|c_i(t)|^2=1
\label{5}
\end{equation}
The {\em probability density} to find the electron at some specific point
$(\vec x_a,t_a)$, after it has passed through the slits and before it ends
up at $(\vec x_f,t_f)$ is given by
\begin{equation}
|\Psi(\vec x_a,t_a)|^2=|\sum_{i=1}^n c_i\Psi_i|^2\ .
\label{6}
\end{equation}
Clearly, this is a standard {\em wave-like behavior} and (\ref{4}) may be
interpreted as describing a quantum state evolving in a {\em coherent way},
or obeying the fundamental quantum mechanical principle of {\em quantum
coherence}, the physical meaning of {\em linear superposition}. Imagine now,
that we would like to find out through which specific slit the electron
went through. Then, we have to make a ``{\em measurement}" or ``{\em
observation}", \ie, to concentrate on those aspects of the quantum system that
can be {\em simultaneously magnified} to the classical level, and from which
the system must then choose. In other words, we have to {\em disturb} the
system (electron in our example) with the {\em magnifying device}, which
results in {\em de-coherence}, thus (\ref{6}) is replaced by
\begin{equation}
\begin{tabular}{ccc}
&``measurement"&\\
$|\Psi(\vec
x_a,t_a)|^2$&$\downarrow\longrightarrow$&$\sum_{i=1}^n|c_i|^2|\Psi_i|^2$\\
&``collapse"&
\end{tabular}
\label{7}
\end{equation}
In other words, we get classical probabilities, highly reminiscent of a
standard {\em particle-like behavior}. The ``measurement"/``observation"
process has caused decoherence of the wavefunction and thus led to its
{\em collapse} to a {\em specific state}. Here are then, in a nutshell, our
basic quantum mechanical rules, that constitute {\em quantum reality}:
\begin{description}
\item (i) A quantum system can, in principle, be in many states {\em
simultaneously} $(\Psi_1,\Psi_2,\ldots,\Psi_n)$ and its quantum state
$\Psi=\sum_{i=1}^n c_i\Psi_i$, a {\em pure state}, evolves {\em coherently}
and according to the quantum equation (\ref{3}), {\em as long as} we don't
{\em disturb} it. This is {\em quantum linear superposition} or {\em quantum
parallelism}, leading to wave-like behavior.
\item (ii) A ``measurement"/``observation" forces the quantum state $\Psi$
to {\em decide} what it wants to be, with probability $|c_i|^2$ that the
$\Psi$ quantum state will turn out to be the $i$-th state (described by
$\Psi_i$), {\em after} the ``measurement"/``observation". This is the
``{\em collapse of the wavefunction}", leading to classical particle-like
behavior.
\end{description}
Incidentally, the famous Heisenberg {\em uncertainty principle} \cite{20} is
nothing else but a quantitative expression of our intuitive statement above
that a  ``measurement"/``observation" disturbs the system in an {\em
uncontrollable way}, entailing always uncertainties in the outcome, \eg,
\begin{equation}
\Delta x\cdot\Delta p\ge\hbar\ .
\label{8}
\end{equation}
Clearly, (\ref{8}) indicates the fact that it is impossible to ``measure"
{\em simultaneously}, at a desirable level, both the position and the momentum
of a particle. Notice that this is a {\em fundamental principle}, and has
nothing to do with the potentially difficult and practical problems that
face experimentalists. Whatever she does, she cannot beat the uncertainty
principle.

The endemic, in the Quantum World, wave-particle duality is responsible for
the necessity of the two-step approach to quantum dynamics discussed above.
This kind of approach is very different from the deterministic approach of
classical dynamics and, in a way, it creates a {\em schism} in our
understanding of the Universe. There is the classical world and there is the
quantum world, each following its own laws which, frankly, do not seem to have
much common ground. It may even, sometimes, lead to some embarrassments
\cite{11}, like \eg, the {\em Schr\"odinger's cat} paradox, a peculiar
situation where a quantum event may
oblige us to treat a cat as 50\% alive and 50\% dead! Furthermore, in the
passage from the quantum to the classical world it is not clear at all who
is there to decide that we crossed the quantum-classical border!

This {\em dualistic} view of the world (classical versus quantum) is
reminiscent
of the ancient needs for heavenly-terrestial dynamics, abolished by Galileo
and Newton for universal dynamics, or for space and time dynamics, abolished by
Einstein for spacetime dynamics, or for electromagnetic and weak interactions,
abolished recently for electroweak interactions. It looks to me that this
classical versus quantum dualistic view of the world {\em cries out}, once
more, for a {\em unified approach} which for many practical purposes would {\em
effectively} look like two separate worlds (classical and quantum). Any
resemblance with the unified approach I discussed in the Introduction for
the brain versus mind problem is {\em not} accidental! A {\em unified approach}
for classical {\em and} quantum dynamics will be attempted in section~\ref{s5},
but
let me prepare the ground here by generalizing a bit the notion of quantum
state and the likes. What we are really after is some kind of formalism that
enables us to express, at least in principle, the two-step process of quantum
dynamics in a more uniform language. Let us represent a given quantum state
$\alpha$ by a state vector $\ket{\alpha}$, while $\bra{\alpha}$ denotes the
complex
conjugate state vector $\alpha^*$, and let us assume that this state vector
has ``length" one: $\vev{\alpha|\alpha}=1$. Consider now a complete set of
orthonormal state vectors $\ket{i}$: $\vev{j|i}=\delta_{ij}$, implying that
any {\em pure state} can be written as $\ket{\Psi^a}=\sum_i c^a_i\ket{i}$,
with $c_i$ complex numbers obeying the conservation of probability condition
$\sum_i|c_i|^2=1$ (see (\ref{4}),(\ref{5})). Then the {\em scalar product}
$\vev{\Psi^b|\Psi^a}=\sum_i c^{*b}_ic^a_i$ expresses the {\em probability
amplitude} that starting with the state vector $\ket{\Psi^a}$ we end up
in the state $\ket{\Psi^b}$. Actually, we can consider all the tensor products
$\ket{\Psi^k}\bra{\Psi^l}$ with the understanding that
$\vev{\Psi^l|\Psi^k}={\rm Tr}(\ket{\Psi^k}\bra{\Psi^l})=\sum_i c^{l*}_ic^k_i$.
It is very convenient to introduce the notion of the {\em density matrix}
$\rho\equiv\ket{\Psi}\bra{\Psi}$ with matrix elements
$\rho_{ij}=c_ic^*_j\ket{i}\bra{j}$ and such that ${\rm Tr}\rho={\rm
Tr}(\ket{\Psi}\bra{\Psi})=\vev{\Psi|\Psi}=\sum_ic_ic^*_i=1$, \ie, the
conservation of probability condition. Notice that, in the case of a {\em pure
state}, the description of a
quantum system by the state vector $\ket{\Psi}$ or by the density
matrix $\rho$ is {\em equivalent}. For example, the measurable quantities
$\vev{\Psi|A|\Psi}$ correspond to ${\rm Tr}(\rho A)$, with $A$ denoting the
quantum operator representing the ``measurable quantity", etc. The quantum
equation
(\ref{3}) becomes in the density matrix approach
\begin{equation}
\dot\rho\equiv{\partial\rho\over\partial t}={i\over\hbar}\,[\rho,H]\ ,
\label{9}
\end{equation}
which is nothing else but the {\em quantum analogue} of the classical
statistical mechanics {\em Liouville equation}, describing the time evolution
of the phase-space  density function. The great advantage of the density
matrix approach is its ability to describe not only {\em pure states}, but also
{\em mixed states}. Imagine that for {\em practical reasons} it is impossible
to know the exact {\em pure state} of our quantum system, \ie, we only know
that we have a combination of different {\em pure states} $\ket{\alpha}=
\sum_i c^\alpha_i\ket{i}$, $\alpha=1,2,\ldots$, each with {\em classical
probability} $p_i$. Clearly, in this case we cannot use the quantum equations
(\ref{3}) or (\ref{9}) because it is only applied for {\em single} pure states,
but we can still use the density matrix approach. Write the density matrix
of the system as a {\em mixed state}
\begin{equation}
\rho=\sum_a p_a \ket{a}\bra{a}\ ,
\label{10}
\end{equation}
then the probability that a ``measurement"/``observation" will find our
system in some pure state $\ket{\Psi}=\sum_i d_i\ket{i}$ is given by
\begin{equation}
P_\Psi={\rm Tr}(\rho\ket{\Psi}\bra{\Psi})=\sum_a p_a|\vev{\Psi|a}|^2\ ,
\label{11}
\end{equation}
which is a sum of products of classical and quantum probabilities! Notice
that in the case of a single {\em pure state}, say $\ket{b}$, all
$p_{a\not=b}=0$ and $p_b=1$ in (\ref{10}), and a ``measurement"/``observation"
causes the ``collapse of the wavefunction" $\ket{b}$, that implies turning a
{\em pure state} $\rho=\ket{b}\bra{b}$ into a {\em mixed state}
$\rho=\sum_i|c^b_i|^2\ket{i}\bra{i}$, which is nothing else but (\ref{7})!
Of course, in the case of a ``measurement"/``observation" we {\em open} the
system under consideration, and clearly (\ref{9}) needs modification, \ie,
addition of extra terms that represent the ``disturbances". On the other hand,
since the ``collapse of the wavefunction" implies loss of quantum coherence,
there is no way to use a wave equation like (\ref{3}), or possible
modifications, to represent the ``disturbances". The notion of description of
a quantum state by state vectors or wavefunctions really gives in to the
density matrix approach, thus the correct approach for a {\em unification}
of classical and quantum dynamics. Usually, when we deal with realistic
quantum systems, composed of different independent or loosely interacting
parts, it helps to express the quantum state of the system as the product of
different independent components. Imagine, for example, a particle called
$\pi^0$ decaying into two photons $\gamma_1$ and $\gamma_2$. Since $\pi^0$
has no spin, the most general description of the system of two photons is
given by
\begin{equation}
\ket{\Psi}=c_1\ket{\gamma_1}_+\ket{\gamma_2}_-
+c_2\ket{\gamma_1}_-\ket{\gamma_2}_+\ ,
\label{12}
\end{equation}
where the subscripts indicate the polarizations of the two photons, always
opposite, such that the whole system has angular momentum zero, corresponding
to the spinless $\pi^0$. Imagine that a ``measurement"/``observation" is done
on the system by measuring say the polarization of $\gamma_2$ and found to
correspond to the $-$ one. After
the ``measurement"/``observation" we know that $c_1=1$ and $c_2=0$, thus
without ``measuring" the polarization of $\gamma_1$, we know it is the $+$ one.
Einstein found it very disturbing, that some ``measurement" on one part of
the system has an ``{\em instantaneous}" effect on some other {\em distant}
part. Sometimes this is referred to as the {\em Einstein-Podolsky-Rosen} (EPR)
puzzle \cite{21,11}, and it is a very clear proof of the {\em non-local} nature
of the quantum world! Experiments done in the mid-80's have confirmed
\cite{22}, beyond any shadow of doubt, the {\em non-local nature} of quantum
mechanics, and the failure of classical spacetime notions to describe quantum
reality.

The Macroscopic Quantum States (MQS), mentioned in section~\ref{s2}, correspond
here
to something like
\begin{equation}
\ket{\Psi}=\sum_i c_i \ket{1}_i\ket{2}_i\cdots\ket{N}_i\ ,
\label{13}
\end{equation}
where $\ket{k}_i$ refers to the quantum state of the $k$-th fundamental
constituent in the $i$-th macroscopic quantum state. Of course, for a MQS $N$
is ${\cal O}(N_{\rm Avogadro}\approx6\times10^{23})$, a rather larger number
and in several occasions the index $i$ can also run into large numbers. For
example, in the case of a ferromagnet, the {\em ordered state} would be
described
by (\ref{13}), and if $\ket{k}_i$ indicates the spin polarization of the $k$-th
electron, then {\em only} one $c_i\not=0$. While in the case of {\em
quasicrystals}, describable also by (\ref{13}), not only is $N$ large
(${\cal O}(N_{\rm Avogadro})$), but {\em also} the linear combinations may
involve a huge number of {\em alternatives}, \ie, the $i$-index can be also
large. {\em Quasicrystals} are rather intriguing physical structures that may
need quantum mechanics in an essential way for their understanding. According
to Penrose \cite{11}, the quasicrystal assembly cannot be reasonably achieved
by the local adding of atoms one at a time, in accordance with the {\em
classical}
picture of crystal growth, but instead there must be a non-local essentially
quantum mechanical ingredient to their assembly. Instead of having atoms coming
individually and attaching themselves at a continually moving growth line
(standard classical crystal growth), one must consider using something like
(\ref{13}), an evolving quantum linear superposition of many different
alternative arrangements of attaching atoms. There is not one {\em single}
thing that happens, many alternative atomic arrangements must coexist! Some
of these linearly superposed alternatives will grow to very large
conglomerates, and at certain point the ``collapse of the wavefunction" will
occur and thus more specific arrangements will be singled out, and so on, until
a good-sized quasicrystal is formed. But why is Nature employing such an
intriguing mechanism? Penrose claims \cite{11} that maybe ``energetics" is the
answer.
Usually, crystalline configurations are configurations of {\em lowest energy},
and the correct arrangement of atoms can be discovered simply by adding one
atom at a time, and solving its own minimizing problem, etc. In quasicrystal
growth, finding the lowest energy state is a very complicated and difficult
problem, because it involves a large number of atoms {\em at once}, and thus,
we have a {\em global}, {\em non-local} problem to solve. Clearly, a quantum
mechanical description, a la (\ref{13}), seems appropriate where many different
combined arrangements of atoms are being ``tried" simultaneously, and
eventually collapsing, through physical environment tangling, to the
``energetically" and ``enviromentally" appropriate arrangements, the {\em
observable} quasicrystal.

It should be stressed that the QM rules have been in place and in successful
use for about 70 years now, and have led to a most deep understanding of the
microworld. Nevertheless, the fundamental mechanism triggering the ``collapse
of the wavefunction" has escaped us, until I believe recently, when string
theory enabled us to put a definite proposal on the table, to be discussed
in section~\ref{s5}. Intriguingly enough, Molecular Biology and Neurobiology in
particular, lies just in the classical-quantum interface and thus very
interesting phenomena may occur. So, let us turn our attention now to the
detailed structure of the brain.

\section{Brain Morphology and Modeling}
\label{s4}
The human brain is the most complicated object, as far as we know, in the
Universe. At a first look, it is amazing that this seemingly amorphous mass
is capable of executing all these miraculous operations that control our
actions and make us aware of the world around. A closer look though points
to a rather recursively hierarchical structure and a very elaborate
organization \cite{23,11}. An average brain weighs about 1.3 kg, and it is made
of: $\sim77\%$ water, $\sim10\%$ protein, $\sim10\%$ fat, $\sim1\%$
carbohydrates,
$\sim0.01\%$ DNA/RNA, and the rest other stuff. The largest part of the human
brain, the {\em cerebrum}, is found on the top and is divided down the middle
into left and right cerebral hemispheres, and front and back into {\em
frontal}, {\em parietal}, {\em temporal}, and {\em occipital lobes}. Further
down, and at the back lies a rather smaller, spherical portion of the brain,
the {\em cerebellum}, and deep inside lie a number of complicated structures
like the thalamus, hypothalamus, hippocampus, etc. It seems that what make
humans more advanced than other animals is not only the largeness of the
cerebrum, but also its proportion of brain as a whole, the largest in the
animal kingdom!

Both the cerebrum and the cerebellum have comparatively thin outer surface
layers of grey matter and larger inner regions of white matter. The grey
regions constitute what is known as the {\em cerebral cortex} and the
{\em cerebellar cortex}. It is in the grey matter where various kinds of
computational tasks seem to be performed, while the white matter consists
of long nerve fibers (axons) carrying signals from one part of the brain to
another. It is the {\em cerebral cortex} that is central to the higher brain
functions, speech, thought, complex movement patterns, etc. On the other hand,
the {\em cerebellum} seems to be more of an ``automaton". It has to do more
with precise coordination and control of the body, and with skills that have
become ``second nature". Cerebellum actions seem almost to take place by
themselves, without thinking about them. They are very similar to the
unconscious reflex actions, \eg, reaction to pinching, which may not be
mediated by the brain, but by the upper part of the spinal column. Thus, it
seems highly likely that the phenomena of consciousness, that we are mainly
concerned here, have much more to do with the cerebrum than with the cerebellum
or the spinal cord. So, from now on, we will concentrate on the cerebral
cortex.

Various parts of the cerebral cortex are associated with very specific
functions. We distinguish several regions. The {\em visual cortex}, a
region in the occipital lobe at the back of the brain, is responsible for
the reception and interpretation of vision. The {\em auditory cortex}, in the
temporal lobe, deals mainly with analysis of sound, while the {\em olfactory
cortex}, in the frontal lobe, deals with smell. The {\em somatosensory cortex},
just {\em behind} the division between frontal and parietal lobes, has to do
with the sensations of touch. There is a very specific mapping between the
various parts of the surface of the body and the regions of the somatosensory
cortex. In addition, just in {\em front} of the division between the frontal
and parietal lobes, in the frontal lobe, there is the {\em motor cortex}.
The {\em motor cortex} activates the movement of different parts of the body
and, again here, there is a very specific mapping between the various muscles
of the body and the regions of the motor cortex. All the above mentioned
regions of the cerebral cortex are referred to as {\em primary}, since they
are the one most directlt concerned with the input and output of the brain.
Near to these primary regions are the {\em secondary} sensory regions of the
cerebral cortex, where information is processed, while in the {\em secondary
motor} regions, conceived plans of motion get translated into specific
directions for actual muscle movement by the primary motor cortex. But the
most abstract and sophisticated activity of the brain is carried out in the
remaining regions of the cerebral cortex, the {\em association cortex}. It is
right here that information from various different sensory regions is analyzed
in a rather complex way, memories are laid down, pictures of the outside world
are constructed, general plans are conceived, etc. This is the anatomic,
morphological structure of the brain, on which my observations of
section~\ref{s2}
were based on! There is a rather well-known and extremely curious phenomenon
that I call {\em X-ism}. It is the {\em right} ({\em left}) cerebral hemisphere
which is concerned {\em exclusively} with the {\em left} ({\em right}) hand
side of the body, so that virtually all nerves must {\em cross over} from one
side to the other as they enter or leave the cerebrum! Furthermore, as I
mentioned above, the vision cortex is right at the back, while the eyes are
at the front, the feet-related region of the somatosensory cortex is at the
top, whereas the feet are at the bottom, and the left (right) auditory
cortex is related to the right (left) ear! It seems that the cerebral
neurosignals prefer to follow the longest possible path, and since this X-ism
is not observed in the cerebellum, whose action appears to be completely
unconscious, it is not inconceivable that the emergence of consciousness is
facilitated by the cerebral X-ism. In our unified scheme, such a strange
correlation between consciousness and X-ism seems to be born out of the
dynamics.

Let us now continue our fascinating trip inside the brain, and let us
concentrate on its basic building blocks, the nerve cells or neurons. Among
the about 200 types of different basic types of human cells, the neuron is
one of the most specialized, exotic and remarkably versatile cell. The neuron
is highly unusual in three respects: its {\em variation in shape}, its
{\em electrochemical function}, and its {\em connectivity}, \ie, its ability
to link up with other neurons in networks. Let us start with a few elements
of neuron microanatomy \cite{23,11}. There is a central starlike bulb, called
the {\em soma}, which contains the nucleus of the cell. A long nerve fibre,
known as the {\em axon}, stretches out from one end of the soma. Its length, in
humans,
can reach up to {\em few cm}, surprisingly long for a single cell! The raison
d'etre of the axon is to transmitt the neuron's output signal, \ie, it acts
like a wire. The axon has the ability of multiple bifurcation, branching out
into many smaller nerve fibers, and the very end of which there is always a
{\em synoptic knob}. At the other end of the soma and often springing off in
all directions from it, are the tree-like {\em dendrites}, along which input
data are carried into the soma. The whole nerve cell, as basic unit, has a cell
membrane surrounding soma, axon, synoptic knobs, dendrites. Signals pass from
one neuron to another at junctions known as {\em synapses}, where a synaptic
knob of one neuron is attached to another neuron's soma or dendrites. There
is very narrow gap, of a few nm, between the synaptic knob and the
soma/dendrite to where the {\em synaptic cleft} is attached. The signal from
one neuron to another has to propagate across this gap. The workings of the
nerve signals are another wonder of Nature!

A nerve fiber is a cylindrical
tube containing a mixed solution of NaCl and KCl, mainly the second, so there
are $\rm Na^+$, $\rm K^+$, and $\rm Cl^-$ ions within the tube. Outside the
tube the same type of ions are present but with more $\rm Na^+$ than $\rm K^+$.
In the {\em resting} state there is an excess of $\rm Cl^-$ over $\rm Na^+$ and
$\rm K^+$ inside the tube, giving it a negative charge, while it has positive
charge outside. A nerve signal is nothing else but a region of {\em charge
reversal} travelling along the fiber. At its head, {\em sodium gates} open to
allow the sodium to flow inwards and at its tail {\em potassium gates} open to
allow potassium to flow outwards. Then, metabolic pumps act to restore order
and establish the {\em resting state}, preparing the nerve fiber for another
signal. Amazingly enough, there is no major material (ion) transport that
produces the signal, just in and out local movements of ions, across the cell
membranes, \ie, a {\em small} and {\em local} depolarization of the cell!
Eventually, the nerve signal reaches the attached synaptic knob, at the very
end of the nerve fiber, and triggers it to emit chemical substances, known as
{\em neurotransmitters}. It is these substances that travel across the synaptic
cleft to another neuron's soma or dendrite. It should be stressed that the
signal here is not electrical, but a chemical one. What really is happening is
that when the nerve signal reaches the synaptic knob, the local depolarization
cause little bags immersed in the {\em vesicular grid}, the {\em vesicles}
containing molecules of the neurotransmitter
chemical (\eg, acetylcholine) to release their contents from the neuron into
the synaptic cleft, the phenomenon of {\em exocytosis}.
 These molecules then diffuse across the cleft to interact
with {\em receptor proteins} on receiving neurons. On receiving a
neurotransmitter molecule, the receptor protein opens a gate that causes a
local depolarization of the receiver neuron. The nerve signal has been
transmitted!

It depends on the nature of the synaptic knob and of the specific synaptic
junction, if the next neuron would be encouraged to {\em fire}, \ie, to start
a new signal along its own axon, or it would be discouraged to do so. In the
former case we are talking about {\em excitory synapses}, while in the latter
case about {\em inhibitory synapses}. At any given moment, one has to add up
the effect of all excitory synapses and subtract the effect of all the
inhibitory ones. If the net effect corresponds to a positive electrical
potential difference between the inside and the outside of the neuron under
consideration, {\em and} if it is bigger than a critical value, then the neuron
{\em fires}, otherwise it stays mute.

For our concerns here, the fundamental dynamical process of neural
communication can be summarized in the following three steps:
\begin{enumerate}
\item The neural axon is an {\em all} or {\em none} state. In the {\em all}
state a signal, called a {\em spike} or {\em action potential} (AP), propagates
indicating that the summation performed in the soma produced an amplitude of
the order of tens of mV. In the {\em none} state there is no signal travelling
in the axon, only the resting potential ($\sim-70$mV). It is essential to
notice that the presence of a travelling signal in the axon, {\em blocks} the
possibility of transmission of a second signal.
\item The nerve signal, upon arriving at the ending of the axon, triggers the
emission of neurotransmitters in the synaptic cleft, which in turn cause the
receptors to open up and allow the penetration of ionic current into the {\em
post synaptic} neuron. The {\em efficacy} of the synapse is a parameter
specified by the amount of penetrating current per presynaptic spike.
\item The post synaptic potential (PSP) diffuses toward the soma, where all
inputs in a short period, from all the presynaptic neurons connected to the
postsynaptic are summed up. The amplitue of individual PSP's is about 1 mV,
thus quite a number of inputs is required to reach the ``firing" threshold,
of tens of mV. Otherwise the postsynaptic neuron remains in the {\em resting}
or {\em none} state.
\end{enumerate}

The cycle-time of a neuron, \ie, the time from the emission of a spike in the
presynaptic neuron to the emission of a spike in the postsynaptic neuron is
of the order of {\em 1-2 msecs}. There is also some recovery time for the
neuron, after it ``{\em fired}", of about {\em 1-2 msecs}, independently of
how large the amplitude of the depolarizing potential would be. This period
is called the {\em absolute refractory period} of the neuron. Clearly, it sets
an upper bound on the spike frequency of 500-1000/sec. In the types of neurons
that we will be interested in, the spike frequency is considerably lower than
the above upper bound, typically in the range of 100/sec, or even smaller in
some areas, at about 50/sec. It should be noticed that this rather exotic
neural communication mechanism works very efficiently and it is employed
{\em universally}, both by vertebrates and invertebrates. The vertebrates have
gone even further in perfection, by protecting their nerve fibers by an
{\em insulating coating} of {\em myelin}, a white fatty substance, which
incidentally gives the {\em white matter} of the brain, discussed above, its
color. Because of this insulation, the nerve signals may travel undisturbed at
about 120 meters/second, a rather high speed!

A very important and significant anatomical fact for our discussion, is that
each neuron receives some $10^4$ synaptic inputs from the axons of other
neurons, usually one input per presynaptic neuron, and that each branching
neural axon forms about the same number ($\sim10^4$) of synaptic contacts on
other, postsynaptic neurons. A closer look at our cortex then would expose
a mosaic-type structure of assemblies of a few thousand densely connected
neurons. These assemblies are taken to be the basic cortical processing {\em
modules}, and their size is about $\rm1(mm)^2$. The neural connectivity gets
much sparcer as we move to larger scales and with much less feedback, allowing
thus for autonomous local collective, parallel processing and more serial
and integrative processing of local collective outcomes. Taking into account
that there are about $10^{11}$ nerve cells in the brain (about $7\times10^{10}$
in the cerebrum and $3\times10^{10}$ in the cerebellum), we are talking about
$10^{15}$ synapses! Counting one synapse per second, you will find yourself
counting past 30 million years after you started! Undoubtedly, the brain is
very special, and it should not be unreasonable to expect it to give rise to
mental properties \cite{24}.

While the dynamical process of neural communication suggests that the brain
action looks a lot like a computer action, there are some fundamental
differences having to do with a basic brain property called {\em brain
plasticity}. The interconnections between neurons are not fixed, as is the
case in a computer-like model, but are changing all the time. Here I am
referring to the synaptic junctions where the communication between different
neurons actually takes place. The synaptic junctions occur at places where
there are {\em dendritic spines} of suitable form such that contact with the
synaptic knobs can be made. Under certain conditions these dendritic spines
can shrink away and break contact, or they can grow and make new contact,
thus determining the {\em efficacy} of the synaptic junction. Actually, it
seems that it is through these dendritic spine changes, in synaptic
connections, that long-term memories are laid down, by providing the means of
storing the necessary information. A supporting indication of such a conjecture
is the fact that such dendritic spine changes occur within {\em seconds}, which
is also how long it takes for permanent memories to be laid down \cite{11}.

Furthermore, a very useful set of phenomenological rules has been put forward
by Hebb \cite{25}, the {\em Hebb rules}, concerning the underlying mechanism of
brain
plasticity. According to Hebb, a synapse between neuron 1 and neuron 2 would be
strengthened whenever the firing of neuron 1 is followed by the
firing of neuron 2, and weakened whenever it is not. A rather suggestive
mechanism that sets the ground for the emergence of some form of {\em
learning}! It seems that {\em brain plasticity} is not just an incidental
complication, it is a {\em fundamental property} of the activity of the brain.
Brain plasticity and its time duration (few {\em seconds}) play a critical
role, as we will see later, in the present unified approach to the brain and
the mind.

Many mathematical models have been proposed to try to simulate ``learning",
based upon the close resemblance of the dynamics of neural communication to
computers and implementing, one way or another, the essence of the Hebb rules.
These models are known as {\em Neural Networks} (NN) \cite{26}.

Let us try to construct a neural network model for a set of $N$ interconnected
neurons. The activity of the neurons is usually parametrized by $N$ functions
$\sigma_i(t),\ i=1,2,\ldots,N$, and the synaptic strength, representing the
synaptic efficacy, by $N\times N$ functions $j_{i,k}(t)$. The total stimulus
of the network on a given neuron ($i$) is assumed to be given simply by the
sum of the stimuli coming from each neuron
\begin{equation}
S_i(t)=\sum_{k=1}^N j_{i,k}(t)\sigma_k(t)
\label{14}
\end{equation}
where we have identified the individual stimuli with the product of the
synaptic strength ($j_{i,k}$) with the activity ($\sigma_k$) of the neuron
producing the individual stimulus. The dynamic equations for the neuron are
supposed to be, in the simplest case
\begin{equation}
{d\sigma_i\over dt}=F(\sigma_i,S_i)
\label{15}
\end{equation}
with $F$ a non-linear function of its arguments. The dynamic equations
controlling the time evolution of the synaptic strengths $j_{i,k}(t)$ are
much more involved and only partially understood, and usually it is {\em
assumed} that the $j$-dynamics is such that it produces the synaptic couplings
that we need or postulate! The simplest version of a neural network model is
the Hopfield model \cite{27}. In this model the neuron activities are
conveniently and
conventionally taken to be ``{\em switch}"-like, namely $\pm1$, and the time
$t$ is also an integer-valued quantity. Of course, this {\em all}($+1$) or
{\em none}($-1$) neural activity $\sigma_i$ is based on the neurophysiology
discussed above. If you are disturbed by the $\pm1$ choice instead of the
usual ``binary" one ($b_i=1$ or 0), replace $\sigma_i$ by $2b_i-1$. The choice
$\pm1$ is more natural from a physicist's point of view corresponding to a
two-state system, like the fundamental elements of the ferromagnet, discussed
in section~\ref{s2}, \ie, the electrons with their spins up ($+$) or ($-$).

The increase of time $t$ by one unit corresponds to one step for the dynamics
of the neuron activities obtainable by applying (for all $i$) the rule
\begin{equation}
\sigma_i(t+{i+1\over N})={\rm sign}(S_i(t+i/N))
\label{16}
\end{equation}
which provides a rather explicit form for (\ref{15}). If, as suggested by
the Hebb rules, the $j$ matrix is {\em symmetric} ($j_{i,k}=j_{k,i}$), the
Hopfield dynamics \cite{27} corresponds to a sequential algorithm for looking
for the minimum of the Hamiltonian
\begin{equation}
H=-\sum_i S_i(t)\sigma_i(t)=-\sum_{i,k=1}^N j_{i,k}\sigma_i(t)\sigma_k(t)
\label{17}
\end{equation}
Amazingly enough the Hopfield model, at this stage, is very similar to the
dynamics of a statistical mechanics {\em Ising-type} \cite{13}, or more
generally a {\em spin-glass}, model \cite{28}! This {\em mapping} of the
Hopfield model to a spin-glass model is highly advantageous because we have now
a justification for using the statistical mechanics language of phase
transitions, like critical points or attractors, etc, to describe neural
dynamics and thus brain dynamics, as was envisaged in section~\ref{s2}. It is
remarkable that this simplified Hopfield model has many {\em attractors},
corresponding to many different {\em equilibrium} or {\em ordered} states,
endemic in spin-glass models, and an unavoidable prerequisite for successful
storage, in the brain, of many different patterns of activities.
In the neural network framework, it is believed that an internal representation
(\ie, a pattern of neural activities) is associated with each object or
category that we are capable of recognizing and remembering. According to
neurophysiology, discussed above, it is also believed that an object is
memorized by suitably changing the synaptic strengths. {\em Associative memory}
then is produced in this scheme as follows (see corresponding (I)-(IV) steps in
section~\ref{s2}): An external stimulus, suitably involved, produces synaptic
strengths such that a specific learned pattern $\sigma_i(0)=P_i$ is ``printed"
in such a way that the neuron activities $\sigma_i(t)\sim P_i$ (II {\em
learning}), meaning that the $\sigma_i$ will remain for all times close to
$P_i$, corresponding to a stable attractor point (III {\em coded brain}).
Furthermore, if a {\em replication signal} is applied, pushing the neurons to
$\sigma_i$ values {\em partially} different from $P_i$, the neurons should
evolve toward the $P_i$. In other words, the memory is able to retrieve the
information on the whole object, from the knowledge of a part of it, or even
in the presence of wrong information (IV {\em recall process}). Of course,
if the external stimulus is very different from any preexisting $\sigma_i=P_i$
pattern, it may either create a new pattern, \ie, create a new attractor point,
or it may reach a chaotic, random behavior (I {\em uncoded brain}).

Despite the remarkable progress that has been made during the last few years
in understanding brain function using the neural network paradigm, it is fair
to say that neural networks are rather artificial and a very long way from
providing a realistic model of brain function. It seems likely that the
mechanisms controlling the changes
in synaptic connections are much more complicated and involved than the
ones considered in NN, as utilizing cytosceletal restructuring of the
sub-synaptic regions. {\em Brain plasticity} seems to play an essential,
central role in the workings of the brain! Furthermore, the ``{\em binding
problem}", alluded to in section~\ref{s2}, \ie, how to {\em bind} together all
the
neurons firing to different features of the same object or category, {\em
especially} when more than one object is perceived during a {\em single}
conscious perceptual moment, seems to remain unanswered.

We have come a long way since the times of the ``{\em grandmother neuron}",
where a {\em single} brain location was invoked for self observation and
control,
indentified with the pineal glands by Descartes \cite{29}! Eventually, this
localized concept was promoted to {\em homunculus}, a little fellow inside
the brain which observes, controls and represents us! The days of this ``{\em
Cartesian comedia d'arte}" within the brain are gone forever!

It has been long suggested that different groups of neurons, responding to a
common object/category, fire {\em synchronously}, implying {\em temporal
correlations} \cite{30}. If true, such correlated firing of neurons may help us
in resolving the binding problem \cite{31}. Actually, brain waves recorded from
the scalp, \ie, the EEGs, suggest the existence of some sort of {\em rhythms},
\eg, the
``{\em $\alpha$-rhythms}" of a frequency of 10 Hz. More recently, oscillations
were clearly observed in the visual cortex. Rapid oscillations, above EEG
frequencies in the range of 35 to 75 Hz, called the ``{\em
$\gamma$-oscillations}" or the ``{\em 40 Hz oscillations}", have been detected
in the cat's visual cortex \cite{32,33}. Furthermore, it has been shown that
these oscillatory responses can become {\em synchronized} in a
stimulus-dependent manner! Amazingly enough, studies of auditory-evoked
responses in humans have shown inhibition of the 40 Hz coherence with {\em loss
of consciousness} due to the induction of general anesthesia \cite{34}! These
remarkable and striking results have
prompted Crick and Koch to suggest that this {\em synchronized} firing on, or
near, the beat of a ``{\em$\gamma$-oscillation}" (in the 35--75 Hz range) might
be the {\em neural correlate} of {\em visual awareness} \cite{35,31}. Such a
behavior would be,
of course, a very special case of a much more general framework where coherent
firing of {\em widely-distributed} (\ie, {\em non-local}) groups of neurons,
in the ``beats" of {\em x}-oscillation (of specific frequency ranges), {\em
bind} them together in a mental representation, expressing the {\em oneness} of
{\em consciousness} or {\em unitary sense of self}. While this is a remarkable
and bold suggestion \cite{35,31}, it is should be stressed that in a
physicist's language it corresponds to a  phenomenological explanation, not
providing the underlying physical mechanism, based on neuron dynamics, that
triggers the synchronized neuron firing. On the other hand, the Crick-Koch
proposal \cite{35,31} is very
suggestive and in compliance with the general framework I developed in the
earlier sections, where macroscopic {\em coherent} quantum states play an
essential role in awareness, and especially with respect to the ``{\em binding
problem}". We have, by now, enough motivation from our somehow detailed study
of brain morphology and modeling, to go back to quantum mechanics and develop
a bit further, using string theory, so that to be applicable to brain dynamics.

\section{Stringy Quantum Mechanics: Density Matrix Mechanics}
\label{s5}
Quantum Field Theory (QFT) is the fundamental dynamical framework for a
successful description of the microworld, from molecules to quarks and
leptons and their interactions. The {\em Standard Model} of elementary particle
physics, encompassing the strong and electroweak interactions of quarks and
leptons, the most fundamental point-like constituents of matter presently
known, is fully and wholy based on QFT \cite{36}. Nevertheless, when
gravitational
interactions are included at the quantum level, the whole construction
collapses! Uncontrollable infinities appear all over the place, thus rendering
the theory inconsistent. This a well-known and grave problem, being with us
for a long, long time now. The resistance of gravitational interactions to
conventionally unify with the other (strong and electroweak) interactions
strongly suggests that we are in for changes both at the QFT front and at the
gravitational front, so that these two frameworks could become eventually
compatible with each other. As usual in science, puzzles, paradoxes and
impasses, that may lead to major crises, bring with them the seeds of dramatic
and radical changes, if the crisis is looked upon as an opportunity. In our
case at hand, since the Standard Model, based upon standard QFT, works
extremely well, we had not been forced to scrutinize further the basic
principles of the orthodox, Copenhagen-like QFT. Indeed, the mysterious
``collapse" of the wavefunction, as discussed in section~\ref{s3}, remained
always
lacking a dynamical mechanism responsible for its triggering. Had gravity
been incorporated in this conventional unification scheme, and since it is
the {\em last known} interaction, any motivation for changing the ground
rules of QFT, so that a dynamical mechanism triggering the ``collapse" of the
wavefunction would be provided, would be looked upon rather suspiciously
and unwarranted. Usually, to extremely good approximation, one can neglect
gravitational interaction effects, so that the standard QFT applies. Once
more, {\em usually} should not be interpreted as {\em always}. Indeed, for
most applications of QFT in particle physics, one assumes that we live in
a {\em fixed, static, smooth} spacetime manifold, \eg, a Lorentz spacetime
manifold characterized by a Minkowski metric ($g_{\mu\nu}$ denotes the metric
tensor):
\begin{equation}
ds^2\equiv g_{\mu\nu} dx^\mu dx^\nu=c^2 dt^2-d\vec x^2
\label{18}
\end{equation}
satisfying Einstein's special relativity principle. In such a case, standard
QFT rules apply and we get the miraculously successful Standard Model of
particle physics. Unfortunately, this is not the whole story. We don't live
exactly in a {\em fixed, static, smooth} spacetime manifold. Rather, the
universe is {\em expanding}, thus it is not {\em static}, and furthermore
unavoidable {\em quantum fluctuations} of the metric tensor $g_{\mu\nu}(x)$
defy the {\em fixed} and {\em smooth} description of the spacetime manifold,
at {\em least} at very short distances. Very short distances here do not
refer to the nucleus, or even the proton radius, of $10^{-13}$cm, but to
distances comparable to the Planck length, $\ell_{Pl}\sim10^{-33}$cm, which in
turn is related to the smallness of $G_N$, Newton's gravitational constant!
In particle physics we find it convenient to work in a system of units where
$c=\hbar=k_B=1$, where $c$ is the speed of light, $\hbar$ is the Planck
constant, and $k_B$ is the Boltzman constant. Using such a system of units
one can write
\begin{equation}
G_N\equiv {1\over M^2_{Pl}}\equiv \ell^2_{Pl}
\label{19}
\end{equation}
with $M_{Pl}\sim10^{19}\GeV$ and $\ell_{Pl}\sim10^{-33}$cm.

It should be clear that as we reach very short distances of ${\cal
O}(\ell_{Pl})$, fluctuations of the metric $\delta
g_{\mu\nu}(x)/g_{\mu\nu}(x)\sim(\ell_{Pl}/\ell)^2\sim{\cal O}(1)$, and thus the
spacetime
manifold is not well defined anymore, and it may even be that the very notion
of a spacetime description evaporates at such Planckian distances! So, it
becomes apparent that if we would like to include quantum gravity as an item
in our unification program checklist, we should prepare ourselves for major
revamping of our conventional ideas about quantum dynamics and the structure
of spacetime.

A particularly interesting, well-motivated, and well-studied example of a
{\em singular} spacetime background is that of a {\em black hole} (BH)
\cite{37}. These
objects are the source of a singularly strong gravitational field, so that
if any other poor objects (including light) cross their ``horizon", they are
{\em trapped} and would never come out of it again. Once in, there is no way
out! Consider, for example, a quantum system consisting of two particles
$a$ and $b$ in lose interaction with each other, so that we can describe its
quantum {\em pure} state by $\ket{\Psi}=\ket{a}\ket{b}$. Imagine now, that at
some stage of its evolution the quantum system gets close to a black hole,
and that for some unfortunate reason particle $b$ decides to enter the BH
horizon. From then on, we have no means of knowing or determining the exact
quantum state of the $b$ particle, thus we have to describe our system not
anymore as a {\em pure} state $\ket{\Psi}$, but as a {\em mixed} state
$\rho=\sum_i|b_i|^2\ket{a}\bra{b_i}$, according to our discussion in
section~\ref{s3}
(see (\ref{10},\ref{11})). But such an evolution of a {\em pure state} into a
{\em mixed state} is not {\em possible} within the conventional framework of
quantum mechanics as represented by (\ref{3}) or (\ref{9}). In conventional
QM {\em purity} is eternal. So, something drastic should occur in order to be
able to accomodate such circumstances related to singularly strong
gravitational fields. Actually, there is much more than meets the eye. If
we consider that our pure state of the two particles
$\ket{\Psi}=\ket{a}\ket{b}$ is a quantum fluctuation of the vacuum, then we
are in more trouble. The vacuum always creates particle-antiparticle pairs
that almost momentarily, and in the absence of strong gravitational fields,
annihilate back to the vacuum, a rather standard well-understood quantum
process. In the presence of a black hole, there is a very strong gravitational
force that may lure away one of the two particles and ``trap" it inside the
BH horizon, leaving the other particle hanging around and looking for its
partner. Eventually it wanders away from the BH and it may even be detected
by an experimentalist at a safe distance from the BH. Because she does not
know or care about details of the vacuum, she takes it that the BH is decaying
by emitting all these particles that she detects. In other words, while
{\em classical} BH is supposed to be stable, in the presence of quantum matter,
BH do decay, or more correctly {\em radiate}, and this is the famous {\em
Hawking radiation} \cite{37,38}. The unfortunate thing is that the Hawking
radiation is
{\em thermal}, and this means that we have lost vast amounts of {\em
information} dragged into the BH. A BH of mass $M_{BH}$ is characterized by a
{\em temperature} $T_{BH}$, an {\em entropy} $S_{BH}$ and a {\em horizon
radius} $R_{BH}$ \cite{37,38,39}
\begin{equation}
T_{BH}\sim{1\over M_{BH}};\quad S_{BH}\sim M^2_{BH};\quad R_{BH}\sim M_{BH}
\label{20}
\end{equation}
satisfying, of course, the first thermodynamic law, $dM_{BH}=T_{BH}dS_{BH}$.
The origin of the huge entropy ($\sim M^2_{BH}$) should be clarified.
Statistical physics teaches us that the {\em entropy} of a system is a measure
of the information {\em unavailable to us} about the detailed structure of the
system. The entropy is given by the number of different possible configurations
of the fundamental constituents of the system, resulting always in the same
values for the macroscopic quantities characterizing the system, \eg,
temperature, pressure, magnetization, etc.  Clearly, the fewer the macroscopic
quantities characterizing the system, the larger the entropy and thus the
larger the lack of information about the system. In our BH paradigm, the
macroscopic quantities that characterize the BH, according to (\ref{20}), is
{\em only} it mass $M_{BH}$. In more complicated BHs, they may posses some
extra ``{\em observables}" like electric charge or angular momentum, but still,
it is a rather small set of ``observables"! This fact is expressed as the
``{\em No-Hair Theorem}" \cite{37}, \ie, there are not many different long
range interactions around, like gravity or electromagnetism, and thus we cannot
``measure" safely and from a distance other ``observables", beyond the mass
($M$), angular momentum ($\vec L$), and electric charge ($Q$). In such a case,
it becomes apparent that we may have a huge number of different configurations
that are all characterized by the same $M,Q,\vec L$, and this the huge entropy
(\ref{20}). Hawking realized immediately that his BH dynamics and quantum
mechanics were not looking eye to eye, and he proposed in 1982 that we should
generalize quantum mechanics to include the {\em pure state} to {\em mixed
state} transition, which is equivalent to {\em abandoning} the quantum
superposition principle (as expressed in (\ref{3}) or (\ref{9})), for some more
advanced quantum dynamics \cite{40}. In such a case we should virtually abandon
the description of quantum states by wavefunctions or state vectors
$\ket{\Psi}$ and use the more accomodating density matrix ($\rho$) description,
as discussed in section~\ref{s3}, {\em but} with a {\em modified} form for
(\ref{9}).
Indeed, in 1983 Ellis, Hagelin, Srednicki, and myself proposed (EHNS in the
following) \cite{41} the following modified form of the conventional
Eq.~(\ref{9})
\begin{equation}
{\partial\rho\over\partial t}=i[\rho,H]+\delta\H\rho
\label{21}
\end{equation}
which accomodates the {\em pure state$\to$mixed state} transition through the
extra
term $(\delta\H)\rho$. The existence of such an extra term is characteristic
of ``{\em open}" quantum systems, and it has been used in the past for {\em
practical reasons}. What EHNS suggested was more radical. We suggested that
the existence of the extra term $(\delta\H)\rho$ is {\em not due} to
{\em practical reasons} but to some fundamental, dynamical reasons having to
do with quantum gravity. Universal quantum fluctuations of the gravitational
field ($g_{\mu\nu}$) at Planckian distances ($\ell_{Pl}\sim10^{-33}$cm) create
a very {\em dissipative} and {\em fluctuating} quantum vacuum, termed {\em
spacetime foam}, which includes virtual Planckian-size black holes. Thus,
quantum systems {\em never} evolve undisturbed, even in the {\em quantum
vacuum}, but they are continously interacting with the spacetime foam, that
plays the role of the {\em environment}, and which ``opens" spontaneously and
dynamically {\em any} quantum system. Clearly, the extra term $(\delta\H)\rho$
leads to a {\em spontaneous dynamical decoherence} that enables the system to
make a transition from a pure to a mixed state accomodating Hawking's proposal
\cite{40}.
Naive approximate calculations indicate that $\VEV{\delta\H}\sim E^2/M_{Pl}$,
where $E$ is the energy of the system, suggesting straight away that our
``low-energy" world ($E/M_{Pl}\le10^{-16}$) of quarks, leptons, photons, etc
is, for most cases, extremely accurately described by the conventional
Eq.~(\ref{9}). Of course, in such cases is not offensive to talk about
wavefunctions, quantum parallelism, and the likes. On the other hand, as
observed in 1989 by Ellis, Mohanty, and myself \cite{42}, if we try to put
together
more and more particles, we eventually come to a point where the decoherence
term $(\delta\H)\rho$ is substantial and decoherence is almost instantaneous,
leading in other words to an instantaneous {\em collapse} of the wavefunction
for large bodies, thus making the transition from quantum to classical {\em
dynamical} and not by decree! In a way, the Hawking proposal \cite{40}, while
leading to a major conflict between the standard QM and gravity, motivated
us \cite{41,42} to rethink about the ``collapse" of the wavefunction, and
it seemed to contain the seeds of a dynamical mechanism for the ``collapse" of
the wavefunction. Of course, the reason that many people gave a ``cold
shoulder"
to the Hawking proposal was the fact that his treatment of quantum gravity
was semiclassical, and thus it could be that all the Hawking excitement was
nothing else but an artifact of the bad/crude/unjustifiable approximations.
Thus, before we proceed further we need to treat better Quantum Gravity (QG).
String Theory (ST) does just that. It provided the {\em first}, and presently
{\em only} known framework for a consistently quantized theory of gravity
\cite{43}.

As its name indicates, in string theory one replaces {\em point like} particles
by {\em one-dimensional}, {\em extended}, {\em closed}, {\em string} like
objects, of characteristic length ${\cal O}(\ell_{Pl})\sim10^{-33}$cm. In ST
one gets an {\em automatic}, {\em natural} unification of {\em all}
interactions {\em including} quantum gravity, which has been the {\em holy
grail} for particle physics/physicists for the last 70 years! It is thus only
natural to address the hot issues of black hole dynamics in the ST framework
\cite{43}.
Indeed, in 1991, together with Ellis and Mavromatos (EMN in the following)
we started a rather elaborate program of BH studies, and eventually, we
succeeded in developing a new dynamical theory of string black holes \cite{44}.
One {\em first} observes that in ST there is an {\em infinity} of particles of
different masses, including the {\em Standard Model} ones, corresponding to the
different excitation modes of the
string. Most of these particles are unobservable at low energies since they
are very massive $M\gsim{\cal O}(M_{Pl}\sim10^{19}\GeV)$ and thus they cannot
be produced in present or future accelerators, which may reach by the year 2005
about $10^4\GeV$. Among the {\em infinity} of different types of particles
available, there is an {\em infinity} of massive ``gauge-boson"-like particles,
generalizations of the $W$-boson mediating the weak interactions, thus
indicating the {\em existence} of an {\em infinity} of {\em spontaneously
broken gauge symmetries}, each one characterized by a specific {\em charge},
generically called $Q_i$. It should be stressed that, even if these stringy
type, spontaneously broken gauge symmetries do not lead to long-range forces,
thus {\em classically} their $Q_i$ charges are unobservable at long distances,
they do become observable at long distances at the {\em quantum level}.
Utilizing the quantum Bohm-Aharonov effect \cite{45}, where one ``measures"
phase shifts proportional to $Q_i$, we are able to ``measure" the $Q_i$ charges
from adesirable distance! This kind of $Q_i$ charge, if available on a black
hole, is called sometimes and for obvious reasons, {\em quantum hair}
\cite{46}. From the infinity of stringy symmetries, a relevant for us here,
specific, {\em closed subset} has been identified, known by the name of
$W_{1+\infty}$ {\em symmetry}, with many interesting properties \cite{47}.
Namely, these $W_{1+\infty}$ symmetries cause the mixing \cite{48}, in the
presence of singular spacetime backgrounds like a BH, between the massless
string modes, containing the {\em attainable localizable low energy world}
(quarks, leptons, photons, etc), let me call if the {\em$W_1$-world}, {\em and}
the massive ($\ge{\cal O}(M_{Pl})$)
string modes of a very characteristic type,
the so-called {\em global states}. They are called {\em global states} because
they have the peculiar and unusual characteristic to have {\em fixed} energy
$E$ and momentum $\vec p$, and thus, by employing the uncertainty type
relations, a la (\ref{8}), they are extended over {\em all} space and time!
Clearly,
while the {\em global states} are as {\em physical} and as {\em real} as any
other states, still they are {\em unattainable} for {\em direct observation}
to a {\em local observer}. They make themselves noticeable through their {\em
indirect} effects, while interacting with, or agitating, the {\em$W_1$ world}.
Let me call the global state space, the {\em$W_2$-world}.

The {\em second} step in the EMN approach \cite{44} was to concentrate on
spherically symmetric 4-D stringy black holes, that can be {\em effectively}
reduced to 2-D (1 space + 1 time) string black holes of the form discussed by
Witten \cite{49}. This
effective dimensional reduction turned out to be very helpful because it
enabled us to concentrate on the real issues of BH dynamics and bypass the
technical complications endemic in higher dimensions. We showed that \cite{44},
as we suspected all the time, stringy BH are {\em endorsed} with $W$-hair, \ie,
they carry an {\em infinity} of charges $W_i$, correponding to the
$W_{1+\infty}$
symmetries, characteristic of string theories. Then we showed that \cite{44}
this $W$-hair was sufficient to establish {\em quantum coherence} and avoid
{\em loss of information}. Indeed, we showed explicitly that \cite{44} in
stringy black holes
there is no Hawking radiation, \ie, $T_{BH}=0$, and no entropy, \ie,
$S_{BH}=0$! In a way, as it should be expected from a respectable quantum
theory of gravity, BH dynamics is not in conflict with quantum mechanics. There
are several intuitive arguments that shed light on the above, rather drastic
results. To start with, the {\em infinity} of $W$-charges make it possible for
the BH to {\em encode} any possible piece of information ``thrown" at it by
making a transition to an altered suitable configuration, consistent with very
powerful selection rules. It should be clear that if it is needed an {\em
infinite} number of {\em observable} charges to determine a configuration of
the BH, then the ``measure" of the unavailable to us information about this
specific configuration should be virtually zero, \ie, $S_{BH}=0$! The {\em
completeness}
of the $W$-charges, and for that matter of our argument, for establishing that
$S_{BH}=0$, has been shown in two complementary ways. Firstly, we have shown
that \cite{44} if we {\em sum over} the $W$-charges, like being unobservables,
we reproduce the whole of Hawking dynamics! Secondly, we have shown that the
$W_{1+\infty}$ symmetry {\em acts} as a phase-space volume (area in 2-D)
preserving symmetry, thus entailing the {\em absence} of the extra
$W_{1+\infty}$ symmetry violating $(\delta\H)\rho$ term in (\ref{21}), thus
reestablishing (\ref{9}), \ie, safe-guarding quantum coherence. Actually, we
have further shown that \cite{44} stringy BHs correspond to ``extreme BHs",
\ie, BH with a harmless horizon, implying that the infinity of $W$-charges
neutralize the
extremely strong gravitional attraction. In such a case, there is no danger of
seducing a member of a quantum system, hovering around the BH horizon, into
the BH, thus eliminating the raison d'etre for Hawking radiation! Before though
icing the champagne, one may need to address a rather fundamental problem. The
low-energy, attainable physical world $W_1$, is made of electrons, quarks,
photons, and the likes, all very well-known particles with well-known
properties, \ie, mass, electric charge, etc. Nobody, though, has ever added to
the identity card of these particles, lines representing their
$W$-charges. In other words, the $W_1$-world seems to be $W$-charge blind.
How is it possible then for an electron falling into a stringy BH, to
excite the BH through $W_{1+\infty}$-type interactions, to an altered
configuration where it has been taken into account all the information carried
by the electron? Well, here is one of the miraculous mechanisms, endemic in
string theories. As discussed above, it has beeen shown \cite{48} the in the
presence of singular spacetime backgrounds, like the black hole one, a mixing,
of purely stringy nature, is induced between states belonging to different
``mass" levels, \eg, between a  {\em Local} ($L$) state ($\ket{a}_L$)
of the $W_1$ world, with the {\em Global} states ($G$) ($\ket{a_i}_G$) of
the $W_2$ world
\begin{equation}
\begin{array}{c}
\ket{a}=\ket{a}_L+\sum_g \ket{a_g}_G\\
{\rm or}\\
\ket{a}_W=\ket{a}_{W_1}\oplus\ket{a}_{W_2}
\end{array}
\label{22}
\end{equation}
Notice that any resemblance between the
symbols in (\ref{22}) and (\ref{2}) is {\em not} accidental and will be
clarified later. Thus, we see that when a low energy particle approaches/enters
a stringy BH, its global state or $W_2$ components while {\em dormant} in flat
spacetime backgrounds, get {\em activated} and
this causes a quantum mechanical coherent BH transition, always satisfying a
powerful set of selection rules. In this new EMN scenario \cite{44} of BH
dynamics, if
we start with a {\em pure state} $\ket{\Psi}=\ket{a}_W\ket{b}_W$, we end up
with a {\em pure state} $\ket{\Psi'}=\ket{a'}_W\ket{b'}_W$, even if our
quantum system encountered a BH in its evolution, because we can {\em monitor}
the $\ket{b}$ part through the Bohm-Aharonov-like $W_i$ charges! So everything
looks dandy.

Alas, things get a bit more complicated, before they get simpler. We face here
a new purely stringy phenomenon, that has to do with the global states, that
lead to some dramatic consequences. Because of their {\em delocalized} nature
in spacetime, the global or $W_2$-states can neither (a) appear as {\em
well-defined asymptotic states}, nor (b) can they be integrated out in a
{\em local path-integral formalism}, thus defying their detection in local
scattering experiments!!! Once more, we have to abandon the language of the
scattering matrix $S$, for the superscattering matrix $\S\not=SS^\dagger$, or
equivalently abandon the description of the quantum states by the wavefunction
or state vector $\ket{\Psi}$, for the density matrix $\rho$ \cite{50}. Only
this time it is for real. While string theory provides us with
consistent and complete quantum dynamics, including gravitational interactions,
it does it in such a way that {\em effectively} ``{\em opens}" our low energy
attainable $W_1$ world. This is not anymore a possible artifact of our
treatment of quantum gravity, this is the {\em effective quantum mechanics}
\cite{50,5,6} that emerges from a consistent quantum theory of gravity. An
intuitive way to
see how it works is to insert $\ket{a}_W$ as given in (\ref{22}) into
(\ref{9}), where $\rho_W\equiv\ket{a}_W\bra{a}_W$, collect all the
$\ket{a}_{W_2}$ dependent parts, treat them as {\em noise}, and regard
(\ref{9}) as describing {\em effectively} some quantum {\em Brownian} motion,
\ie,
regard it as a {\em stochastic differential equation}, or {\em Langevin
equation} for $\rho_{W_1}=\sum_i p_i\ket{a_i}_{W_1}\bra{a_i}_{W_1}$ (see
(\ref{10})), where the $p_i$'s depend on $\ket{a}_{W_2}$ and thus on the
{\em$W_2$ world} in a {\em stochastic way} \cite{51}. In the EMN approach
\cite{50,51,5,6} the emerging equation, that reproduces the EHNS equation
(\ref{21}) with an explicit form for
the $(\delta\H)\rho$ term, reads (dropping the $W_1$ subscripts)
\begin{equation}
{\partial\rho\over\partial t}=i[\rho,H]+i G_{ij}
[\alpha_i,\rho]\beta^j
\label{23}
\end{equation}
where $G_{ij}$ denotes some {\em positive definite ``metric"} in the string
field space, while $\beta^j$ is a characteristic function related to the field
$\alpha_j$ and representing {\em collectively} the agitation of the {\em$W_2$
world} on the $\alpha_j$ dynamics and thus, through (\ref{22}), one expects
$\beta^j\approx{\cal O}((E/M_{Pl})^n)$, with $E$ a typical energy scale in
the $W_1$-world system, and $n=2,3,\ldots$.

Before I get into the physical interpretation and major consequences of
(\ref{23}), let us collect its most fundamental, {\em system-independent}
properties, following {\em directly} from its specific structure/form
\cite{50,5,6}
\begin{description}
\item I) {\em Conservation of probability} $P$ (see (\ref{5}) and discussion
above (\ref{9}))
\begin{equation}
{\partial P\over\partial t}={\partial\over\partial t}({\rm Tr}\rho)=0
\label{24}
\end{equation}
\item II) {\em Conservation of energy, on the average}
\begin{equation}
{\partial\over\partial t}\VEV{\VEV{E}}\equiv{\partial\over\partial t}[{\rm
Tr}(\rho E)]=0
\label{25}
\end{equation}
\item III) {\em Monotonic increase in entropy}/{\bf microscopic arrow of time}
\begin{equation}
{\partial S\over\partial t}={\partial\over\partial t}[-{\rm Tr}(\rho\ln\rho)]
=(\beta^iG_{ij}\beta^j)S\ge0
\label{26}
\end{equation}
due to the positive definiteness of the metric $G_{ij}$ mentioned above, and
thus {\em automatically} and {\em naturally} implying a {\em microscopic arrow
of time}.
\end{description}
Rather remarkable and useful properties indeed.

Let us try to discuss the physical interpretation of (\ref{23}) and its
consequences. In conventional QM, as represented by (\ref{9}), one has a
{\em deterministic, unitary} evolution of the quantum system, and it is {\em
only when} one feels compelled to ``measure"/``observe" the system, that the
probabilistic element of QM {\em emerges}. One, of
course, tacitly assumes the existence of a fixed, smooth spacetime background
that does not ``disturb" the system, acting simply as the {\em arena} in which
things are happening, and thus leaving the system ``{\em closed}". The
characteristics of such ``closed" systems include, of course, conservation of
energy and no {\em definite arrow
of time} or no {\em flow of time}, which is reflected in the forms of
(\ref{9}), (\ref{18}), which are invariant under $t\to -t$! When we decide
to ``open" the system we basically perform a ``measurement", \ie, we force the
system to ``decide" what it wants to be, by choosing a very specific state, out
of many coexisting possible ones, \ie, we are talking about the ``collapse" of
the wavefunction. That's in a nutshell the Copenhagen interpretation of QM,
leaving too much to be desired, and too much on the ``eye" of the ``observer"!
We need to do better. In the {\em density matrix mechanics}, as represented by
(\ref{23}), and as emerged, in one {\em interpretation} from string theory, one
has a {\em stochastic},
{\em indeterministic} evolution of the quantum system, {\em ab initio}, due to
the {\em unavoidable} existence of {\em spacetime foam}. The {\em
uncontrollable,
universal} quantum fluctuations of the spacetime metric at very short distances
(${\cal O}(\ell_{Pl})$), containing creation and annihilation of virtual
Planckian-size BH, {\em agitate} through the global or
{\em$W_2$-world} states, our low-energy quantum system, rendering it {\em
dynamically} and {\em spontaneously} ``{\em open}". This is an {\em objective},
{\em universal} mechanism, independent of any ``observer", that is always
``up and working", thus {\em eroding} the quantum coherence and eventually
leading to a {\em dynamical}, {\em spontaneous collapse}. It should be clear
that the {\em natural} ``{\em opening}" of our quantum system is due to our
{\em inability} to take into account all the detailed effects of the global
states, because of their delocalized nature, and thus we do {\em truncate}
them, arriving at the {\em Procrustean Principle}, a {\em new universal
principle} \cite{6} that goes beyond the standard uncertainty principle
(\ref{8}).
Furthermore, since this new dynamical mechanism of the ``collapse" of the
wavefunction, as emerged in the EMN approach \cite{50,5,6}, is an {\em
objective spontaneous}, {\em time-ordered}, and thus an {\em orchestrated} one,
I propose
here to call it {\em synchordic collapse}.\footnote{chord=string in greek;
synchordia something like symphonia.} Schematically, one can represent this new
mechanism of the ``collapse" of the wavefunction, by using (\ref{22}), as
follows
\begin{equation}
\begin{tabular}{cccccccc}
&&&&&&{\em synchordic}&\\
$W$&$\supset$&$W_1$&$\otimes$&$W_2\to$&cause&$\longrightarrow$&$W_1$\\
$\vert\vert\vert$&&$\vert\vert\vert$&&$\vert\vert\vert$&&{\em collapse}&\\
{\small Physical World}&&{\small Attainable}
&&{\small Global}&&&\\
{\small(including all local}&&{\small Physical World}&&
{\small States}&&&\\
{\small and global states)}&&{\small(including all local,}&&{\small World}&\\
&&{\small low-energy states)}&&&&
\end{tabular}
\label{27}
\end{equation}
which makes it apparent that the {\em global} or {\em$W_2$-world} states are
the {\em agents} of the {\em synchrodic collapse}, as being the raison d'etre
of
{\em stochasticity} in quantum dynamics. Also, notice the similarity between
(\ref{2}) and (\ref{27}), rather remarkable and very suggestive! The most
{\em amazing} and {\em astonishing} thing is that, despite the well-known fact
that usually open, dissipative systems defy quantization and energy
conservation, our naturally ``open" system, as represented by (\ref{23}) and as
explicitly indicated in (\ref{24}), (\ref{25}), and (\ref{26}), is {\em
different} \cite{52,53}. It is susceptible to quantization, it {\em conserves
energy} in the mean,
and {\em monotonically increases} its entropy, leading to loss of information,
quantitatively expressed as {\em quantum decoherence}, and thus supplementing
us with a very {\em natural}, {\em universal}, {\em objective microscopic arrow
of time}! In the EMN approach \cite{50,5,6}, {\em time is a statistical measure
of the interactions (quantum gravitational friction) between the local,
low-energy
world $W_1$ and the global or $W_2$-world states, in the presence of singular
spacetime backgrounds (spacetime foam).} The strong emerging correlation
between loss of information, quantum decoherence leading to wavefunction
collapse and the dynamical appearance of {\em flowing} time, I believe is
unprecedented in physics.

Clearly, the role of the {\em magic} extra term proportional to $\beta^j$ in
(\ref{23}), is multifunctional, as exemplified by making use of the
{\em dissipation-fluctuation theorem} of statistical mechanics \cite{13}. It
can be viewed as a {\em dissipative term}
that destroys quantum coherence, by damping the off-diagonal elements and
{\em also} it can be seen as a {\em noise term able} to drive the system away
from its equilibrium position and, after some time, bring it back to the same
position or bring it to some other equilibrium position. In other words, we
may interpret (\ref{23}) as a renormalization group equation (RGE), as
discussed in section~\ref{s2}, describing the evolution of the system between
different {\em phases}, each corresponding to one of the {\em infinite}
spontaneously broken $W_{1+\infty}$ symmetries. Clearly, at an {\em equilibrium
position}, or at a {\em critical point}, all $\beta^j$ do vanish, thus
recovering naturally (\ref{9}) from (\ref{23}), or equivalently recovering
standard QFT as applied to particle physics for the past 70 years. In
principle, in fixed, smooth spacetime backgrounds, hopefully corresponding to
{\em critical points} in our new stringy language, there is a {\em decoupling}
of the global states from the local, low-energy states in (\ref{22}), \ie, all
$c_g$'s do vanish, and thus implying vanishing $\beta^j$ in (\ref{23}). Before
though, we are carried away from the highly promising stringy big quantum
picture that emerges here, it should pay to have a closer look at some
numerical details, if not for any other reason, just as a {\em reality check}!
Indeed, one can work out, using (\ref{23}), the time that it takes for quantum
decoherence, or equivalently the quantum coherence lifetime $\tau_c$, as
defined by the off-diagonal elements damping factor \cite{42}:
$\exp[-Nt(m^6/M^3)(\Delta X)^2]$, for a system of $N$ constituents of mass $m$,
assuming that its center of mass gets finally pinned down within $\Delta X$,
and is given by
\begin{equation}
\tau_c={M^3\over Nm^6(\Delta X)^2}
\label{28}
\end{equation}
where $M$ stands for $M_{SU}\approx(1/10)M_{Pl}\approx10^{18}\GeV$, the
characteristic string scale \cite{54}. What about the value of $m$? The most
natural
value for it would be $m\approx m_{\rm nucleon}\approx1\GeV$ for the following
reason. Our attainable low-energy world, as far as we know is made up of
electrons, protons, and neutrons: that is what constitute us, \ie, our cells,
our proteins, our DNA, etc, and also that is what everything else we use, \ie,
the ``apparatus", is made of. Of course, protons and neutrons are mainly made
of up ($u$) and down ($d$) constituent quarks, but for my arguments they are of
comparable mass and thus would give the same results. Now, since the bulk
of matter is due to nucleons, and not to electrons ($m_{\rm nucl}\approx1836
m_e$), the shortest coherence lifetimes that we are interested in would
be provided by $m\approx m_{\rm nucl}$. Furthermore, independent of the
complicated structure that you may consider, \eg, a complicated protein polymer
structure, a la Microtubules (MTs), the virtual Planckian BHs have such high
energy that they ``see" and interact/agitate with the most fundamental
constituents of the complicated structure, \ie, up and down quarks and
electrons, thus as explained above, justifying the identification $m\approx
m_{\rm nucl}\approx1\GeV$ in (\ref{28}). Thus, using $M\sim10^{18}\GeV$,
$m\sim1\GeV$, and $(\Delta x)\sim1{\rm nm}\equiv10^{-7}$cm, (\ref{28}) yields
\begin{equation}
\tau_c={10^{16}\over N}\, {\rm sec}
\label{29}
\end{equation}
a rather suggestive formula. In the case of a single ($N=1$) hydrogen atom,
(\ref{29}) becomes $\tau_H\sim10^{16}$sec, the present age of the universe!
In other words, standard QM applies extremely accurately in microsystems, as
of course, we want, because of the spectacular successes of QM in the
microworld. On the other hand, if we take a piece of ice, containing say
$N\sim N_{\rm Avogadro}\approx10^{24}$ nucleons, then we get $\tau^{\rm
ice}_c\approx10^{-8}$ sec, a rather short-lived quantum coherence implying that
for macroscopic objects ($N\sim N_{\rm Avogadro}$) QM rules fail and classical
physics emerges {\em naturally, dynamically, spontaneously}, and {\em
objectively}! The {\em Schr\"odinger's cat} paradox is automatically resolved:
within ${\cal O}(10^{-8}{\rm sec})$ the cat would be dead or alive, not the
fifty/fifty stuff anymore. Furthermore, the ``measurement"/``observation"
problem gets a similar satisfactory resolution. Indeed, performing a
``measurement"/``observation" on a quantum system implies bringing it in ``{\em
interaction}" with some suitable {\em macroscopic apparatus} ($N_{\rm
macr}\sim{\cal O}(N_{\rm Avog}))$, thus triggering an almost {\em
instantaneous}
``collapse" of the wavefunction of the quantum system, as suggested by
(\ref{29}) with $N\approx N_{\rm macr}+N_{\rm quant.syst}\sim{\cal O}(N_{\rm
Avog})$. The {\em magic step}, as indicated in (\ref{7}), and which constitutes
basically the one-half of quantum mechanics {\em it does need not to be
postulated},
but it comes out from the stochastic dynamics, as provided by the agitating
global or {\em$W_2$-world} states. It should not escape our notice that there
is no quantum-classical border, but a continous and smooth transition.
Furthermore, as (\ref{28}) indicates, the
Avogadro number, a measure of the macroscopicity of the system, is basically
dynamically determined to be the inverse of the dimensionless product of the
gravitational strength ($\sqrt{G_N}$) times the characteristic strong
interaction scale ($\Lambda_{\rm QCD}\sim{\cal O}(0.1\GeV))$ times the
electromagnetic fine structure constant ($\alpha=1/137$)
\begin{equation}
N_{\rm Avogadro}\sim{1\over\sqrt{G_N}\ \Lambda_{\rm QCD}\,\alpha}
\label{30}
\end{equation}
I do hope that I have convinced the reader that the performed {\em reality
check} has been rather successful and illuminating.

It is highly remarkable that stringy modified QM or {\em density matrix
mechanics} is offering us, see ((\ref{23}),(\ref{27})), a new {\em unified
approach} to quantum dynamics, by turning a {\em deterministic} wave-type
equation into a {\em stochastic differential}
equation able to successfully describe {\em both} evolution and ``measurement"
of quantum systems. At the same time, a {\em unified picture} of the quantum
and classical world is emerging, as
promised in section~\ref{s3}, without the need of raising artificial borders
between the quantum and the classical, the transition between them is dynamical
and smooth. The fundamental property of string theory that allows all these
``miraculous events" to occur is its defining property, \ie, the need of
2-dimensions (1 space + 1 time) to describe a 1-dimensional (1-D) {\em
extended} object and its accompanying {\em infinity} of excitation
modes/particles, due
exactly to its extended nature. While a pointlike particle ``runs" on a {\em
world-line}, a string sweeps a {\em world-sheet}. Eventually, all 4-D spacetime
physics would be mappings of corresponding physics in the 2-D stringy
world-sheet. The existence of the $W_{1+\infty}$ symmetry was first established
in 2-D ``world sheet" physics and then mapped into 4-D spacetime physics. The
{\em infinity} of spontaneously broken stringy gauge symmetries, and the very
existence of the {\em global states}, {\em somehow} can trace back their origin
to the {\em 2-dimensionality} of the world-sheet! In other words, the stringy
nature of the modified quantum mechanics prevails, as should be apparent at
each and every turn!

The alert reader may have already noticed the stunning similarity between the
string dynamics in singular spacetime backgrounds, like black holes and
spacetime foam, and the brain mechanics presented in section~\ref{s2}. Presence
or lack of quantum coherence and its cause, the existence of an {\em infinite}
number of possible equilibrium or critical points corresponding to an {\em
infinite}
number of spontaneously broken ``gauge" (stringy) symmetries with {\em
appropriate selection rules}, the possibility of ``running" away from one
equilibrium point, and eventually coming back to it, or end up at another
equilibrium point, in a {\em timely} manner, etc, etc. If we could only find a
structure in the brain  that it renders the EMN string dynamics
\cite{44,50,51,5,6} applicable,
we would then be able to provide a rather explicit answer to most of the
problems raised in sections~\ref{s2} and \ref{s4}. Namely, the {\em binding
problem}; how
the brain represents a {\em physical, objectively real, flowing time}? {\em
free will}, etc, etc.

Well, these brain structures {\em do exist} and they are called

\section{MicroTubules (MT) I: The biochemical profile}
\label{s6}
Living organisms are collective assemblies of cells which contain collective
assemblies of organized material, including membranes, organelles, nuclei, and
the {\em cytoplasm}, the bulk interior medium of living cells. Dynamic
rearrangements of the cytoplasm within {\em eucaryotic cells}, the cells
of all animals and almost all plants on Earth, account for their changing
shape, movement, etc. This extremely important cytoplasmic structural and
dynamical organization is due to the presence of networks of inteconnected
protein polymers, which are referred to as the {\em cytosceleton} due to their
boneline structure \cite{1,2}. The cytosceleton consists of {\em Microtubules}
(MT's), action microfilaments, intermediate filaments and an {\em organizing
complex}, the {\em centrosome} with its chief component the {\em centriole},
built from two bundles of microtubules in a separated {\bf T} shape.
Parallel-arrayed MTs are
interconnected by cross-bridging proteins ({\em MT-Associated Proteins}: MAPs)
to other MTs, organelle filaments and membranes to form {\em dynamic networks}
\cite{1,2}. MAPs may be contractile, structural, or enzymatic. A very important
role is played by contractile MAPs, like dynein and kinesin, through their
participation in cell movements as well as in intra-neural, or
axoplasmic transport which moves material and thus is of fundamental importance
for the {\em maintenance} and {\em regulation} of {\em synapses}.
The structural bridges formed by MAPs stabilize MTs and prevent their
disassembly. The MT-MAP ``complexes" or {\em cytosceletal networks} determine
the cell architecture and dynamic functions, such a {\em mitosis}, or {\em cell
division}, {\em growth}, {\em differentiation}, {\em movement}, and for us here
the very crucial, {\em synapse formation and function}, all essential to the
living state! It is usually said that {\em microtubules} and {\em ubiquitous}
through the entire biology! \cite{1,2}

Microtubules \cite{1,2,3} are hollow cylindrical tubes, of about 25~nm in
diameter on the
outside and 14~nm on the inside, whose walls are polymerized arrays of {\em
protein subunits}. Their lengths may range from tens of nanometers during early
assembly, to possible centimeters (!) in nerve axons within large animals. The
protein subunits assemble in longitudinal strings called {\em protofilaments},
thirteen (13) parallel protofilaments laterally allign to form the hollow
``{\em tubules}". The protein subunits are ``barbell" or ``peanut" shaped {\em
dimers} which in turn consists of two globular proteins, {\em monomers}, known
as {\em alpha ($\alpha$)} and {\em beta ($\beta$)} {\em tubulin}. The $\alpha$
and $\beta$ tubulin monomers are similar molecules with identical orientation
within protofilaments and tubule walls. In the polymerized state of the MT, one
monomer consists of 40\% $\alpha$-helix, 31\% $\beta$-sheet and 29\% random
coil. The $\alpha$-tubulin consists of four $\alpha$-helixes, four
$\beta$-sheets, and two random coils, while the $\beta$-tubulin has six
$\alpha$-helixes, one $\beta$-sheet, and seven random coils. Each {\em monomer}
consists of about 500 aminoacids, is about 4nmx4nmx4nm, and weighs
$5.5\times10^4$ daltons or equivalently its atomic number is $5.5\times10^4$,
and has a {\em local polarity}. Each {\em dimer}, as well as each MT, appears
to have an electric polarity or dipole, with the negative end oriented towards
the $\alpha$-monomer and the positive end towards the $\beta$-monomer.
The dipole character of the dimer originates from the 18 Calcium ions ($\rm
Ca^{++}$) bound within each $\beta$-monomer. An equal number of negative
charges required for the electrostatic balance are localized near the
neighboring $\alpha$-monomer. Thus, MTs can be viewed as an example of {\em
electret} substances, \ie, oriented assemblies of dipoles, possessing {\em
piezoelectric} properties, pretty important in their functions including their
assembly and disassembly behavior. The dimers are held together by relatively
weak Van der Waals hydrophobic forces due to dipole coupling. Each dimer has
6 neighbors which form slightly skewed {\em hexagonal lattices} along the
entirety of the tube, with a ``leftward" tilt, and several {\em helical
patterns} may be ``seen" in the relations among dimers. Imagine a MT slit along
its length, and then opened out flat into a strip. One then finds that the
tubulins are ordered in sloping lines which rejoin at the opposite edge 5 or
8 places displaced (5+8=13), depending on the line slope, it is to the right or
to the left. The crystal-like symmetry packing of the tubulin in MTs is very
suggestive for a possible use of MTs as ``{\em information processors}". It
should be rather obvious that such a delicate, fine MT organization is there
for some good reason.

Further evidence for the very special role that MTs are made to play is
provided by the very interesting assembly and disassembly behavior. Dimers
self-assemble in MTs, apparently in an {\em entropy-driven} process which can
{\em quickly} change by MT disassembly and reassembly into {\em another
orientation}. It seems that Guanosine TriPhosphate (GTP) hydrolysis to
Guanosine DiPhosphate (GDP) provide the energy that binds the polymerizing
tubulin dimers, while biochemical energy can also be pumped into MTs by
phosphorylation/dephosphorylation of MAPs. In fact, each tubulin dimer, as a
whole, can exist in two different geometrical configurations or {\em
conformations}, induced, \eg, by the GTP-GDP hydrolysis. In one of these they
bend $29^\circ$ to the direction of the microtubule. It seems that these two
conformations {\em correspond} to two different states of the dimer's electric
polarization, where these come about because an electron, centrally placed at
the $\alpha$-tubulin/$\beta$-tubulin junction, may shift from one position to
another, the textbook, gold-platted case of a {\em quantum-mechanical two-state
system} \cite{19}!  Several
``on-off" functions linked to $\rm Ca^{++}$ binding could do the job. The
$\rm Ca^{++}$ concentration changes could alter the conformational states of
certain tubulin subunits, which may be pre-programmed to undergo conformational
changes in the presence of $\rm Ca^{++}$, through GTP, glycosylation, etc.
Furthermore, a calcium-calmodulin complex could facilitate charge and/or energy
transfer, similar to the way acceptor impurities act in semiconductors! The
$\rm Ca^{++}$ may delocalize an electron from its orbital spin mate, both
electrons belonging to an aromatic aminoacid ring within a hydrophobic pocket,
resulting in an unstable {\em electron ``hole"}, and thus enhancing the
probability for either a charge transfer from an adjacent subunit, and/or
transfer of energy to an adjacent subunit. Tubulins in MTs may also be modified
by binding various ligands, MAPs, etc. Then, given the fact that the genes for
$\alpha$ and $\beta$ tubulins are
rather complex, providing a varying primary tubulin structure, \eg, at least
17 different $\beta$-tubulins can exist in mammalian brain MTs, one easily sees
that the number of different possible combinations of tubulin states and thus
the {\em information capacity} within MTs may be very large indeed! It should
be stressed that proteins undergo conformational motions over a wide range of
time and energy scales. However, significant conformational changes related to
protein function generally occur within the $(10^{-9}-10^{-12})$ sec time
scale. The conformational changes are related to cooperative movements of
protein sub-regions and charge redistributions, thus {\em strongly linked} to
{\em protein function} (signal transmission, ion channel opening, enzyme
action, etc) and may be triggered by factors including phosphorylation, GTP
hydrolysis, ion fluxes, electric fields, ligand binding, and neighboring
protein conformational changes. In the case of MTs, the {\em programmable} and
{\em adaptable} nature of the tubulin conformational states can be easily used
to {\em represent}
and {\em propagate} information. Further evidence for some of the extraordinary
tasks that may be undertaken by the MTs, due to their specific fine structure,
is their fundamental role in {\em mitosis}, or {\em cell division}. The {\em
centriole}, as we discussed above, consists basically of two cylinders of nine
triplets of MTs each, forming a kind of separated {\bf T}. At some point, each
of the two cylinders in the {\em centriole} grows another, each apparently
dragging a bundle of MTs with it, by becoming a focal point around which MTs
assemble. These MT fibers connect the centriole to the separate DNA strands
in the nucleus, at the {\em centromeres}, and the DNA strands separate, thus
initiating cell division. Another, indeed extraordinary mechanism from the many
contained in Nature's magic bag of tricks! The interelation and parallelism
between MTs and DNA goes much further. The {\em centriole}, a rather critical
part of the {\em centrosome} or {\em MT's organizing center}, seems to be a
kind
of {\em control center} for the cytosceleton. Thus, it seems that we have two
{\em strategic centers} in a single cell: the {\em nucleus}, where all the
fundamental genetic material of the cell resides, controlling the cell's
heredity and governing the production of proteins, of which the cell itself
is composed! On the other hand, the {\em centrosome}, with the MT-composed
{\em centriole} as its chief component seems to control the cell's movements
and its organization. As DNA is the common genetic database containing
hereditary information, {\em microtubules} are {\em real time executives}
of dynamic activities within living cells. One may wonder at this point, that
while DNA's very suggestive double-helical structure enables it to possess a
code, the {\em genetic code} \cite{9}, nothing of similar caliber occurs within
microtubules. This is a false alarm! So, let us take things from the beginning.
One nucleotide of DNA is composed of three elements: a {\em base}, {\rm
ribose}, and {\em phosphate group}. Four types of bases are present:
Adenine (A), Thymine (T), Guanine (G), and Cytosine (C), belonging to two
basic categories, a purine base (A,G) and a pyrimidine base (T,C). Nucleotides
are inteconnected by hydrogen bonds organizing them in a specific double-helix
structure (A=T, G$\equiv$C). From the aspect of organization of structure, one
such double-helix may be considered as an {\em aperiodic crystal}. ``Aperiodic"
signifies the irregular interchange of bases {\em inside} the helix, while the
phospates and riboses are located on the outside making up a {\em periodic
crystal} structure. The irregular repetition of bases within the helix
represents properties of the living beings which make sense, from an
information
point of view, only as {\em code system}. In the genetic code, one triplet of
bases, the {\em codon}, codes one aminoacid. The basic genetic code is coded
by 20 aminoacids and there exists a ``stop" as three more codons. Thus, there
exist 61 codons which code 20 aminoacids, from the $4^3=64$ possible
combinations of four bases of triplets. Then, the messenger RNA (mRNA) is
synthesized from the one strand of the DNA double helix, while the other strand
of the double helix remains in the nucleus making possible the synthesis of
another chain of DNA. The complete genetic information is preserved and remains
inside the nucleus. From mRNA through carrier RNA (tRNA) to ribosomal RNA
(rRNA) there is a continual transmission of the genetic information message,
making in effect proteins, the other side of the genetic code. One crucial
point to emphasize here is the following \cite{55}: it is well known that the
protein's catalytic or other functions strongly depends on its {\em exact
3-dimensional}
structure, thus making it a Tantalian job to try to exactly reproduce
genetically a protein! Nature, though, is more subtle. All a gene has to do is
to get the {\em sequence} of the aminoacids correct in that protein. Once the
correct polypeptide chain has been synthesized, with all its side chains in the
right order, then following the laws of quantum mechanics, called Chemistry in
this particular case, the protein would fold itself up correctly into a unique
3-D structure. A difficult 3-dimensional (reproduction) problem has been recast
as a much easier attractible 1-dimensional one! A very good lesson to be
appreciated and remembered and maybe to be used in other similar circumstances.

Until recently, it was widely believed that MTs were just base elements of the
cytosceleton and that they played a role in the mitotic spindle and active
transport. More careful study of the MT's structure, notably by Koruga
\cite{56}, showed
that MTs possess also a {\em code system}! One should not be surprised by such
a finding. Recall that the two different {\em conformational states} of a
tubulin dimer can switch from one to the other, due to alternative
possibilities for their {\em electric polarization}. Clearly, the state of each
dimer would be influenced by the polarization states of each of its six
neighbors, due to the Van der Waals forces between them, thus giving rise to
certain specific rules governing the conformation of each dimer in terms of
the conformations of its neighbors. This would allow all kind of messages to
be propagated and processed along the length of each microtubule. These
propagating signals appear to be relevant to the way that microtubules
transport various molecules alongside them, and to the various interconnections
between neighboring microtubules through MAPs. The repetitive geometric lattice
array of MT units may serve as a {\em matrix} of directional transfer and
transduction of biochemical, conformational, or electromagnetic energy.
It seems highly plausible that the
continuous grids of intramural MT could function as programable switching
matrices capable of information processing. Within neurons, transfer of MT
conformational charge or energy state could be driven by travelling nerve
action potentials and/or associated transmittance $\rm Ca^{++}$ flux.
Such a view is supported by the fact that velocities of action potentials and
accompanying $\rm Ca^{++}$ flux ${\cal O}(10-100)$m/sec would result in time
intervals for 4nm tubulin subunit transfers of about $10^{-10}$ sec, consistent
with the observed nanosecond range of protein conformational oscillations
\cite{57}! Taking into account the intraneural MT density, the neural fraction
of the brain, and average neural firing rates, {\em parallel computing} in MT
coupled to action potentials could reach $10^{28}$ transfers/sec ({\em bits})
in the human brain!

Koruga observed \cite{56} that the hexagonal packing \cite{58} of the $\alpha$
and $\beta$ tubulin
subunits in MT with {\em 13 protofilaments} corresponds to {\em information
coding}. He noticed that hexagonal packing and face-centered cubic packing of
spheres have equal density and thus he used both to explain MT organization.
It is known that the $\rm Oh(\bar6/4)$ symmetry group describes
face-centered-cubic sphere packing and {\em derives} information coding laws
\cite{59}.
In the case of hexagonal packing, the centers of the spheres should lie on the
surface of a cylinder (with radius equal to the $\rm Oh(\bar6/4)$ unit sphere)
and the sphere values in the axial direction (lattice) of the cylinder by
order of sphere packing is the {\em same} as in the dimension in which
face-centered-cubic packing is done. There should be two kinds of spheres
(white and black) on the cylinder surface, but linked such that they have the
dimension value in which the face-centered-cubic packing is done, leading to
an ``helical symmetry". Amazingly enough, the MTs satisfy all these desiderata!
Thus, the MTs possess one of the best known \cite{59} {\em binary
error-correcting} codes, the 6-binary dimer $K_1[13,2^6,5]$, where the distance
between spheres in order of packing is 5 and with $2^6=64$ {\em words}!!! It
should be noticed that
information theory suggests that the optimal number of spheres (white and black
corresponding to, say, $\alpha$ and $\beta$ {\em monomers}) for
information processing is 11, 12, or 13! A rather amazing result, supported
further by the fact that 13 (=5+8) seems to be almost universal amongst
{\em mammalian} MTs. Thus 13 is our lucky number!
 In addition, symmetry theory suggests that on the surface
of a circular cylinder in axial direction of the MT, there must be a code of
length of 24 monomer subunits (or 12 dimers), the code $K_2[24,3^4,13]$
corresponding to a 4-dimer ternary sequence \cite{56}. It is under the
influence of the above discussed $\rm Ca^{++}$-calmodulin ``complex" that
6-binary dimers of $K_1$ code
give 4-dimer ternary sequence of $K_2$ code, corresponding to biophysical
transfer of information from one point to other in MT, by transforming the
hexagonal surface organization into a new cubic state. Undoubtedly, microtubule
symmetry and structure are {\em optimal} for information processing. Thus
microtubules along with DNA/RNA are {\em unique cell structures} that possess
a {\em code system}, signifying their singularly important position. Like in
the case
of DNA/RNA, the specific structure of MTs led to the conclusion that they
possess code systems which can be utilized in the neuron dynamic information
activities, and other dynamical biological activities as well. It is very hard
to believe that the detailed, fine, paracrystalline MT structure, which,
among the many other useful functions, enables MTs to possess the
$K$-codes, is just accidental and parochial. It is not very hard to speculate
that, since the MTs are strongly involved in {\em exocytosis}, which is the
most fundamental
process that may somehow transform intentions/feelings/etc into neural action,
the $K$-codes may be used as a dictionary translating {\em psychological
``orders"} into {\em physiological actions}! In other words, the DNA/RNA
provide the {\em genetic code}, while the MTs provide the {\em mental code} or
{\em$K$-code}. As such, MTs become primary suspects for further investigations
concerning their possible role as the {\em microsites of consciousness}. One
should not worry that, at this stage of our investigation, the mechanism of
``{\em real time}" {\em regulation} and {\em control} by MT or other
cytosceletal filaments seems to be missing, because it will be provided soon,
once we study their physics in the light of {\em density matrix mechanics},
presented in the previous section.
Before we get to this fascinating subject, let us provide some further
phenomenological/experimental evidence that indeed neural MTs have to do a lot
with learning, memory, cognition, and thus, eventually, with consciousness ...

Our story starts thousands of millions of years ago, when the then popular
cytosceleton-less {\em procaryotic} cells became entangled with spirochetes
possesing whiplike tail composed of cytosceletal proteins. This, fortunate for
us, symbiosis produced the {\em eucaryotic} cells, possessing cytosceletons
\cite{60,3}. All this is well, but it has led to the following puzzle. {\em
Single} eucaryotic cell organisms, the protozoa, like the amoeba and the
paramecium, without possessing
a single neuron or synapse, still appear able of cognitive and adaptive
activities. Amoebae have been seen to hunt for food and paramecia to avoid
obstacles! How is this possible? The only logical explanation left is that the
{\em key structure} is the cystosceleton, including MTs, that act as the {\em
nervous system} of single cells, as has been observed almost half a century
ago, by the famous neuroscientist C.~S.~Sherrington \cite{61}. Indeed, the
paramecium seems to use its cytosceleton for coordinated action, in the form of
{\em metachronal waves}. Furthermore, metachronal waves of ciliary beating in
paramecea are {\em reversibly inhibited} by the general anesthisogon,
chloroform \cite{62}. In addition, it has been shown that signal transduction
in sensory cilia is due to propagating {\em conformational changes} along
ciliary microtubule subunits \cite{63}!

Further evidence, in modern times, that links the cytosceleton with cognitive
function is provided by the following findings:
\begin{enumerate}
\item Experiments with trained goldfish show that the drug colchicine produces
retrograde amnesia, by affecting memory fixation, through interference with the
MTs responsible for the structural modification of certain synapses \cite{64}.
\item Production of tubulin and MT activities correlate with peak learning,
memory and experience in baby chick brains \cite{65}.
\item Experiments with baby rats show that when they first open their eyes,
neurons in their visual cortex begin producing vast quantities of tubulin
\cite{66}.
\item Selective dysfunction of animal brain MTs by the drug colchicine causes
defects in learning and memory which mimic the symptons of Alzheimer's disease
(AD). It has been reported that in rats, continuous MT disruption induced by
chronic colchicine administration results in a dose-dependent learning deficit,
and retention is also impaired. It has also been stressed that these
colchicine-induced cognitive defects resemble those of AD, \ie, amnesia of
recent learning and loss of formerly established memories \cite{67}.
\item It has been hypothesized \cite{68}, and very recently supported by
detailed experimental studies \cite{69}, that impairment of MTs, leading to
tangled and dysfunctional neural cytosceleton, may be one explanation for the
pathogenesis of Alzheimer's disease (AD) \cite{70}.
\item In specific hippocampal regions of the brain of schizophrenic patients,
neuronal distorted architecture found due to a lack of 2 MAPs (MAP-2 and MAP-5)
\cite{71}.
\end{enumerate}

Arguably, we have plenty of evidence that, the cytosceleton, and in
particular the microtubules, have been rather instrumental through the whole
{\em natural evolution}, from the amoeba and paramecium to humans, and they
even helped or were deeply involved in {\em natural selection}. All these
facts,
I believe, make it difficult to justify the rather popular attitude of taking
the neuron as the fundamental, structureless unit and try to explain the brain
function from there on. An analogous attitude would be to try to understand
Chemistry by {\em only} accepting the existence of {\em structureless} a-toms,
in their original Democritean form. We can make a bit of progress but we cannot
go that far! The Pauli exclusion principle, of pure quantum mechanical origin,
seems
to play a rather fundamental role in understanding the periodic table, ... We
should come to terms with the {\em complexity} of the neuron, and we should not
treat it just as a {\em switch}. It will be wiser to concentrate on the
{\em nervous system} of the {\em neuron}, namely the {\em microtubule network}
\cite{1,3}. By avoiding taking this rather natural step, we are vulnerable to
the accusations of being micro-behaviorists or micro-functionalists, by
treating the whole neuron as a black box. Personally, I don't feel comfortable
with such an accusation!

So, let us concentrate now on the detailed structure of the neural MTs. Each
individual neuron, as being an eucaryotic cell, has its cytosceleton. Due to
the unfortunate for us, fact that neurons do not multiply after the brain is
fully formed, there seems to be no role for a {\em centriole} in the neural
cell. Indeed, centrioles seem to be absent in the neuron's centrosome, which as
usual, is found close to the neuron's nucleus. Neural MTs can be very long
indeed, in comparison with their diameter, of order of ${\cal O}(10{\rm nm})$
and can reach lengths of mms or more! There are about 450 MTs/$\mu^2$ or about
$7\times10^5$ tubulins/$\mu^3$, along the neural axon. Furthermore, as we
mentioned above, the potential {\em computing} brain power increases
substantially if the tubulin dimers (of characteristic two-state conformational
frequency of $10^{10}$Hz) are taken to be the basic computational units.
Indeed, in the case of the ``neuron unit", we get something like $10^{14}$
basic operations per sec (= $10^{11}$ neurons x $10^3$ signals/(neuron sec)),
while in the case of the ``tubulin dimer unit" we get something like $10^{28}$
basic operations per sec (= $10^{11}$ neurons x $10^7$ tubulin/neuron
x $10^{10}$ signals/(tubulin sec))! A rather remarkable gain on brain power
by replacing ``neuron-type" switches with ``{\em microtubular information
processors}", even if we reduce it for efficiency, non-participation, etc. down
to, say, $10^{25}$ ``bits". The neural MTs can grow or shrink, depending on the
circumstances, they transport neurotransmitter molecules, they are running
along the lengths of the axons and dendrites and they do form communicating
networks by means of the connecting MAPs. Neural MTs seem to be responsible
for {\em maintaining} the synaptic strengths, while they are able to effect
{\em strength-alterations} when needed. It also seems that neural MTs play a
fundamental role in organizing the growth of new nerve endings, piloting them
towards their connections with other neuron, thus contributing or being mainly
responsible for the formation of neural networks {\em in vivo}. Neural MTs
extend from the {\em centrosome}, near the nucleus, all the way up to the
presynaptic
endings of the axon, as well as in the other direction, into the dendrites and
dendritic spines, the postsynaptic end of the synaptic cleft. These dendritic
spines are subject to growth and degeneration, a rather important process for
{\em brain plasticity}, in which the overall interconnections in the brain are
suffering continuous and subtle changes, and as we discussed in
section~\ref{s4}, out of reach for the conventional neural networks (NN)
approach to brain function.
As a further indication for the involvement of neural MTs in {\em exocytosis},
or the release of
neurotransmitter chemicals from the presynaptic vesicular grid, Penrose has
emphasized \cite{3} the existence and role of certain substances, called {\em
clathrins}, found in the presynaptic endings of axons, and {\em associated}
with MTs. Clathrins are built from protein trimers, known as {\em triskelions},
which form three-pronged structures. The clathrin triskelions fit together
in an incredulous way, to form very beautiful configurations, basically
identical in general organization to the carbon molecules known as
``fullereness" or ``bucky balls" \cite{72}, but much bigger, since the single
carbon atoms are replaced by an entire clathrin triskelion involving several
aminoacids. Thus, clathrins have a very fascinating geometrical structure, of
a truncated icosahedron, that should be related to their important role in the
release of neurotransmitter chemicals.

If what is happening in the synaptic clefts, involving always microtubule
networks in a rather fundamental way both at the presynaptic and postsynaptic
stage, reminds you of the quasicrystals discussed at the end of
section~\ref{s3}, you
are right. Brain plasticity shares some similarities with quasicrystal growth
\cite{11}. Also, I do hope that I have presented significant evidence
indicating the direct
involvement of MTs in the control of brain plasticity, and thus coming to a
point, where the physics of MTs needs to be discussed.

\section{MicroTubules (MT) II: The physical profile}
\label{s7}
The remarkable biological/physiological properties of MTs discussed in the
previous section
is a typical example of the amazing {\em high degree of order} present in
biological
systems. Usually, bioscientists pay more attention to the functional
organization rather than to
the spatial/physical structure, but we should always remember that, if we would
like to
understand function we should study structure \cite{55}. The DNA story is a
good example at hand,
emphasizing the strong structure-function correlation \cite{17}. The basic
physical framework
for understanding biological order was put forward by Fr\"ohlich \cite{73}. As
we discussed in
the previous section, proteins are vibrant, dynamic structures in physiological
conditions.
A variety of recent techniques have shown that proteins and their component
parts
undergo conformational transformations, most significantly in the
``nanosecond"
$10^{-9}-10^{-10}$ sec range, as predicted by Fr\"ohlich. It should be stressed
that these
motions are {\em global changes} in protein conformation {\em rather than}
rapid
{\em thermal fluctuations} of side chains or local regions. About 25 years ago,
Fr\"ohlich
suggested \cite{73} that such {\em global} protein changes are {\em completely
triggered} by
charge redistributions such as dipole oscillations or electron movements within
specific {\em hydrophobic} regions of proteins. Hydrophobic regions within
proteins
are comprised of non-polar side chains of aminoacids which exclude water.
Incidentally,
and for later use, general anesthesia gas molecules apparently act there to
prevent
protein conformational responsiveness \cite{74}.  Fr\"ohlich's basic conjecture
was that
quantum-level events such as the movement of an electron within these
hydrophobic
regions act as a {\em trigger/switch} for the conformational state of the {\em
entire} protein.
The movement of an electron among resonant bond orbitals of aminoacid and side
chains such as aromatic rings of tyrosine, is a good example of Fr\"ohlich's
electrons. Fr\"ohlich considered an ensemble of high-frequency oscillators that
can be subjected to an external electric field and allowed to strongly interact
among themselves.
He conjectured that, if biochemical energy such as ATP or GTP hydrolysis were
supplied to the dipolar system, a new state would be formed that is
characterized by
a {\em long-range coherence}, as manifested by a macroscopic occupation of a
single
mode. He provided some physical evidence, that coherent excitation frequencies
in
the range $10^9-10^{10}$ Hz were possible in such biological systems. He
further
predicted metastable states (longer-lived conformational state patterns
stabilized
by local factors) and travelling regions of dipole-coupled conformations. Such
global
protein conformations appear suitable for computations: finite states which can
be
influenced by dynamic neighbor interactions. There is some experimental
evidence
for Fr\"ohlich's excitations in biological systems that include observations of
GHz-range
phonons in proteins \cite{75}, sharp-resonant non-thermal effects of GHz
irradiation on living cells \cite{76},
GHz-induced activation of microtubule pinocytons in rat brains \cite{77}, Raman
detection of
Fr\"ohlich frequency energy \cite{78} and the demonstration of propagating
signals in
microtubules \cite{79}. Fr\"ohlich's basic physical ideas \cite{73} seem to
make a lot of sense, but is
there any structure(s) that may realize them, or is it another theoretical
pipedream?
Lo and behold, microtubules just fit the bill. The entire MT may be viewed
within the
context of the Fr\"ohlich framework, as a regular array of coupled dipole
oscillators
interacting through resonant long-range forces. Furthermore, as we discussed in
the
previous section in detail, in the case of MTs we have an explicit mechanism
involving
the calcium-calmodulin ``complex" for the electron movement in the hydrophobic
pocket.
In addition, coherent vibrations within regions of an MT may take the form of
{\em kink-like
excitations} separating adjacent regions with opposite polarization vectors,
with the
dipole orientations in the direction of the MT axis. The extra energy needed
for the
creation of kink-like excitations may be provided by GTP hydrolysis, as
discussed
in the previous section. It is known that the energy produced during GTP
hydrolysis
is delivered to assembled MTs, although the precise manner in which this energy
is
utilized is still not understood. Amazingly enough, the free energy released in
GTP
hydrolysis is about 10Kcal/mole (0.42eV/molecule), or about the energy content
of
a kink-like excitation! Recently rather detailed and interesting studies of the
physics
of microtubules, at the {\em classical level}, have been undertaken by several
groups \cite{80,81,82},
as it is discussed next.

Microtubules are viewed as polymers of subunit proteins, the tubulins, and as
such they
may be considered as lattices of oriented dipoles. There are three types of
arrangements of dipoles in lattices: (i) random, (ii) parallel-aligned or {\em
ferroelectric},
and (iii) regions of locally frozen orientations or {\em spin-glass} \cite{28}.
As discussed in
section~\ref{s2}, depending on the values of the parameters characterizing the
system
(temperature and external electric field look the most relevant here) the
system may
exhibit different phases. In the ferroelectric phase, there is a long range
order (global
dipole alignment), encouraging the propagation of kink-like excitations and
thus able of
MT signaling and assembly/disassembly. On the other hand, the spin-glass phase
with its
locally frozen dipole orientations seems to be useable for efficient
information processing
and computations. So, it seems that the MTs organize cell activities by
operating in two
different phases, accessible by slightly changing the temperature and the
external
electric field. A rather remarkable operational biological system
\cite{1,56,80,81,82}.

The basic characteristics of the physical MT model, put forwad in
Ref.~\cite{80,81}, is that the
MT's strong uniaxial dielectric anisotropy align the dipole oscillators so that
they can
be effectively described by {\em only one degree of freedom}! In fact,
experiments
 have shown \cite{83} that a tubulin undergoes a conformational change induced
by GTP-GDP
hydrolysis in which one monomer shifts its orientation by $29^\circ$ from the
dimer's
vertical axis, as we discussed in the previous section.
Thus, the relevant degree of freedom, identifiable with an ``order parameter",
is the projection on the MT cylinder's axis of the monomer's displacement
from its equilibrium position. The mobile electron on each dimer, as discussed
in the previous section, can be localized either more toward the
$\alpha$-monomer or more toward
the $\beta$-monomer. The latter possibility is associated with changes in dimer
conformation, and thus we should identify the ``order" parameter with the
amount of $\beta$-state distortion when the latter is projected on the MT
longitudinal axis. Using the language of Quantum Mechanics (see
section~\ref{s3})
I will denote the two {\em conformational states} of the dimer as
$\ket{\alpha}$ and $\ket{\beta}$ referring respectively to the cases of the
mobile electron being on the $\alpha$- or $\beta$-{\em court} and with
$\ket{\alpha}\leftrightarrow\ket{\beta}$ the {\em quantum transition} triggered
by the movement of the electron from the one court to the other. The archetypal
of a two-state quantum system indeed!
 The remarkable inherent symmetry of a MT enables one to
view it effectively as nearly perfect {\em one-dimensional crystal}, and
thus including time, as a {\em highly symmetric 2-dimensional physical system}.
Furthermore, one should take into account the fact that the whole MT cylinder
represents a ``{\em giant dipole}". When the cross section of a MT is viewed
using electron microscopy, the MT's outer surfaces are surrounded by a ``clear
zone" of several nm which apparently represents the {\em oriented} molecules
of cytoplasmic water called sometimes ``{\em vicinal}" water, and enzymes.
It seems that the MT produces an electric field. Therefore, it is assumed that,
together with the polarized water surrounding it, a MT generates a nearly
uniform intrinsic electric field parallel to its axis. The existence of a
solvent in the environment of the MT, assumed for simplicity to be just water,
has some further consequences. The water provides a dielectric constant
($\epsilon\sim80$) that reduces the long-range electrostatic energy between
the dimer dipoles, and at the same time, it provides a viscous medium that
damps out vibrations of dimer dipoles.

All the above detailed physical structure is taken into account in a {\em
classical} mean field theory approach to the dynamics of the MT \cite{80,81}.
One mimics
the overall effect of the surrounding dimer-dipoles on a {\em chosen site $n$},
by qualitatively describing it by a double-well quartic potential, a standard
method, applied in the past rather successfully in similar systems, \eg, in
dipolar excitations of ferroelectrics \cite{84}. The potential then, for the
$\beta$-displacement $u_n(t)$ along the longitudinal symmetry ($x$) axis of
the MT cylinder, in the continuous limit $u_n(t)\to u(x,t)$, where $u(x,t)$
represents a {\em 1+1 dimensional classical field}, takes the form
\begin{equation}
V(u)=-\coeff{1}{2}Au^2(x,t)+\coeff{1}{4}Bu^4(x,t)
\label{31}
\end{equation}
with $B>0$ and $A=-(+{\rm const})(T-T_c)$, where $T_c$ denotes the critical
temperature of the system. The equation of motion then reads
\begin{equation}
M{\partial^2 u\over\partial t^2}-kR^2_0{\partial^2u\over\partial x^2}
-Au+Bu^3+\gamma{\partial u\over\partial t}-qE=0
\label{32}
\end{equation}
where $M$ denotes the mass of the dimer, $k$ is a stiffness parameter, $R_0$
is the equilibrium spacing between adjacent dimers, $\gamma$ is the viscous
water damping coefficient, and $E$ is the electric field due to the ``giant"
MT dipole discussed above, with $q$ the effective mobile charge of a single
dimer. Detailed studies \cite{80,81} of the dynamical equation (\ref{32}), in
the appropriate parameter range, have revealed very interesting
results/properties.
Indeed, for temperatures below the critical temperature $T_c\approx300^\circ$K,
the coefficient $A$ in (\ref{31}) is positive, thus putting the system into
the ferroelectric phase, characterized by long-range order, \ie, all dipoles
aligned along the MT longitudinal direction. In this phase, Eq.~(\ref{32})
admits travelling waves in the form of displaced {\em classical} kink-like
solitons with {\em no energy loss} \cite{85}. The kink-like soliton propagates
along the
protofilament with a fixed velocity $v$, which for $T<T_c$, \ie, in the
ferroelectric phase is well approximated by \cite{80,81}
\begin{equation}
\vec v\approx 2\times10^{-5}{\rm(m/sec)} \vec E/(1{\rm V/m})
\label{33}
\end{equation}
implying, for a characteristic average value of $E\approx10^5$V/m,
$v\approx2$m/sec and thus a propagating time, from one end to the other of
an ${\cal O}(1\mu)$ MT, $\tau\approx5\times10^{-7}$ sec. As (\ref{33})
suggests, the kink-like soliton travels preferentially in the direction of the
intrinsic electric field, thus transferring the energy that created it, \ie,
chemical GTP-GDP hydrolysis type energy, towards a specific end where it can
be used to detach dimers from the MT, in accordance with experimental
observations \cite{86}, concerning the assembly/disassembly of MTs \cite{87}.
The role of MAPs,
the lateral cross-bridging proteins, as MTs {\em stabilizers} becomes clearer
now. From the physical point of view, these bridges represent {\em lattice
impurities} in the MT structure, and it is well-known that impurities play a
very important role in soliton propagation. Kinks may be totally reflected
by an attractive impurity, for a specific range of the kink propagating
velocities, thus MAPs may significantly reduce the MT disassembly. Furthermore,
the addition of an {\em external} electric field introduces a new control
mechanism in the MT dynamics. As (\ref{33}) suggests, depending on the relative
direction and sign of the two fields (external versus internal) the kink-like
solitons may travel faster or stop altogether! Here we have a mechanism that
turns MTs to artificial information strings \cite{80,81,82}. Each kink-like
soliton can be
viewed as a bit of information whose propagation can be controlled by an
external electric field! Nevertheless, while the ferroelectric phase can be
useful for signaling and the assembly/disassembly of MTs, it is to ``straight"
for {\em information processing} and {\em computation}! For such operations one
has to move to the spin-glass phase \cite{28}. Detailed studies show \cite{81}
that as we increase
the temperature above the critical one $T_c$, while keeping the electric field
at appropriate small values, the coefficient $A$ in (\ref{31}) becomes
negative, signaling the formation of a {\em metastable} phase, the spin-glass
phase, before eventually reaching the naively expected {\em random} phase,
where all dipoles are distributed randomly. To understand the existence and
properties of the spin-glass phase better, it helps to notice that an MT, as
a regular array of coupled local dipole states, can be mapped to an anisotropic
two-dimensional {\em Ising model} \cite{13} on a {\em triangular lattice}, so
that the effective Hamiltonian is
\begin{equation}
H=-\sum j_{ik}\sigma_i\sigma_k
\label{34}
\end{equation}
with the effective spin variable $\sigma_i=\pm1$ denoting the dipole's
projection on the MTs longitudinal axis, and the exchange constants $j_{ij}$,
representing the interaction energy between two neighboring lattice sites, are
given by
\begin{equation}
j_{ij}={1\over4\pi\epsilon}\left({3\cos^2\theta-1\over r^3_{ij}}\right)\ p^2
\label{35}
\end{equation}
In (\ref{35}), $p$ is the dipole moment $p=qd$, where $d\approx4$nm; $r_{ij}$
is the distance between sites $i$ and $j$, and $\theta$ is the angle between
the dipole axis and the directions between the two dipoles. Explicit
calculations using MT X-ray diffraction data, have succeeded to determine all
relevant parameters ($j_{ij}$, $\theta$, and $r_{ij}$) relevant to the MT
system and be found in Ref.~\cite{81}. As is well-known \cite{28}, such a
system is able to exhibit {\em frustration} in its ground state, \ie, there
will always be a
conflict between satisfying all the {\em energy requirements} for the ``+"
bonds (two-parallel dipoles) and ``--" bonds (parallel-antiparallel dipoles).
That leads to the {\em spin-glass phase} where spin orientations are locally
``frozen" in random directions due to the fact that the ground state has a
multitude of equivalent orientations. For each triangle, reversing the spin
on one side with respect to the remaining two leads to an equivalent
configuration. In a MT with about $10^4$ dipoles or dimers the degeneracy of
the ground state is of the order of $6^{10,000}$, a very large number indeed!
Small potential barriers separating the various equivalent arrangements of
spins play a fundamental role. Relaxation times are very long for the various
accessible states giving them {\em long-term stability}! All these properties
of the spin-glass phase makes it optimal for computational applications. The
spin-glass phases allow easy formation of {\em local} ordered states, each of
which
carries some information content and is relatively stable over time, thus the
perfect candidate for information processing and computation. It is highly
remarkable that tubulin subunits in closely arrayed neural MTs (450 MT/$\mu^2$)
have a density of about $10^{17}$tubulins/cm$^3$, very close to the theoretical
limit for charge separation \cite{88}. Thus, cytosceletal arrays have {\em
maximal} density for information storage via charge, and the {\em capacity} for
{\em dynamically coupling} that information to mechanical and chemical events
via {\em conformational
states} of proteins. Furthermore, the switch between the different phases ({\em
ferroelectric}, {\em spin-glass} and {\em random}) is achieved through various
physical
means, \eg, temperature or electric field changes, both within easily
attainable physiological conditions! For example, as the intrinsic electric
field is raised above, about $10^4$V/m, easily attainable in MTs, the MT state
switches from the spin-glass to the ferroelectric phase. While the similarities
between the equations (\ref{34}) and (\ref{17}) as well as between the brain
function phases of section~\ref{s2} and the MT phases discussed here, are
striking and
rather suggestive, some further steps are needed before shouting {\em eureka}.

The treatment of MT dynamics \cite{80,81} presented above is based on {\em
classical}
(mean) {\em field theory}. For some physical issues this is an acceptable
approximation, given the fact that MTs may sometimes have macroscopic
dimensions. On the other hand, our main purpose would then evaporate, since the
central issue of quantum coherence and its loss would remain mute and its
relevance or not to brain function would remain unaswered. Usually, after the
classical treatment of a system, one goes directly to {\em quantize} the
dynamics of the system in a standard way. Alas, things here are not so easy.
We have seen that there are very important, {\em dissipative, viscous} forces,
due, for example, to the existence of water molecules that play a very
important role in the support and propagation of classical kink-like solitons,
but on the other hand, as is well-known, render the possible quantization of
the dynamical system, rather impossible! Amazingly enough, very recently
\cite{89} together with N.~Mavromatos
we have been able to map the 1+1 dimensional MT physical model discussed above
to a 1+1 dimensional {\em non-critical} string theory \cite{90,91}, the
precursor of the 1+1 black-hole model \cite{49} discussed in section~\ref{s5}.
Should we be surprised by
such a mapping? Probably, not that much. To start with, there are not that many
different theories in 1+1 dimensions, and even seemingly completely different
theories may belong to the same {\em universality class}, discussed in
section~\ref{s2}, implying very similar physical, ``{\em critical}" properties.
In fact, the possibility of casting the 1+1 dimensional MT dynamics in the,
rather
simple, double-well quartic potential form (\ref{31}), stems from the
well-known equivalence \cite{13}between such a quartic potential and the
one-dimensional
{\em Ising model}, \ie, interacting one-dimensional ``spin" chains, similar to
the MTs protofilaments! Furthermore, one can ``derive" \cite{92} a
1+1-dimensional
non-linear $\sigma$ model (resembling the 1+1 dimensional, {\em non-critical}
string theory \cite{90,91}) as the infrared limit of the Heisenberg
(anti)ferromagnet model
(resembling the 1+1-dimensional MT electret). The consequences of such a
mapping of the 1+1 MT dynamics on to a suitable 1+1 {\em non-critical} string
theory are rather
far-reaching. All the interesting and novel results discussed in
section~\ref{s5},
when appropriately translated, hold also true for the MT system, including the
construction of a {\em completely integrable} 1+1 dimensional model for the
MTs, admitting {\em consistent} (mean-field) {\em quantization}. Furthermore,
the {\em completely integrable} nature of the MT system, implying the existence
of an ``infinity" of quantum numbers labelling the states of the system (like
the Black-Hole $W_{1+\infty}$ {\em hair} discussed in section~\ref{s5}), make
it possible
to store and eventually retrieve information in a {\em coherent way}. The
practically infinite dimensional degeneracy of the spin-glass ground state,
discussed above with its remarkable information processing/computation
abilities, is, of course, due to the available ``infinity" of quantum numbers,
characterizing the system. In any case, the {\em consistent quantization} of
the MT-dynamics/system, make the possible appearance of large-scale coherent
states, the MQS of section~\ref{s2}, not only plausible, but also feasible.
But, as
we discussed in detail in section~\ref{s5}, there is no ``closed" system in
Nature.
Because of the {\em Procrustean Principle} \cite{6}, a concise, synoptic
expression of the
spacetime foam effects, all physical systems are rendered necessarily ``open",
and thus eventually ``collapse". The MT system is no exception to the rule.
On the contrary, the above discussed mapping of the 1+1-dimensional MT dynamics
to a 1+1-dimensional {\em effective non-critical} string theory, as observed by
N.~Mavromatos and myself \cite{89}, simplify things considerably in this
context too! After all, the central issue of section~\ref{s5} was basically how
to take into account space-time foam effects in string theory and their
possible consequences, as coded in the EMN approach \cite{50,5,6}. Let us give
here a {\em very simplified} physical picture of what is going on.

More specifically, in the case of the MT system the {\em conformational,
quantum} transitions of the dimers ($\ket{\alpha}\leftrightarrow\ket{\beta}$)
create abrupt distortions of spacetime, thus enhancing the possibility of
creation and annihilation of virtual, Planck-size black holes. The Planckian
black holes interact with the MT system, through the global string
states\footnote{It should be remarked that the effective non-critical string
picture advocated in Ref.~\cite{89}, applies more generally to the case where
the {\em $W_2$-world} does not correspond {\em necessarily} to Planckian
states but describes complicated, yet unknown, biological effects in the
brain.} (the {\em$W_2$ world} of (\ref{27})), which {\em agitate} the MT system
in a
{\em stochastic way}, as described by (\ref{23}), but with a monotonic increase
in entropy (\ref{26}) supplying the MT system with a {\em microscopic arrow of
time}, badly needed specifically in biological systems, while allowing for
loss-free energy propagation (\ref{25}). Furthermore, the $W_2$ global states
lead to {\em synchordic collapse} (\ref{27}) with a time period $\tau_c$
(\ref{29}). While all these facts start painting a rather fascinating picture,
one may justifiably wonder that the brain, being a hot, wet, noisy environment,
is the complete {\em antithesis} of what is really needed for quantum effects
to develop! In other words, even if we could be able to produce a macroscopic
quantum state (MQS), would not be that {\em environmental effects} take over
and ``destroy" everything before any ``useful" quantum effects take place?
There are different ways/levels of answering this question in our framework
here. The MT-dynamics, including viscous water and all, can be {\em mapped}
to a {\em non-critical} string theory and as such MTs may be viewed as ``open"
systems obeying consistent quantum dynamics as contained in (\ref{23}). One
then is entitled, if so desired, to ignore completely the mapping, and just use
(\ref{23}) as a successful phenomenological equation describing the MT system,
but with all parameters entering (\ref{23}) determined appropriately
by the {\em physical environment}. One then hopes to reproduce most of the
interesting results mentioned above, without reference to the rather specific
and detailed quantum gravitational framework used above. In principle, I don't
see anythig wrong with such an {\em agnostic approach}, beyond losing some
predictive power. Nevertheless, it should be stressed that the amazing
shielding of the whole neuronal axon through the insulating coating of {\em
myelin}, as discussed in section~\ref{s4}, and the whole
astonishing fine paracrystalline structure of the MT network provide just the
right environment for the fluorishing of quantum effects. One may even wonder
if Nature, or more precisely {\em natural selection} supported throughout
evolution, all these fine structures in a random, parasitic way or, as I
believe, because they were needed to perform useful work. {\em Survival of the
finest}!

It is encouraging that further studies of the MT dynamics strongly
indicate that the MT's filamentous structure may be due to spontaneous symmetry
breaking effects, a la superconductivity, and provided further evidence for the
MTs' usefulness to support and sustain quantum coherence. Indeed, considering
the layer of ordered water outside and inside MTs, Del Giudice, \etal \cite{93}
proposed that the formation of MT's cylindrical structure from tubulin subunits
may be understood by the concept of self-focusing of electromagnetic energy
by ordered water. Like the Meissner (symmetry breaking) effect for
superconducting media, electromagnetic energy would be confined inside
filamentous regions around which the tubulin subunits gather. Del Giudice,
\etal \cite{93} showed that this self-focusing should result in filamentous
beams of radius 15nm, precisely the inner diameter of microtubules!
Furthermore, Jibu, \etal \cite{94}, have
proposed that the quantum dynamical system of water molecules and the quantized
electromagnetic field confined inside the hollow MT core can manifest a
specific collective dynamical effect called {\em superradiance} \cite{95} by
which the MT can transform {\em any incoherent, thermal} and {\em disordered}
molecular, atomic or electromagnetic energy into {\em coherent photons} inside
the MT. Furthermore, they have also shown \cite{94} that such {\em coherent
photons} created by superradiance penetrate perfectly along the internal hollow
core of the MT as if the optical medium inside it were made ``transparent" by
the propagating photons themselves. This is the quantum phenomenon of {\em
self-induced transparency} \cite{96}. {\em Superradiance} and {\em self-induced
transparency} in cytosceletal MTs can lead to {\em ``optical" neural
holography} \cite{1}. Neurons (and maybe other cells) may contain microscopic
coherent optical supercomputers with enormous capacity. Thus Jibu, \etal
\cite{94}, suggest that MTs can behave as {\em optical waveguides} which result
in coherent photons. They estimate that this quantum
coherence is capable of superposition of states among MT spatially distributed
over {\em hundreds of microns}! These in turn are in superposition with other
MTs hundreds of microns away in other directions and so on...

It seems to me that we have accumulated enough evidence to safely assume that
the MT structure and dynamics are not only, strongly supportive of the onset
of long-range quantum coherence, but they are also very protective of quantum
coherence, shielding it from standard physical environmental effects, modulo,
of course, the menace of the spacetime foam. So, finally we have in place all
the physical and biological facts needed to put forward our thesis about a {\em
unified theory} of the Brain-Mind dynamics promised in the Introduction.

\section{Microtubules and Density Matrix Mechanics (I): Quantum Theory of Brain
Function}
\label{s8}
Let us assume that an ``external stimulus" is applied to the brain. This, of
course, means that some well-defined physical signal, presumably representing
some form of information, interacts with the brain. The physical content of the
signal (energy content, ...) starts to ``straigthen up" the {\em relevant}
regions of the brain, as analyses of EEGs, discussed in section~\ref{s2}, have
shown
\cite{15}. In our picture, the detailed microstructure, both physical and
biological, of the MT network entails that this ``external stimulus" would
initially {\em trigger/cause coherent} vibrations of the {\em relevant} part of
the MT network. Eventually, it is most probable that the ``{\em prepared}", by
the external stimulus, quantum state of the system
$\Psi$ would be a quantum {\em superposition} of many states or many {\em
alternatives}, all taking place at {\em once}. This is extremely likely to
occur in the {\em spin-glass} phase with its huge degeneracy, thus basically
allowing the {\em relevant part} of the MT network to perform many-many
quantum (parallel) computations {\em at once}, while processing the data
contained in the ``external stimulus". After some time $\tau_c$, as given by
(\ref{28}) or (\ref{29}), and because of the {\em global} or {\em $W_2$-world
states} the {\em relevant} MT wavefunction would ``{\em collapse}" to one
specific state. The {\em $W_2$-world states} have forced the system to
``decide" what it wants to be, by triggering it to {\em choose one} among many
alternative states. Notice that since the MT network is rather {\em extensive},
from the {\em ``sensory" cortex} to the {\em association cortex} to the
{\em motor cortex} (see section~\ref{s4}), the whole process of
{\em input$\to$processing$\to$output} is well-coordinated/correlated through
the {\em magic} properties of the {\em chosen} quantum state. The dynamically
{\em emerging}, due to {\em synchordic collapse} (see (\ref{22})) {\em chosen}
state has all the desirable properties (see section~\ref{s2}), like long-term
stability and non-locality, as being one of the many possible states of the
spin-glass phase, to be of primary importance in {\em brain function} and
{\em ``decision making"}! Indeed as we have stressed numerous times by now,
one of the most important functional roles of the MTs, is their strong
involvement in {\em brain plasticity} and {\em exocytosis} (see
sections~\ref{4},\ref{6}).
MTs control the shrinkage or growth of dendritic spines ({\em brain
plasticity}) and by triggering the ``{\em unlocking}" of the presynaptic
vesicular grid, thus allowing {\em one} vesicle to ``fire" or emit its content
of neurotransmitters towards the synaptic cleft, they control exocytosis.
Certainly MTs are the {\em masters} of the {\em neurophysiological game}. The
whole {\em neurophysiological response} to the ``external signal" {\em depends}
on the specific form of the {\em chosen} state of the {\em relevant part} of
the MT network, which in turn, at least {\em partially} depends on the {\em
$W_2$-world states} in a {\em stochastic way} (see sections~\ref{5},\ref{7}).
That is how,
finally they may lead to {\em learning} or {\em memory recall} or, through
the motor cortex, to {\em action}, or {\em nothing}, as discussed in sections
\ref{s2},\ref{s4},\ref{s6}. It should be stressed that the biological/physical
properties of the
MTs, as discussed in sections~\ref{s6} and \ref{s7}, are rather suggestive of
their important
role in the brain function. The very existence of the $K$-codes \cite{56},
related to the MT {\em conformational states}, which in turn are {\em strongly
correlated} to {\em protein function} (see section~\ref{s6}) make it apparent
that
{\em everything}, from bioinformation transmission to memories lay down, to
decision making, to movement, is {\em MT-driven}, and thus, as mentioned above,
at least {\em partially}, {\em global states} or {$W_2$-world states}
dependent! Actually, I cannot refrain from recalling here the analogy between
{\em brain plasticity} and {\em quasicrystal growth} discussed in sections
\ref{s3},\ref{s4},\ref{s6}. In the case of quasicrystals, the ground state,
\ie, the state with
{\em minimal energy}, is determined by employing many-many alternatives {\em at
once}, \ie, parallel ``computations" of energy considerations {\em at once},
depending, of course, on the physical environment, \eg, solvent, etc, until the
quasicrystal grows enough, so that {\em synchordic collapse} occurs, with {\em
one} final macroscopic state possible, the one that the experimentalists look
at \cite{11}! In the case of {\em brain plasticity}, including dendritic spine
growth and shrikage, a very similar situation occcurs, though now we are
dealing with a much more involved situation where many {\em minimization
conditions} have to be satisfied simulataneously, corresponding to the very
complex nature of the brain, and thus in a way, make {\em imperative} the
posssibility of {\em quantum computation}, as provided by the {\em MT network}
in a {\em stringy modified quantum mechanics} or {\em density matrix mechanics
framework}!

While the above emerging quantum theory of brain function has several
suggestive and {\em qualitatively} sound features, it would be nice to be able
to make some {\em quantitative statements} as well, in other words work out
some predictions or even postdictions! Indeed, this is possible. To start with,
in order for this new dynamical theory to ``hold water" at all, we first
have to check whether the very phenomenon of {\em exocytosis} is of quantum
nature, as we claim, or whether it can be explained on the basis of statistical
or thermal fluctuations. Well, the answer is on our favor. Ecccles \cite{8} and
Beck and Eccles \cite{97} have shown that {\em exocytosis} is a {\em quantum
phenomenon} of the presynaptic vesicular grid. They noticed that the synaptic
vesicles forming an hexagonal array, are packaged between the presynaptic dense
projections in a triangular array, composing the presynaptic vesicular grid,
having {\em paracrystalline properties} \cite{98}. Any similarity with the MT
hexagonal paracrystalline structure is not accidental, since the {\em boutons}
are the end points of MTs! There are about {\em 10,000 vesicles} per synaptic
unit or {\em bouton}, of which only (30--80) belong to the ``{\em firing zone}"
of the paracrystalline presynaptic vesicular grid \cite{98} and of which,
{\em only one} (1) ``fires" about (5,000--10,000) neurotransmitter molecules,
in a {\em probabilistic} ($\sim0.3-0.4$) way. Thus, the probability of quantum
(vesicular) emission is a {\em holistic} property of the presynaptic vesicular
grid of the bouton! Actually, they further noticed \cite{8,97} that this
probability {\em is not a fixed number}, but can be increased or decreased by
physiological or pharmacological treatment \cite{99}! This is exactly what the
doctor ordered. Indeed, one can {\em schematically} identify the {\em prepared
state} $\Psi$, discussed with the one represented by (\ref{13}), where the
$\ket{k}_i$ refers now to the specific $\ket{\alpha}$ {\em or} $\ket{\beta}$
conformational state of the $k$-th tubulin dimer in the $i$-th relevant
macroscopic (MT-network) quantum state, and $N$ is the number of tubulins
involved. Then, the system suffers {\em synchordic collapse}
\begin{equation}
\ket{\Psi}=\sum_i c_i\ket{1}_1\ket{2}_i\cdots\ket{N}_i\longrightarrow
\rho_{W_1}=\sum_{i=1} p_i \ket{1}_i\cdots\ket{N}_i\,
{}_i\bra{N}\cdots{}_i\bra{1}
\label{36new}
\end{equation}
where $\rho_{W_1}$ has been discussed in Section~\ref{s5} (just above
(\ref{23})), with the $p_i$ denoting {\em probabilities} depending in an
{\em stochastic way} on the {\em $W_2$-world states}. Since, the MT network
extends all the way to the relevant vesicular grids, it becomes apparent that
we expect a {\em synchordic}, {\em simultaneous} (EPR-like \cite{21,11}),
{\em probabilistic} ``firing" of all the {\em boutons} involved, triggering
thus the appropriate standard neurophysiological action!
Thus, not only do we expect {\em quantum exocytosis} to occur, but we
also do expect to be able to {\em influence}, through the {\em stochasticity}
brought by the {\em global} or {\em $W_2$-world states}, the probabilistic
outcome, allowing thus for (see below) {\em free-will}! And indeed it is
happening \cite{8,97,98,99}. So far so good. Another immediate prediction or
natural expectation, that one has in this new dynamical theory, concerns the
{\em time difference} between say an ``{\em external order}" and ``{\em
action}". According to our new picture advanced here, there is some time-lapse
between the input and the output, characterized {\em mainly} by $\tau_c$, the
quantum coherence lifetime, as given by (\ref{28}),(\ref{29}), \ie, the time
that takes for information processing and quantum computation. The way that
(\ref{29}) has been derived should make it clear that it was meant to apply
in the MT network system! The only thing we are missing is the value of $N$.
It seems to be a consensus, very rare in Brain Science, that the basic
{\em module} of $10^4$ neurons, discussed in section~\ref{s4}, should be able
to ``decide" something useful! In this case
\begin{equation}
N\approx 10^4{\rm neurons\over module}\cdot 10^8 {\rm tubulins\over neuron}
\cdot 10^5 (Z_{\rm tubulin}) (10\%\ {\rm efficiency})\approx10^{16}
\label{36}
\end{equation}
implying, when inserted in (\ref{29})
\begin{equation}
\tau_c^{\rm``Brain"}\approx{\cal O}(1\,{\rm sec})\ ,
\label{37}
\end{equation}
a rather long time compared to the neuron cycle-time of about {\rm (1--2)
msecs} and to {\em neurosignal} velocity of about 100 m/sec, as discussed in
section~\ref{s4}. Let me stress at this point that the rather long time of
${\cal O}(1\,{\rm sec})$ should not be compared with {\em cerebellum} guided
reflections, as discussed in section~\ref{s4}, of much smaller reaction time,
since they have become of second nature and there is no ``{\em thinking}" or
``{\em decision making}" involved. For the skeptical reader, who may feel
queasy with our philosophy to use the nucleon mass ($m_{\rm nucleon}\approx 1$
GeV) as the fundamental mass unit ($m$) in (\ref{28}) and thus yielding
(\ref{29}), we offer the following hopefully soothing remarks. It has been
noticed in \cite{89} that it is reasonable, in the case of an assembly of
tubulin dimers as in microtubules, to assume that the pertinent moving mass is
the effective mass $M^*$ of the kink background. This effective mass $M^*$
has been estimated to be \cite{80} $M^*\approx 3m_{\rm nucleon}$! By inserting
now $M^*$ as the fundamental mass unit in (\ref{28}), where $N$ denotes, in
this interpretation, the number of tubulin dimers $N_T\approx10^{12}$, as
provided by (\ref{36}), we get {\em again} $\tau^{\rm ``Brain"}_c\approx{\cal
O}(1\,{\rm sec})$! For yet another way, the third way of reproducing (\ref{37})
see \cite{89}. So, we feel kind of confident that (\ref{37}) provides indeed
a rather indicative, {\em canonical} value of the time lapse needed, in our
scheme,
for an ``{\em event}" to be perceived {\em consciously}, under normal
circumstances. Clearly, (\ref{28}),(\ref{29}) and (\ref{36}) spell out
{\em explicitly} the dependence on different parameters involved in getting
{\ref{37}) and thus enabling us to derive estimates for $\tau^{\rm ``Brain"}_c$
in circumstances different that the normal/canonical one discussed above.
Individual {\em conscious events} may occur at different time scales depending
on the number ($N$), effective mass ($M^*$), etc, of the tubulin dimers
involved in the {\em prepared} coherent state $\Psi$ (\ref{36new}). For
example, the ``$\gamma$-{\em oscillations}" (or ``40 Hz {\em oscillations}")
\cite{32,33}
discussed at the end of section~\ref{s4}, {\em may be} due to the successive,
synchordic collapses of an {\em extended} MT-network. Indeed, it is plausible
that the {\em relevant} MT-network involves either a bigger number of, or
longer, neurons than the {\em canonical} values used in (\ref{28}),(\ref{29}),
(\ref{36}) to yield (\ref{37}), thus enabling us to get in this case
$\tau^{\rm ``Brain"}_c\approx{\cal O}(1/50\,{\rm sec})$, without much sweat
and pain. It is too early yet, to get down to such specifics, and would be
foolhardy to claim that everything has been explained! Simply, it does not
seem inconceivable to be able to accomodate such ``$\gamma$-{\em oscillations}"
in our scheme, thus providing a microscopic, physical explanation to the
phenomenological Crick-Koch proposal \cite{35,31} that synchronized firing
in the ``$\gamma$-range" might be the neural correlate of visual awareness.
Generalizing this notion to other ``x-oscillations" we may naturally lead to
the solutionof the {\em binding problem} or {\em unitary sense of self}! It is
highly remarkable and astonishing the {\em synergy}, in our scheme, between
{\em Planck scale physics}, {\em atomic and subatomic physics}
providing the relevant parameters in (\ref{28}), thus leading to (\ref{29}),
{\em and} {\em Neurobiology} (\ref{36}), to eventually yield the estimate
(\ref{37}), seemingly in the right ballpark!
Indeed, as discussed in sections \ref{s4} and \ref{s6}, {\em learning} or {\em
memory laydown}, closely related to {\em brain plasticity}, involving shrinkage
or growth of dendritic spines are supposed to occur \cite{100} within ${\cal
O}(seconds)$, in amazing agreement with our prediction (\ref{37})!

Further evidence that our prediction (\ref{37}), and more generally, that our
new quantum theory of brain is making sense relies upon rather complicated
experiments, including clinical studies, that have been discussed in detail by
Penrose \cite{11,3}, so I will be rather concise. These are experiments
that have been performed on {\em human subjects}, and have to do with the time
that {\em consciousness} takes to act and to be enacted, \ie, they are
concerned with the {\em active} and {\em passive} role of {\em consciousness}
respectively. In the first one, performed by Kornhuber, \etal \cite{101} on a
number of human subjects volunteered to have electrical signals recorded at a
point on their heads, \ie, EEGs, and they were asked to flex their index finger
of their right hands suddenly at various times, at {\em free-will}. Averaged
over many trials, Kornhuber's experiments showed that the decision to flex the
finger appears to be made a {\em full second} or even {\em 1.5 seconds} {\em
before} the finger is actually flexed. Furthermore, if {\em free-will} is
replaced by reponse to the flash of a light signal, then the reaction time for
finger flexing is, at least, five times shorter than the {\em free-will} one!
In the second experiment, by Libet, \etal \cite{102}, subjects who had to have
brain surgery consented to having electrodes placed at points in the brain,
in the somatosensory cortex. The upshot of Libet's experiment \cite{102} was
that when a stimulus was applied to the skin of the patients, {\em skin-touch},
it took about ${\cal O}(second)$ before they were {\em consciously aware}
of that stimulus, despite the fact that the brain itself would have received
the signal of the stimulus in about {\em 0.01 sec}, and a pre-programmed
``reflex" response to such a stimulus could be achieved by the brain in about
{\em 0.1 sec}! Furthermore, cortical stimuli of less than ${\cal O}(\rm sec)$
are not perceived at all, and a cortical stimulus over ${\cal O}(\rm sec)$ is
perceived from ${\cal O}(\rm sec)$ onwards! It is even possible that a cortical
stimulus can even ``{\em backward mask}" an earlier skin stimulus, indicating
that {\em awareness} of the skin stimulus had actually not yet taken place by
the time of cortical stimulus. The {\em conscious perception} can be prevented
(``masked") by a later event, provided that the event occurs within ${\cal
O}(\rm sec)$. In addition, when a cortical stimulation lasting for more than
${\cal O}(\rm sec)$ is followed by a skin stimulation, within the original
${\cal O}(\rm sec)$, {\em both signals} were perceived, but in {\em reversed
order}! The subject would think that first was the skin-touch, followed by the
cortical stimulation, \ie, a {\em referal backwards in time} for the skin
stimulus of about ${\cal O}(\rm sec)$. Though for the cortical stimulation,
assumed to occur this time after the skin-touch, there is no {\em referal
backwards in time}, implying that this is {\em not} simply an overall error in
the internally perceived time. These are rather dramatic results with
far-reaching
consequences for the understanding of {\em consciousness} \cite{103,11,3}.
In our new dynamical theory they admit a rather simple and straightforward
explanation. Indeed, the Kornhuber type experiments \cite{101}, concerning
{\em active} consciousness, imply that indeed there is a time-lapse between
input$\to$output of about ${\cal O}(\rm sec)$ as suggested by (\ref{37}),
and not in the naively expected ${\cal O}(\rm msec)$ range from simplistic
``neurosignal" analysis. One may imagine, as discussed in detail above, that
the external stimulus, flex the finger at {\em free-will} in this particular
case, sets the {\em relevant preconscious state} ``{\em in gear}", and
eventually, through the involvement of {\em global or $W_2$-world states},
the ``collapse" of the wavefunction occurs, leaving only one specific state,
the {\em conscious state}, that carry the order to {\em physiologically} flex
the index finger! The strong correlation between {\em free-will} and the
{\em global or $W_2$-world states} should be apparent. Clearly, if {\em
free-will} is replaced by {\em reflective response} to an external stimulus,
then we expect much smaller reaction time, since basically there is no
conscious thinking involved and thus the situation is very similar to
cerebellum reflective actions. Concerning the Libet type experiments
\cite{102}, involving {\em passive consciousness}, again we can provide simple
explanations. Since it takes about ${\cal O}(\rm sec)$ for {\em conscious
perception} in our new dynamical theory, if the cortical stimulus is removed
in time less than ${\cal O}(\rm sec)$, we feel nothing, since presumably it
did not succeed to ``prepare" the {\em preconscious states}, thus it acts
simply like {\rm random} noise. On the other hand, if it lasts about ${\cal
O}(\rm sec)$,  then it is able to ``straighten" the {\em relevant} states up,
and thus it is able
to create {\em conscious perception}, that we ``feel" it! On the other hand,
the skin-touch, as more ``{\em real}" and {\em effective}, would always felt
after ${\cal O}(\rm sec)$, {\em except} when, during the ${\cal O}(\rm sec)$,
a {\em relevant} cortical stimulus is applied that {\em eliminates} the
skin-touch's efforts to ``prepare" a {\em preconscious state and} let it ``{\rm
run}"
or ``{\em compute}", to be more specific. In a way, since the cortical stimulus
is applied before the ``collapse" of the skin-touch related wavefunction,
quantum superposition, even if it is approximate, suggests that indeed
something like $|\Psi_{\rm skin-touch}+\Psi_{\rm cort.stim.}|^2\approx0$ is
possible, thus providing a possible quantum explanation to the ``{\em backwards
masking}" effect! Concerning the {\em referal backwards in time} puzzle, one
should recall that a {\em microscopic arrow of time}, presumably responsible
for the {\em consciously perceived time ordering}, past, present, future, is
{\em only} present in the EMN approach \cite{5,6,50}, and as such is strongly
correlated with the spontaneous collapse. The skin-touch case as more
``effective", involving more ``mass"/``energy" movement in its process ({\em
longest way}) may have a ``collapse" characteristic time $\tau_c$, as given in
(\ref{29}), smaller than the cortical stimulus case ({\em shortest way}), thus
because $(\tau_c)_{\rm s-t}<(\tau_c)_{\rm c-s}$, {\em independently} of the
time of their application, we always feel that the skin-touch occurred always
first! A rather interesting application of the EMN approach \cite{5,6,50}.
Incidentally, if this new approach to brain dynamics is right, one may
understand the famous {\em X-ism} phenomenon, referred to in section~\ref{s4}.
The neurons seem to follow the principle of the {\em longest possible
path}, because in such a case they {\em activate} the most ``mass"/``energy"
movement possible, thus {\em shortening} the ``decision" time $\tau_c$ given
by (\ref{28}) or (\ref{29}), thus contributing better to hierarchical and
non-local actions of the brain. This kind of microphysical explanation is, of
course, supportive of an evolutionary natural selection, where in this case
survival of the fittest reads survival of the longest neuron ... It should not
be very surprising that the modern man is around only 50,000 years and that the
dawn of civilizations was about 10,000 years ago! It is a lot of
fine-complicated structure to put together, starting from the very simple
amoeba or paramecium and eventually evolving to humans with their extremely
long microtubule networks.

Another very suggestive key feature, supporting further the eminent direct
connection between coherent MT conformational oscillations and the emergence
of {\em consciousness}, is the fact that {\em absence} of conformational
oscillations, as caused by general anethesia molecules, leads to {\em loss}
of consciousness \cite{74,3}. We have already discussed in section~\ref{s6}
the case of {\em reversible inhibition} of of paramecium's methachronal
waves by chloroform \cite{62}. {\em Metachronal waves} are paramecium's best
shot for a conscious event! What about higher organisms? It is rather
well-known
that brains of patients under general anethesia are commonly quite active:
EEG, evoked potentials and other brain functions persist despite lack of
consciousness. In a way, general anethesia, at the right level, implies {\em
absence} of consciousness. It has been suggested \cite{74,3} that, as
{\em anesthetic gas} diffuse into individual nerve cells, their electric
dipole properties (unrelated, in principle, to their ordinary chemical
properties) can interrupt the actions of MTs. They interfere through weak
Van der Waals forces, with the normal switching actions of the tubulins,
``blocking" the crucial tubulin electrons, as discussed in section~\ref{s6}.
It should be stressed that although there seems to be no generally accepted
{\em detailed} picture of the action of anesthetics, it is widely believed
that it is the Van der Waals interactions of these substances with the
conformational dynamics of the brain proteins that do the job. Here, the
relevant brain proteins are identified with the tubulin dimers consisting the
MT network. Such a detailed scenario for the workings of general anesthesia
seem to explain easily some of its key features. For example, it is a rather
remarkable fact that general anethesia can be induced by a large number of
completely different substances of no chemical affinity whatsoever, \eg, from
ether to chloroform to xenon! In our case it is just the electric dipole
properties of these substances that need to be similar and not necessarily
their chemical properties. Furthermore, if the general anesthesogon
concentrations are not too high, complete reversibility or recovery of
consciousness is achieved, indicating that the temporary Van der Waals
``blocage" of the crucial tubulin electron has ended and conformational
oscillations reoccur. On the other hand, general anesthetics, which are
known to bind to microtubules, at high enough concentrations cause their
depolymerization \cite{104}, implying in our picture partial or total
{\em irreversible} loss of consciousness. It is also known that anesthetics
may disrupt hydrophobic links among MAPs which interconnect MTs into functional
coherent networks \cite{105}. These, rather simple, in our framework,
explanations of certain {\em puzzling} features of general anesthesia provide
further positive evidence and credibility to our central thesis here, that
MTs are the microsites of consciousness. We have argued before that quantum
coherence in MT networks leads eventually, through synchordic collapse, to
conscious events, while we see here that {\em systematic, organized,
prevention} of quantum coherence, a la general anesthesia, leads to {\em loss}
of consciousness!

It is remarkable how well the MT's biological/physical structure  fits within
the density matrix mechanics framework. We were able not only to derive several
qualitatively interesting results, but as I showed above, we were able to get
some highly desirable {\em numbers} too! Nevertheless, we should not be carried
away and we should also not lose perspective of what we want to achieve, \ie,
how the {\em whole} brain works and what is {\em consciousness}, etc. There is
a cognitive hierarchy, and what we have showed is that the MT information
processing may provide the basement level, implying that everything else is
build upon it. The neuron synapse is the next layer up leading to yet another
layer, the {\em neural synaptic network} or {\em module}, that it is able to
operate cooperatively by utilizing dense interconnectedness, parallelism,
associative memory and learning due to synaptic plasticity, as we explained
above. At intermediate cognitive levels the motor and sensor {\em maps}
represent the body and the outside world, while the next to highest cognitive
level appears to be comprised of anatomically and functionally recognizable
brain systems and centers (\ie, respiratory center, ...). The highest cognitive
level is {\em global brain function}, which {\em correlates} with awareness,
thought or {\em consciousness}. Clearly, this hierarchical structure is {\em
susceptile} to {\em quantum treatment}, because of the very special dynamics
that characterize the MT network. In a way, one may consider the conformational
($\ket{\alpha}$ or $\ket{\beta}$) states of the tubulin dimers assembled in
microtubules, as the basic units of the quantum system. While the more
evolved hierarchical structures comprised of neurons, modules, modules of
modules, and, eventually the whole brain, may be viewed as the ``measuring
apparatus" providing the bulk of the ``mass"/``energy" needed in {\em
synchordic collapse}. Recall that, in the case if quantum mechanics discussed
in section~\ref{s3} (around (\ref{7})), it is only after the ``collapse" of
the wavefunction has occured that we are able to discuss with certainty,
``observable" properties of the system. Likewise, in our case here, it is
only after the {\em synchordic collapse} has occured that we can ``{\em feel}"
{\em consciously an event}. As we discussed above, it depends on the individual
conscious event, \ie, on the specifics of the {\em relevant} MT-network
involved, of how long is going to take before we ``{\em feel}" it.
Thus, we get in our scheme a dynamically organized {\em time-ordered}
appearance of conscious events, corresponding to the synchordic collapse
of the {\em relevant} MT-network involved, representing the very nature of the
event under consideration. At each instant, and in a {\em cohesive way}, the
``{\em sum}" of the conscious events consists of what we call {\em
consciousness}! If $c_i(t)$ refers to the $i$-th conscious event at time $t$,
then consciousness $C$ at time $t$ may {\em symbolically} be represented by
$C(t)=\sum_i c_i(t)$. This is how {\em consciousness emerges hierarchically}
in our dynamical scheme. It looks like, at each moment, we ``{\em read}" the
outputs ($c_i(t)$) of the different ``{\em microscopic measuring apparati}",
we ``{\em decide}" ($C(t)$) and we {\em proceed} accordingly, and so on, ad
infinitum, meaning here our lifetime span! A very simplistic analogy would be
the way we use the panel of our cars, with all its numerous indicators, showing
us, at each moment, how we are doing with gas, oil, temperature, water, etc,
and thus, ``forcing" us to ``decide" if we have to stop or not for gas, etc.
As I mentioned above, while discussing the phenomenon of ``{\em backwards
masking}" and ``{\em referal backwards in time}", {\em conscious time}, \ie,
past, present, future make sense only when it refers to {\em conscious events}.
In our scheme, conscious events are due to synchordic collapse which, as
discussed in section~\ref{s5}, introduces a microscopic arrow of time,
providing thus, naturally, time-ordering! It is amazing that the mechanism
that we have proposed \cite{51} to explain the origin and arrow of cosmic time,
applies all the way down to the MT-networks, explaining the origin and arrow
of consciousness. Putting it differently, in our scheme, the notions of cosmic
and conscious time are naturally identified as one may naively expect, and as
it was, since long, suspected.

So, we expect to see a kind of {\em fractal phenomenon} occuring in which we
have quantum coherence (and synchordic collapse)
extended over a MT, over hundreds of MTs comprising the neuron, over thousands
of neurons comprising the module, over tens of modules (incidentally explaining
the ``40 Hz oscillations" discussed above and in section~\ref{s4}), etc.
Actually, there is enough space in our dynamical, hierarchical scheme to
accomodate {\em neural networks} \cite{27,26}, attempts to use {\em
synchronized neural firing} \cite{32} in explaining the {\em binding problem}
\cite{35,31}, {\em and} eventually Neural Darwinism \cite{24}.
Eventually the whole brain is involved, one way or another, but coherently and
in a correlated way, subjected to {\em synchordic collapse}, thus explaining
the ``{\em binding problem}" or the ``{\em unitary sense of self}" problem.
Furthermore, the {\em stochastic nature} of the synchordic collapse, due to the
existence of the {\em global or $W_2$-world states}, provides a very plausible
explanation of {\em free-will}.

In order to see how our new dynamical theory of brain function, spelled out
in a rather detailed manner above, would work in practice, it would be
interesting and perhaps amusing to present a very simple example. Let us
consider (\ref{36new}), in the admittedly very unrealistic, case of only two
superimposed quantum states: $\Psi=c_1(t)\Psi_1+c_2(t)\Psi_2$, where $\Psi_i$
stands for $\ket{1}_i\ket{2}_i\cdots\ket{N}_i$, and with
$c_1(0)=c_2(0)={1\over\sqrt{2}}$. Then, if we denote by $\gamma$
the synchordic collapse frequency ($\gamma\equiv 1/\tau_c$ (\ref{28})), and
assume that the finally {\em chosen} state will be, say $\Psi_1$, then one
may deduce that \cite{41,42} $|c_1|^2=(1+e^{-2\gamma t})^{-1}$. In
Fig.~\ref{f1}, $|c_1|^2$ is plotted against time ($t$), for different values
of $\gamma$, corresponding, in our scheme, to rather indicative {\em
psychological} or {\em personality} states, providing thus our {\em
psychological} or {\em personality profile}! Depending on the value of
$\gamma$, the curves are {\em schematically} denoted as ``{\em visible}",
``{\em violet}", ``{\em ultraviolet}", and ``{\em infrared}". A common feature
of all these curves is the increase with time of $|c_1|^2$, until it reaches
some rather big (close to 1) value (say $\approx0.9$), at which point one
safely may assume that synchordic collapse is occuring. At this moment, we
pass from the {\em superimposed ($c_1\Psi_1+c_2\Psi_2$)} quantum state},
identified here with the {\em preconscious state}, to the {\em chosen
($\Psi_1$) state}, identified here with the {\em conscious state} or {\em
event}, \ie, we ``{\em feel}" it! Fig.~\ref{f1}(a) indicates a {\em normal
psychological state}, in which things happen in a straightforward way as
represented by the canonical, standard (``{\em visible}") value of
$\gamma=1\Hz$, corresponding to $\tau_c^{\rm``Brain"}\approx{\cal O}(1\,{\rm
sec})$ (\ref{37}). Fig.~\ref{f1}(b) indicates {\em excitement} (``{\em
violet}"), in other word things are happening quicker by involving, maybe,
more tubulins (increase $N$ in (\ref{36}) and thus (\ref{28},\ref{29})
increasing $\gamma$, say $\gamma=2\Hz$ or $\tau_c^{\rm``Brain"}\approx{\cal
O}(0.5\,{\rm sec})$. Clearly, in this case there is less time for quantum
computations, and maybe, not enough time for very wise ``decisions", thus we
may start acting a bit incoherent in the social sense! This case gets much
worse in the presence of ``stimulants", where maybe many more than the usual
tubulins get involved and thus the synchordic collapse frequency gets much
bigger (``{\em ultraviolet}") disrupting, eventually, complete ``collapse",
as schematically indicated in Fig.~\ref{f1}(c). In this ``{\em high}" {\em
state} \cite{106}, while we are ``closer" to a coherent quantum superposition,
we clearly act in a completely incoherent, and thus unacceptable, social way.
On the other end of the {\em synchordic collapse} frequency sector, in the
``{\em infrared}" limit, lies the {\em dream state} as indicated in
Fig.~\ref{f1}(d). Indeed, during our sleep, basically by definition, the
brain is working in a very slow, subnormal mode entailing thus rather small
values of $\gamma$ (see Fig.~\ref{f1}(d)). In such a case, a quantum
superposition, initiated presumably in a {\em parasitic} way, may last much
longer than a {\em normal state} case, and thus, eventually, may get lost in
the {\em environmental} background, one way or another, before
suffering our specific {\em synchordic collapse}, the agent of {\em conscious
events}. That is why in most cases, we don't {\em remember} our dreams!
Furthermore, as we all know, when we dream of someone, the person in the dream
is usually a {\em mixture} of two or three rather similar people, read
quantum superposition of {\em relevant} quantum states in our scheme, and
eventually disappear without leaving any strong imprint in our memory, read
absence of complete synchordic collapse in our scheme! Of course, it may
happen, as in the case of not being quite asleep, that $\gamma$ gets close to
its ``normal" value (\eg, $\gamma\approx0.9$ in Fig.~\ref{f1}(d)), in which
case complete synchordic collapse is achievable and we do, then, vividly
remember our dream or nightmare! It is amazing and worth mentioning,that a
similar, but phenomenologically postulated picture explaining the {\em Dream
states}, or {\em Rapid-Eye-Movement} (REM) {\em sleep state}, has been put
forward in Ref.~\cite{107},\cite{31}(p.161-2). There, words like ``disturbed",
``superimposed", ``condensation" are used to describe {\em Dream states} in a
generic way, without any reference to Quantum Physics. Here we see that such an
explanation \cite{107,31} seems to emerge naturally from the quantum aspects of
our dynamical scheme.

It should be strongly emphasized that in order to be able to provide positive
evidence or refute our scheme, further experiments are badly needed and their
results eagerly awaited. MT dynamics have to be studied in {\em vivo} and
in {\em vitro}. We need to have a clear experimental picture about their
assembly and disassembly properties, including their growth; we also need to
have experimental information on which specific mechanism, if any, of the ones
that have been suggested, is responsible for sustaining quantum coherence of
the conformational states. We need further clinical studies of the ``{\em
funny}" time related phenomena. We also need to understand experimentally
{\em and} theoretically, the role played by the K-code(s) in bioinformation
processing, and their connection to the genetic code. Is it accidental that
both codes have 64 words? Is it accidental that MT-networks look suspiciously
similar to ``quantum computers"? Can we use them in {\em vitro} for quantum
computing? Is it accidental that microtubules, as participants in centrioles,
are partially responsible for {\em mitosis} or cell division, thus ``{\em
interacting}" directly with the DNA, maybe thus being able to bring in
{\em environmental information}, since MT-networks extend all the way to the
cell membrane? Is it accidental that both {\em DNA} and MTs, the unique
cellular structures known to posses a code system, are {\em effectively}
1+1 dimensional? Is it accidental that as we move from micro-organisms to
macro-organisms, the amount/length of {\em normal} and {\em selfish or junk}
DNA and the length of MTs do increase? Probably not, but we have to, and we
are going to find out.

\section{Microtubules and Density Matrix Mechanics (II): Quantum Psychophysics}
\label{s9}
Any scientifically sound theory of brain function, by its very nature, has not
only to provide a credible picture of what is happening at the very microscopic
(basic) level but it should also accomodate naturally all phenomena observed
at the very macroscopic (top) level, \ie, {\em personality} level as described
by {\em psychology}. Psychology is usually defined as the science of mental
life, where
the latter includes feelings, desires, intentions, cognitions, reasonings,
decisions, and the like. It is advisable and useful, for our purposes here, to
distinguish between {\em Jamesian psychology} \cite{7}, or psychology of the
{\em conscious}, and {\em Freudian}\footnote{Sigmund Freud (1856-1939), founder
of psychoanalysis and arguably the single most important figure in pointing
out the role of {\em unconscious processes} in our behavior and feelings.}
{\em psychology} \cite{108,109} or psychology of the {\em unconscious}. I use
here the term {\em Freudian psychology} instead of the, maybe, more proper
one {\em psychoanalysis} for the following reasons. As defined by Freud
\cite{108}, {\em psychoanalysis} falls under the head of psychology, not of
medical psychology, nor of the psychology of the morbid processes, but simply
psychology. Psychoanalysis is certainly not the whole psychology, but its
substructure and perhaps its entire foundation (unconscious$\to$conscious)!
But, {\em psychoanalysis} is also a method of psychotherapy, \ie, it consists
of techniques for treating emotionally disturbed people. Since this last
property of psychoanalysis is, commonly, the prevailing one, and since the
therapy shouldn't swallow up the science, I prefer to stick to the term
Freudian psychology, as the theoretical system, background of psychology, and
view psychoanalysis strictly as a method of psychotherapy. We describe next
the essentials of Jamesian psychology \cite{7} and how they fit in (or are
explained) within our scheme, which also seems able to accomodate the basics
of Freudian psychology \cite{108,109}, \ie, we will move from the {\em
conscious} to the {\em preconscious} to the {\em unconscious}! The relevance
of the connection of Jamesian views of consciousness to Copenhagen Quantum
Mechanics has been {\em repeatedly} and {\em forcefully} emphasized by H.~Stapp
\cite{12}.

The brain-mind interaction is of central importance in Jamesian thought
\cite{7}. James opposed, vigorously, sterile, (pseudo)scientific, prevailing
at his time, views purporting that feelings, no matter how intense that may
be present, can have {\em no causal efficacy} whatever. He counterattacked
by making a positive argument for the efficacy of consciousness by considering
its distribution. For James, consciousness is at all times primarily a
{\em selecting agency}, being present when choices must be made between
different
possible courses of action \cite{7}. Clearly, such distribution makes sense
{\em only} if consciousness plays a role, one way or another, in making these
selections. James went even further, developing his principal claim about the
unity of each conscious thought \cite{7}. It is the whole thoughts, he argued,
that are the proper fundamental elements of psychology, not some collection
of elementary components out of which thoughts are assumed to be formed by
aggregation. In other words, even if each thought has components, these
component thoughts are experienced together in a particular way that makes the
experienced whole an essentially new {\em emerging} entity! He even had the
courage to speculate that if all these properties were not to be born out
of his contemporary physics (what we now call Classical Physics), physics has
to be modified! All this activity was taking place in the 1890's!! \cite{7}
What a {\em wise} man, indeed. Coming back to the 1990's, it is stricking to
notice that James' views of consciousness are {\em mapped}, almost {\em
one-to-one} to our dynamical theory of brain function. Our {\em central
thesis} suggest, that every {\em conscious event} is the {\em psychological
counterpart} of a {\em related}, specific {\em synchordic collapse event} in
the brain, that triggers a specific neutral activity, described here by
MT-dynamics, strongly correlated and {\em quantum computably}, responding to
stimuli. An {\em isomorphism}, or a {\em one-to-one mapping} seems to emerge
between {\em conscious events}, in a generic sense, and {\em specific neural
patterns}, described by {\em specific} MT-networks, generated by, and thus
strongly dependent on, synchordic collapse. By, just, recalling that it is
synchordic collapse that causes the quantum MT-system to ``{\em decide}" its
course of action in a fundamentally {\em integrative character}, EPR-like
\cite{21,3} way, and using the isomorphism available in our scheme, one should
be able to reproduce, almost {\em verbatim}, the Jamesian views of
consciousness. If, James' proposal about consciousness is not the mental or
psychological version,
or counterpart, of our physical/physiological views about consciousness,
frankly, I don't know what would ever be. However, in order to complete our
{\em isomorphism} between mental events and neural patterns described by
MT-network states, we clearly have to discuss the preliminary phase that
``{\em prepares}" the specific set of superimposed MT quantum states, of
which {\em only one} is going to be selected or {\em chosen}. But then, we
naturally have been led to the domains of the other great master of modern
psychology.

Freud \cite{108} felt that consciousness was only a {\em thin slice} of the
total mind, that like an iceberg, the larger part of it existed below the
surface of awareness. He said that scientific work in psychology will consist
in {\em translating} unconscious processes into conscious ones, and thus
filling the gaps in conscious perceptions! He argued that the {\em personality}
is a complex and intricate {\em energy system} \cite{109}. The form of energy
that operates the personality and enables it to perform work is called  {\em
psychic energy}. He assumed that {\em psychic energy} comes from the energy
of the body, but he was agnostic on how this transformation takes place. He
insisted, though, that there is nothing mystical, vitalistic or supernatural
about the concept of {\em psychic energy} \cite{109}. It performs work as does
any other form of energy, but in this case is psychological work, thinking,
perceiving, and remembering. There is a continuously transformation taking
place of bodily energy to psychic energy and viceversa.

A mental event is conscious or not, according to Freud \cite{108,109},
depending upon the magnitude of energy invested in it and the intensity of the
resisting force! A person feels pain or pleasure when the magnitude of the
pain or pleasure exceed a certain {\em cathexis} value which is called the
{\em threshold value}. Likewise, (s)he perceives an object in the world when
the perceptual process is energized beyond a threshold value. Sometimes even
when the {\em cathexis} exceed the treshold, the feeling or perceptions may not
become {\em conscious} because of the inhibiting effects of an {\em
anti-cathexis} which prevents it from becoming conscious! Freud \cite{108,109}
differentiated between two qualities of {\em unconsciousness}, the {\em
preconscious} and {\em unconscious proper}. A {\em preconscious state} is one
which can become {\em conscious} quite easily because of weak resistance, and
in sharp contrast, to an {\em unconscious proper state} where the opposing
force is rather strong! Actually, there is a continuous {\em spectrum of
unconsciousness}. At the one end, ending at the {\em unconscious proper state},
there is memory that can never become conscious, because it has no association
with language, while at the other end, including the {\em preconscious state},
there is memory which is ``on the tip of the tongue".

Freud assumed that, since a relatively large concentration of energy in a
mental process is required in order for it to become conscious, we can be
conscious of only one thing at a time \cite{109}. However, the rapid shifting
of energy from one idea, memory, perception or feeling to another provides for
a wide range of conscious awareness within {\em a short time-lapse}! The
perceptual system is like a radar mechanism which rapidly scans and takes many
quick pictures of the world. When the perceptual system discovers a needed
object, or apprehends potential danger in the external world, it comes to rest
and focuses its attention upon the object or danger. Ideas and memories, \ie,
mental representations of past experiences, are summoned form the {\em
preconscious} to help the person adjust to the situation confronting him. When
the danger is past or the need is satisfied, the mind turns its attention to
other matters \cite{108,109}.

Concerning the nature of the ``{\em unconscious proper}", Freud
suggested \cite{109} that ``{\em threatening}" {\em events} could be {\em
repressed} in memory so that they were not ordinarily available for conscious
recall. Freud's analysis of {\em repression}, the selective inability to
recall, is a form of Darwinism (survival of the fittest) as applied to the
mental world to become the Freudian {\em suppression of the ``threatening"}.
``Threatening" events belong to the set of the ``{\em unconscious proper
events}". Freud developed further \cite{108,109} a theory about the fate of
the repressed events, connecting them, partially, to {\em dreams}! Dreams
are filled with {\em disguised} or {\em symbolic} representations of repressed
desires. When the disguise becomes too transparent, the dreamer usually wakes
up. Anxiety dreams and nightmares, for example, are caused by the emergence of
repressed desires which makes the person anxious. He noticed that somatically,
sleep is an act which reproduces intra-uterine existence,\footnote{There is,
presently, evidence to suggest that in the {\em womb}, especially in the
third trimester, {\em Dream or REM sleep} occurs more than 8 hours a day
\cite{107}.} fulfilling the
condition of repose, warmth and absence of stimulus. The feature characterizing
the mind of a sleeping person is an almost complete withdrawal from the
surrounding world and the cessation of all interest in it. Freud pictured
\cite{109} the situation which leads to dream formation as follows: the
{\em preconscious} dream-wish is formed, which expresses the {\em unconscious}
impulse in the material of the {\em preconscious} day-residues. This dream-wish
must be sharply distinguished from the day-residues, it need not have existed
in waking life and it may already display the {\em irrational character}, \eg,
a person in the dream is the {\em mixture} of two or three rather similar
people, etc, noticeable in all that is unconscious when we come to translate
it into terms of consciousness! The logical validity, freshness, and stunning
resemblance to our presently holding views about brain function, characterizing
{\em Freudian psychology} \cite{108,109}, are properties very hard to miss.
Since his time, ample evidence has accumulated from the study of neurosis,
hypnotism, and parapraxes to show that his basic views about the action of the
unconscious and its role in behavior, were essentially correct.

After our, hopefully, enjoyable and useful excursion to Freud-land, we have all
that is needed to complete the above-discussed {\em isomorphism} between mental
events and neural patterns, described by MT-quantum states. Freud's {\em
psychic energy} as opposed to {\em bodily energy} and the transformation into
each other, corresponds to the exchange of energy/interactions between the
{\em $W_1$-world} or {\em attainable physical world localizable states} and the
$W_2$-world or {\em global states}, as explicitly indicated in (\ref{27}).
Notice, as
(\ref{25}) explicitly shows, that there is {\em conservation of energy} in
our scheme! Furthermore, the ``preparation" of the {\em relevant superimposed}
MT-quantum states depends on the {\em nature} and {\em intensity} of the
stimulus, as discussed in sections~\ref{s2},\ref{s4},\ref{s7},\ref{s8}, \ie, if
it can ``easily" ``straigthen up" the {\em relevant} states, corresponding to a
{\em preconscious state}, or if, it can ``hardly" have any effect on the
states, corresponding to {\em ``unconscious proper" states}. In the case of
{\em
preconscious states}, identifiable with the {\em relevant superimposed quantum
states}, synchordic collapse follows easily, turning it into a {\em conscious
state}! In the case of {\em ``unconscious proper" states}, identifiable with
either isolated, not easily reproduced, or random states, nothing happens!
Clearly, there is a continous spectrum of quantum states from the ``{\em
preconscious}" to the ``{\em unconscious proper}". In the case that an
{\em ``unconscious proper" stata} gets ``prepared", then synchordic collapse
leads to Freud's ``threatening events". For example, while we sleep being
``off guard", ``{\em unconscious proper}" states may be partially and {\em
parasitically}
prepared, even in disguised form, and may lead to nightmares! On the other
hand, synchordic collapse, of variable effectiveness, of presumably partially
{\em parasitically prepared preconscious dream-states}, as discussed in the
previous section (see Fig.~\ref{f1}(d)) reproduces Freud's basic views about
dream formation discussed above. If our dynamical theory of brain function,
with its now completed {\em isomorphism} between mental events and MT-quantum
dynamics states has not reproduced, almost {\em verbatim} the basic elements
of Freudian psychology, I don't know what would ever do. Needless to say,
Freud's terms are psychological, while ours are structural. It is in this sense
that we consider the {\em mental world} somehow {\em isomorphic} to the
{\em $W_2$-world of physical global states} that help to ``{\em prepare}"
and eventually dismantle, by ``{\em synchordic collapse}", the {\em relevant
superimposed} MT-quantum states of the $W_1$-attainable physical world
constituted by tubulin-dimer conformational states, as depicted clearly in
(\ref{2}) and (\ref{27}). It should be mentioned here (see relevant discussions
in sections~\ref{s6},\ref{s8}) the rather fundamental role played by the
K-code(s) \cite{56} possessed by the microtubules, in advancing and completing
our {\em isomrphism} between the mental world and the $W_2$-world, by acting
as a dictionary translating {\em psychological orders} into {\em physiological
actions}. It is in this sense that I propose to call the K-code(s), the
{\em Mental Code}, playing in a way the role of the {\em genetic code},
but in the mental world. It should be stressed once more here
(see the appropriate discussion in section~\ref{s5}, between (\ref{22}) and
(\ref{23}), that there is nothing mystical or supernatural about the $W_2$
{\em world global states}, or the way they interact with the {\em $W_1$-world
attainable physical states}, except that, due to
their {\em delocalized nature}, sometimes, a bit different than normal, novel
properties may emerge! Through the above mentioned {\em isomorphism}, these
novel properties are transmitted to the mental world, which thus is an ({\em
emerging}) part of the physical world, but with ({\em inherited}) distinct
qualities. Notice further, that in particle physics at very high energies,
we only talk about electroweak interactions, and only at low energies we may
talk about ``{\em effective}" electromagnetic and weak interactions. Similarly
here and in a {\em unified theory sense}, we should talk only about the
{\em physical world} ($W$) when {\em all states}, {\em localized} and {\em
delocalized} are accounted for (\ref{2},\ref{27}), and only talk about the
{\em attainable physical world} ($W_1$) and the {\em mental world} and their
interactions, \ie, an {\em effectively emerging dual world} (\ref{1}), {\em
only} when
the delocalized states get truncated, which happens realistically most of the
time! Incidentally, if all these kind of (post) modern views sound pretty
drastic, let me remind you that {\em Empedocles} (490-430 B.C.), the famous,
ancient greek, presocratic philosopher, in his ``cosmic phantasy", ascribed to
the whole universe the same animistic principle as is manifested in each
individual organism! If he was not describing, in his way, the $W_2$-world
global, delocalized states, I don't know whatever would do better. He certainly
was the first complete {\em effective} dualist! Hopefully, this emerging
compromising
resolution of the age-old problem concerning the brain-mind relation, will
bring peace, once and for all, to the different quarters of {\em dualists}
and {\em non-dualists}, and avoid further {\em duels}! Nevertheless, as I
already mentioned in the Introduction (just after (\ref{2})), hard-core
{\em materialists} may, if they so wish, concentrate their attention on the
physical relation/transition between the {\em $W$-physical world} and the
{\em $W_1$-attainable physical world}. It is {\em immaterial} to me!

The interface between psychology and physics ({\em psychophysics}) has always
been rather interesting, though-provoking, challenging, sometimes
controversial, but certainly not dull. Before Darwin, man was set apart from
the rest of the animal kingdom by virtue of having a soul. The evolutionary
doctrine made man a part of nature, an animal among other animals. Man became
an object of scientific study, no different save in complexity, from other
forms of life. Literally at the same time (1860), Fechner founded the science
of psychology, by showing that the mind could be studied scientifically and
that it could be measured quantitatively. At about the same time, the physical
formulation of the {\em principle of conservation of energy}, notably by
Helmholtz, stating that energy is a quantity that can be transformed, but
it cannot be destroyed, had rather far reaching consequences for biology and
psychology. It made possible an even more radical view of man. This is the
view that man is an {\em energy system} which obeys the same physical laws
that regulate, say, the fall of an apple or electromagnetic phenomena. Thanks
to Freud's genius, the physical dynamics extended to apply to man's {\em
personality}, and not only to her/his body. This really amazing visionary step,
as taken by Freud, led to {\em dynamical psychology} \cite{108}, \ie, one that
studies transformation and exchanges of energy within the {\em personality}, as
well as between the {\em personality} and the body. It is an amusing
coincidence to
notice that Freud's chef d'oevre ``{\em The Interpretation of Dreams}"
\cite{108}, and Planck's revolutionary paper on energy quantization, {\em both}
appeared in 1900 (!), and {\em both} after considerable hesitation and
self-doubt!!! The dynamical scheme presented here, is nothing more than a
supermodest attempt to continue the psychophysical tradition described above,
by combining the most recent advances in quantum dynamics, as described in
non-critical superstring theory \cite{5,6,89}, with the amazing progress in
microtubules and their dynamics \cite{1}--\cite{4}. A {\em unified scheme} of
brain-mind dynamics emerges, consistent with all known laws of physics, notably
including the law of conservation of energy, and at the same time, providing
satisfactory answers to age-old problems such as what is {\em consciousness},
the {\em binding problem} or {\em unitary sense of self}, {\em free-will} and
the like, involving parts or the entire activity of the brain. Indeed, {\em
conscious thoughts} seem
to correspond to metastable states of the brain associated with particular
integrated patterns of neural excitations, that are {\em selected} by
{\em synchordic collapse}, from among a plethora of such neural patterns
described by MT-network states (quantum) mechanically generated according to
(\ref{3},\ref{4}). Since {\em synchordic collapse} is due to the truncation of
{\em global delocalized} states, our {\em consciousness} is nothing else but a
{\em localized} aspect of a global, integrative process. There is a new image
of man emerging, in which human  consciousness is placed in the inner workings
of a non-local global process that link the whole universe together, defying
classical physics and observations of usual everyday life. It seems, that we
are {\em intimately} and {\em integrally} connected into
the same global process that is actively creating the form of the universe,
as we suggested in \cite{51}, thus providing a whole new meaning to the,
presently fashionable, expression {\em global village}. There seems to be a
{\em central organizing principle} at work, essentially what I called {\em the
Protean Principle} at the end of my review ``As time goes by ..." \cite{6}.
This new view of man's place in the universe is an essential ``{\em paradigm
shift}". We are not {\em just} small, irrelevant, struggling for survival
creatures in a meaningless universe, but through our dynamically created
consciousness, {\em we participate actively} in the intrinsically global
process that forms the world around us. {\em We are brains with strings
attached!} I do believe that this, scientifically geared, ``paradigm shift" in
our {\em Weltanschauung}, or ``world view", is bound to have a tremendous
impact, but mostly presently unimaginable, in all forms of human behavior form
the individual to the social level. Some visionary people have already started
talking about the dawn of the {\em brain man}, at the dawn of {\em third
wave} \cite{110} of civilization, characterized by strongly declining muscle
work and fastly increasing brain work, that succeeds the ``{\em second wave}"
related to the industrial revolution of 300 years ago, and which in turn
succeeded the ``{\em first wave}" related to the agricultural revolution of
10,000 years ago,  This is just the {\em dawn} of the {\em Homo Quantum} ...

\section*{Acknowledgements}
It is a great pleasure to thank: my collaborators, John Ellis and Nick
Mavromatos for stimulating discussions on some of the topics discussed here and
the latter also for reading the manuscript; Jorge Lopez for discussions,
reading the manuscript, and help with the figure; Steve Kelley, Stuart
Hameroff, and E.~Roy John for informative discussions; David Norton for
discussions, encouragement and support, throughout this work; Skip Porter for
continuous interest in this project, and Gil Marques for the warm hospitality
extended to my wife and myself during our visit to Brazil. This work has been
supported in part by DOE grant DE-FG05-91-ER-40633.

\begin{figure}[p]
\vspace{6in}
\includegraphics{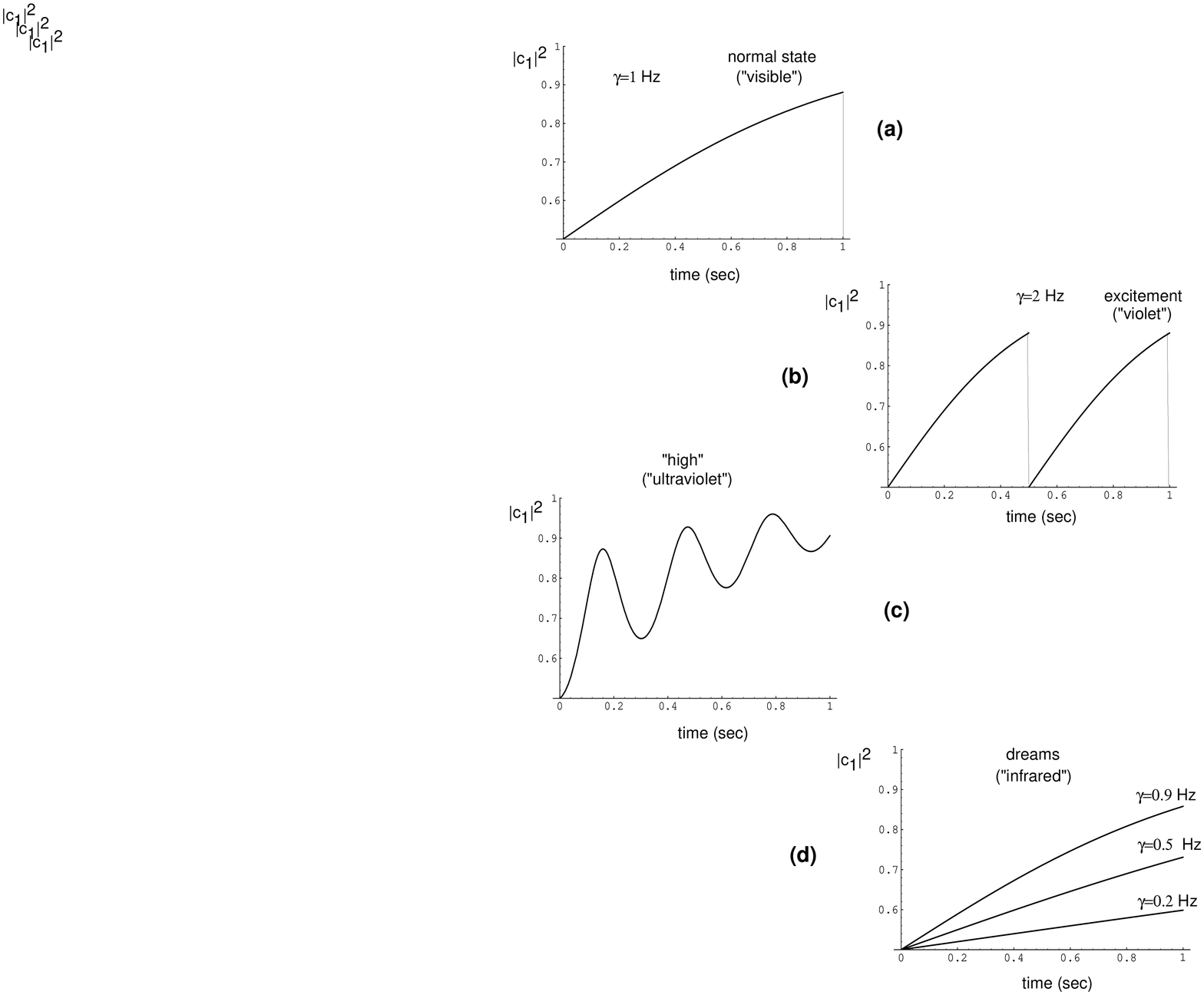}
\vspace{5cm}
\caption{{\em Psychological} or {\em Personality} profile as a function of
time, parametrized by different values of the MT-network synchordic collapse
frequency $\gamma$ ($\equiv1/\tau^{\rm``Brain}_c$), as indicated in (a) through
(d).}
\label{f1}
\end{figure}
\clearpage

\end{document}